\documentclass[aps,superscriptaddress,twocolumn,pre,longbibliography,floatfix]{revtex4-1}

\usepackage{amsmath,amsthm,amsfonts,amssymb,bm,graphicx,color,mathpazo,times, braket}
\usepackage[colorlinks={true}, citecolor={blue}, filecolor={blue}, linkcolor={blue}, urlcolor={blue}]{hyperref}
\usepackage[caption=false]{subfig}
\usepackage{graphicx}
\usepackage{graphicx}
\usepackage{soul}
\usepackage{amsmath}
\usepackage{soul}

\usepackage{orcidlink}
\newcommand{\mar}[1]{\textcolor{black}{#1}}

\newcommand{\abs}[1]{\left\vert#1\right\vert}

\allowdisplaybreaks
\begin{document}
\title{Single-qubit probes for temperature estimation in the presence of collective baths}
\author{Asghar Ullah}
 \email{aullah21@ku.edu.tr}
\affiliation{Department of Physics, Ko\c{c} University, 34450 Sar\i yer, Istanbul, T\"urkiye}
\author{Marco Cattaneo}
\email{ marco.cattaneo@helsinki.fi}
\affiliation{QTF Centre of Excellence, Department of Physics,
University of Helsinki, P.O. Box 43, FI-00014 Helsinki, Finland}
\author{\"Ozg\"ur E. M\"ustecapl\i o\u glu}	
\email{omustecap@ku.edu.tr}
\affiliation{Department of Physics, Ko\c{c} University, 34450 Sar\i yer, Istanbul, T\"urkiye}
\affiliation{T\"UB\.{I}TAK Research Institute for Fundamental Sciences, 41470 Gebze, T\"urkiye}

 \date{\today} 
%*******************************************************************************%
\begin{abstract}
We study the performance of single-qubit probes for temperature estimation in the presence of collective baths. We consider a system of two qubits, each locally dissipating into its own bath while being coupled to a common bath. In this setup, we investigate different scenarios for temperature estimation of both the common and local baths. First, we explore how the precision of a single-qubit probe for estimating the common bath temperature can be enhanced by collective effects arising from the shared bath itself, particularly when the second qubit is in resonance with the probe. Interestingly, we find that the presence of local baths on each qubit can either jeopardize or, if these baths are sufficiently cold, enhance this precision. 
Next, we demonstrate a remote temperature sensing scheme in which one qubit acts as a probe to estimate the temperature of a local bath affecting the other qubit, by leveraging their indirect interaction through the common bath. This approach enables remote temperature sensing without directly coupling the probe to the target qubit or its local environment, thereby minimizing potential disturbances and practical challenges. Notably, we show that the collective Lamb shift, induced by the common bath, plays a crucial role in enabling remote temperature sensing by generating qubit-qubit correlations, even in the case of non-interacting qubits.
%********************************************************************************%
\end{abstract}

\maketitle
%********************************************************************************%
%==================================================================================%
\section{Introduction}
%==================================================================================%
Temperature directly influences the behavior and performance of quantum systems, making its precise control fundamental to the development and applications of modern quantum technologies~\cite{Mehboudi_2019,Giovannetti2011,Dedyulin_2022}.
Directly measuring the temperature of quantum systems provides an immediate way to access the state of the system. However, direct measurements can also be invasive and may disturb the system, leading to backaction effects~\cite{PhysRevA.108.062421}. Therefore, remote temperature sensing can offer significant advantages where direct access to a quantum system is difficult or undesirable. One of the most important benefits is the ability to extract temperature information without disturbing the system~\cite{Jacobs_2014}. A traditional approach involves a small quantum probe—a thermometer, such as a qubit, to infer the temperature of a sample~\cite{PhysRevLett.114.220405,Mehboudi_2015,PhysRevA.86.012125,PhysRevA.84.032105,PhysRevLett.118.130502,PhysRevA.99.062114,PhysRevResearch.2.033394,PhysRevX.10.011018,PhysRevA.91.012331}. The theory of equilibrium thermometry—where a probe reaches equilibrium before it is measured—is well-established and has been extensively studied over the years~\cite{Mehboudi_2019,binder2018thermodynamics,Campbell_2018,Campbell_2017,PhysRevA.99.052318,PhysRevLett.118.130502}. 

In contrast, nonequilibrium quantum thermometry~\cite{Razavian2019,PhysRevA.109.L060201} is a rapidly advancing field, offering unique opportunities to leverage quantum coherence~\cite{PhysRevA.82.011611,PhysRevA.107.063317,PhysRevLett.129.120404,Zhang2022,PhysRevResearch.5.043184,PhysRevA.110.032605}, quantum correlations~\cite{PhysRevLett.123.180602,PhysRevLett.128.040502,PhysRevA.104.012211}, and non-Markovian effects~\cite{PhysRevApplied.17.034073,PhysRevA.103.L010601,PhysRevA.108.022608,PhysRevResearch.3.043039,PhysRevE.110.024132} to achieve higher precision in temperature estimation. By employing ancillary qubits as intermediaries~\cite{PhysRevA.98.042124,Ullah_2025, PhysRevA.109.042417}, it becomes possible to measure temperature without requiring direct access to the sample. This approach expands the operational temperature range while markedly enhancing the precision. This has been widely studied in the context of ultracold gases~\cite{PhysRevLett.122.030403,PhysRevLett.125.080402,Sabín2014,PhysRevResearch.4.023069,PhysRevResearch.4.023191,PhysRevA.107.063317,PhysRevA.109.023309}. Numerous techniques have leveraged fundamental quantum properties to push forward the limits of low-temperature estimation. To this aim, critical quantum thermometry~\cite{PhysRevResearch.5.013087,PhysRevLett.133.120601,Aybar2022criticalquantum,PhysRevA.104.022612,PhysRevA.103.023317}, thermometry with impurity probes~\cite{Mihailescu_2024,PhysRevA.107.042614,PhysRevLett.125.080402,PhysRevResearch.6.033102}, the use of entanglement~\cite{PhysRevLett.114.220405,TIAN20171,PhysRevResearch.2.033498}, and driving techniques~\cite{Glatthard2022bendingrulesoflow,Mukherjee2019} have been explored, to name a few. Temperature estimation enhanced by the collective effects generated by a common bath has already been recently investigated in different setups~\cite{PhysRevLett.128.040502,segalMultispin,PhysRevA.107.042614,PhysRevA.109.023309}.

\begin{figure}[t!]
    \centering
    \includegraphics[scale=0.67]{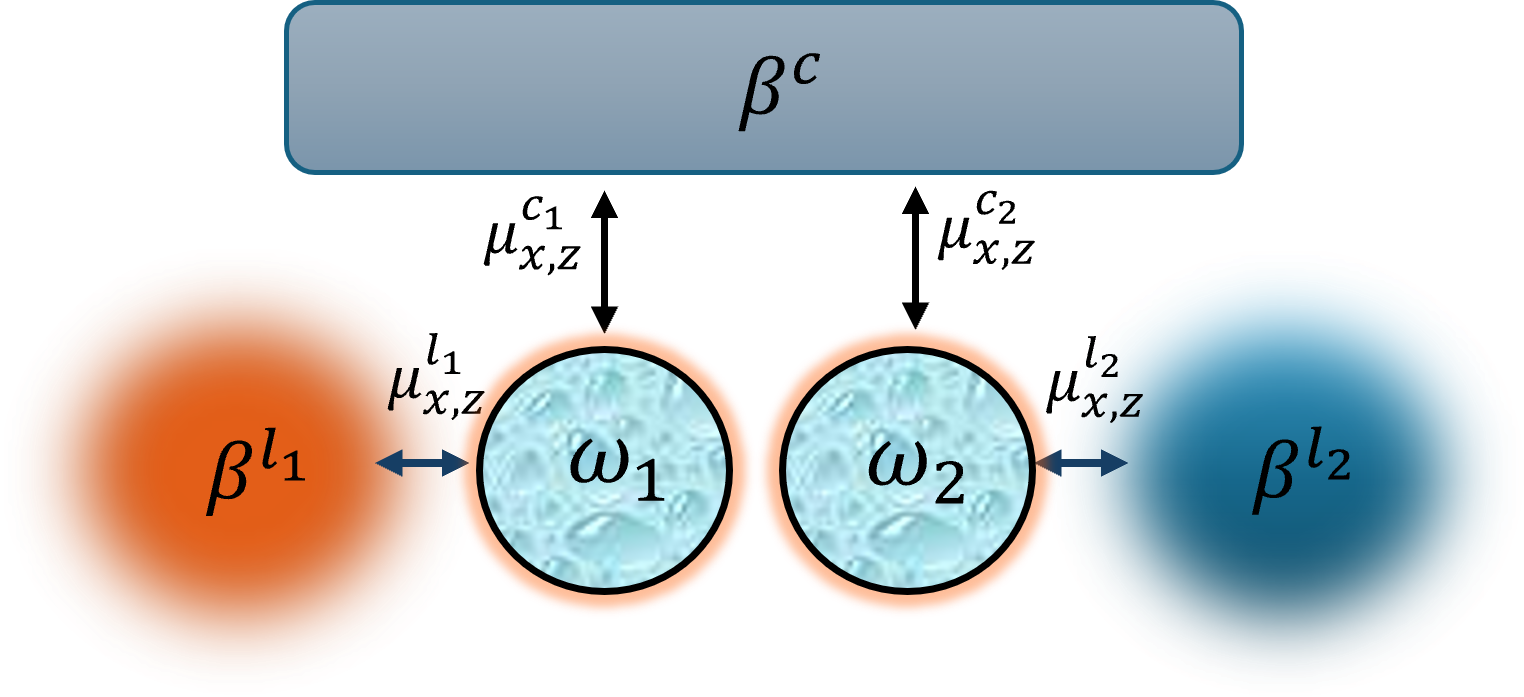}
\caption{ Two qubits interacting with thermal baths labeled by the inverse temperatures $\beta^c$, $\beta^{l_1}$, and $\beta^{l_2}$, corresponding to the common bath, the local bath on the first qubit, and the local bath on the second qubit, respectively. There is no direct coupling between the two qubits, but an effective coupling arises from the interaction with a common bath. The dissipative and dephasing interactions of qubits with common and local baths are given by the coupling strengths $\mu_{x,z}^{c_i}$, and $\mu_{x,z}^{l_i}$ (with $i=1,2$), respectively.}
    \label{fig1}
\end{figure}

In this paper, we explore the performance of single qubit probes for the temperature estimation in different scenarios involving local and collective baths (see Fig.~\ref{fig1}). 
In particular, we consider a system consisting of two qubits, each coupled to its own dissipative bath, while both are also coupled to a shared bath.  One of these qubits serves as the probe for temperature estimation. This scenario is of particular interest because there are various physical situations in which one qubit is experimentally accessible---and therefore measurable---while the other remains inaccessible. We study two main different protocols: either the probe qubit is measured to estimate the temperature of the common bath, or to measure the temperature of the local bath acting on the other qubit.

In the first scenario, we show that the correlations induced by the common bath can actually enhance the performance of single-qubit temperature estimation. Similar correlations are used in Refs.~\cite{PhysRevLett.128.040502,segalMultispin}, with a different setup and approach. We also investigate the robustness of this protocol against the presence of local baths on each qubit, and how the results may vary depending on the bath temperatures and couplings. %\textcolor{blue}{It is worth mentioning that, we find that the precision of common bath temperature estimation can be enhanced in the steady-state regime when the local baths are very cold, especially in contrast to the transient regime where no local baths are present.}

In the second scenario, we are putting forward a protocol for remote temperature sensing: instead of attaching a probe qubit directly to the local bath or to the qubit whose environment we aim to monitor---an approach that may introduce disturbances or be technically challenging---we couple the probe to the shared bath, assuming it has a larger spatial extent and a distant monitoring qubit would be more accessible for local measurements.
%the actual system of interest, operating between a hot and a cold environment, with its performance critically dependent on the \textcolor{blue}{temperature of the local environment and qubit-bath couplings}.  \textcolor{blue}{Specifically, a method for remote temperature sensing that avoids directly coupling a probe to the target qubit or its hot environment, which could introduce unwanted disturbances or practical challenges.} 
This setup not only minimizes potential interference with the environment but also leverages correlations induced by the common bath to enhance measurement precision in the transient regime. Specifically, the Lamb shift plays a critical role in generating these bath-induced correlations, even in the absence of direct qubit-qubit coupling. Indeed, as is well known, a shared bath can induce correlations~\cite{PhysRevLett.128.040502,PhysRevA.109.023309,Braun2002,Benatti2003a} that yield metrological advantages~\cite{Aybar2022criticalquantum,PhysRevA.85.022322} and effectively couple the qubits. These correlations enable remote and precise temperature monitoring of a local bath of a distant qubit, even when the probe has no direct physical access to the qubit and its environment.

We employ a Lindblad master equation approach formulated under the partial secular approximation to study the system's transient dynamics and steady state~\cite{Cattaneo_2019}. To compare the thermometric precision of our scheme, we use the quantum Fisher information (QFI) \cite{paris2009,Liu2020}, computed for the reduced density matrix of the probe system.

We remark that our model for temperature sensing using single-qubit probes may be realized in different platforms for quantum technologies. A prominent example is circuit quantum electrodynamics \cite{Blais2020}, as non-interacting superconducting qubits can be easily coupled to the same common bath. The role of the common bath can be played, for example, by a resistive element engineered on the chip \cite{Cattaneo2021engineering}, or by the qubit waveguide itself \cite{Lalumiere2013}, which has been shown to generate collective effects such as superradiance \cite{vanLoosuper,Mlynek2015}. Thermometry in such systems, with the presence of both common and local baths, has been recently tested experimentally \cite{Sharafiev2024}, and our proposals may inspire future experimental directions. Moreover, note that ``remote sensing'' may still apply for systems with a direct qubit-qubit coupling. For instance, it may be possible to realize a capacitive coupling between superconducting qubits that are actually ``distant'' on the same chip while at the same time interacting with a common bath \cite{Cattaneo2021engineering}.

The remainder of the paper is structured as follows. In Sec.~\ref{SP}, we provide a brief overview of single-parameter estimation theory. % \mar{Our model and objectives are introduced in Sec.~\ref{mod}.}  
In Sec.~\ref{PSA}, we outline the derivation of the master equation and examine the validity of the secular approximations. Our results are presented in Sec.~\ref{res}, beginning with the case of a single qubit and then extending the analysis to two qubits. We further explore mutual information and quantum discord before concluding this section with a discussion on the steady-state QFI. The coefficients of the master equation for a single qubit are provided in Appendix~\ref{apA}. Finally, we summarize our findings in Sec.~\ref{conc}.
%==================================================================================%
\section{Single Parameter Estimation}\label{SP}
%==================================================================================%
The goal of any quantum estimation protocol is to estimate the unknown value of a parameter $\theta$, encoded in the parametrized quantum state, say $\rho_{\theta}$, by performing specific measurements with positive operator-valued measures (POVMs) elements. The accuracy of the parameter measurement is bounded by the quantum Cramér-Rao bound, such that~\cite{paris2009}
\begin{equation}
    \Delta \theta^2\ge\frac{1}{m\mathcal{F}_\theta},
\end{equation}
where $m$ represents the number of measurements, which is set to $m=1$ for single-shot measurement, and $\mathcal{F}_\theta$ is the quantum Fisher information maximized over all POVMs~\cite{PhysRevLett.72.3439}, which quantifies the amount of information encoded into a state $\rho_{\theta}$ and is defined as
\begin{equation}
    \mathcal{F}_\theta=\text{Tr}[\rho_{\theta}L^2_{\theta}],
\end{equation}
where $L_{\theta}$ is a \textit{symmetric logarithmic derivative}, which implicitly satisfies the following Lyapunov equation:
\begin{equation}
    2\partial_{\theta}\rho_{\theta}=(L_{\theta}\rho_{\theta}+\rho_{\theta}L_{\theta}).
\end{equation}

Our probe is a single qubit to measure the temperature of a thermal bath. Therefore, we will focus on the QFI for a single qubit, which for any parametrized density matrix $\rho_{\theta}$ is given as~\cite{Dittmann_1999,PhysRevA.87.022337}
\begin{equation}\label{qfi}
    \mathcal{F}_\theta=\text{Tr}\Big[(\partial_{\theta}\rho_{\theta})^2\Big]+\frac{1}{|\rho_{\theta}|}\text{Tr}\Big[\big(\rho_{\theta}\partial_{\theta}\rho_{\theta}\big)^2\Big],
\end{equation}
where $\partial_{\theta}:=\partial/\partial \theta$ is the partial derivative with respect to parameter $\theta$ and $|\rho_{\theta}|$ stands for the determinant of the density matrix $\rho_{\theta}$. 

In this work, we study the single-parameter QFI with respect to the inverse temperature $\beta$, $\mathcal{F}_\beta$. Therefore, to be more precise, in this paper, we study ``inverse temperature estimation.'' For the sake of simplicity, we employ the term ``temperature estimation'' in the rest of the work. Using the chain rule for derivatives and Eq.~\eqref{qfi}, the QFI with respect to the temperature $T=1/\beta$ (we set $k_B=1$) can be obtained as
\begin{equation}\label{qfi_wrt_T}
    \mathcal{F}_T= \frac{\mathcal{F}_\beta}{T^4}\Big|_{\beta=\frac{1}{T}}.
\end{equation}
%==================================================================================%

%\section{Model and Objectives}\label{mod}
%==================================================================================%

%--------------------------------------------------------------%
%-----------------------------------------------------------------------%

%We begin by considering the simple case of a single qubit attached to a bosonic bath to measure its temperature. Subsequently,  we extend our analysis to a model consisting of two non-interacting and interacting qubits interacting with a common bath and their respective local baths, as illustrated in Fig.~\ref{fig1}. These qubits serve as probes for estimating the temperatures of the thermal baths and our objectives are as follows:
%\begin{itemize} 
%\item First, we use one of the qubits (the left hot qubit with frequency $\omega_1$ in Fig.~\ref{fig1}) to estimate the temperature of the common bath, $\beta^c$, while keeping the local baths turned off. 
%\item Second, in the presence of the common bath, we turn on the local baths of both qubits, with inverse temperatures $\beta^{l_1}$ and $\beta^{l_2}$ corresponding to the hot and cold baths, respectively. We then utilize the hot qubit (with frequency $\omega_1$) to estimate the inverse temperature $\beta^{l_2}$ of the cold qubit, assuming that $\beta^c$ of the common bath and $\beta^{l_1}$ of the hot bath are known. 
%\item Finally, we use the cold qubit (with frequency $\omega_2$) to estimate the temperature $\beta^{l_1}$ of the hot qubit, assuming that the inverse temperatures $\beta^c$ of the common bath and $\beta^{l_2}$ of the cold bath are known. 
%\end{itemize}

\section{Master equation in the partial secular approximation}\label{PSA}
%==================================================================================%

\subsection{General formalism} \label{generalFormalism}

Consider the total Hamiltonian of a system interacting with the bath (we set $\hbar=1$) \cite{Breuer},
\begin{equation}
    \hat{H}_T=\hat{H}_S+\hat{H}_B+\mu\sum_\alpha\hat{A}_\alpha\otimes\hat{B}_\alpha,
\end{equation}
where $\hat{H}_s$ is the free Hamiltonian of a quantum system, $\hat{H}_B=\sum_k\omega_k\hat{b}_k^\dagger\hat{b}_k$ is the free Hamiltonian of the bath, $\hat{A}_\alpha$ and $\hat{B}_\alpha$ are the system and bath operators, and $\mu$ is the system-bath coupling constant. The timescale of the system relaxation is typically $\tau_R\propto \mu^{-2}$.

Let \(|e_n\rangle\) be the eigenstates of the free Hamiltonian of a quantum system \(\hat{H}_S = \sum_n \epsilon_n |e_n\rangle \langle e_n|\), with \(\epsilon_n\) being the corresponding eigenenergies. The jump operators of the system Hamiltonian between the eigenstates $|e_m\rangle$ and $|e_n\rangle$ with the energy gap $\epsilon_m-\epsilon_n=\omega$ can be defined as~\cite{Breuer}

\begin{equation}\label{jump}
    \hat{A}_\alpha(\omega)=\sum_{\epsilon_m-\epsilon_n=\omega}|e_n\rangle\langle e_n|\hat{A}_\alpha|e_m\rangle\langle e_m|.
\end{equation}

We assume that the coupling of the quantum system to the bath is weak and that the autocorrelation functions of the bath decay sufficiently fast in time, so as to guarantee the validity of the Born-Markov approximations \cite{Breuer, Cattaneo_2019}. After applying these approximations, we obtain the Bloch-Redfield master equation, which in the interaction picture can be written as
\begin{equation}\label{GKLS}
\begin{aligned}
    \Dot{\hat{\rho}}_S(t) =& \sum_{\omega,\omega^\prime}\sum_{\alpha,\beta} \Big[e^{i(\omega^\prime-\omega)t} \Gamma_{\alpha\beta}(\omega) \\
    &\times \Big(\hat{A}_{\beta}(\omega)\rho_S(t)\hat{A}_{\alpha}^\dagger(\omega^\prime) 
    - \hat{A}_{\alpha}^\dagger(\omega^\prime)\hat{A}_{\beta}(\omega)\rho_S(t)\Big)\\
    &+ \text{H.C.}\Big],
\end{aligned}
\end{equation}
where the one-sided Fourier transform of the bath correlation function is given by
\begin{equation}
    \Gamma_{\alpha\beta}(\omega)=\mu^2\int_{0}^\infty dt^\prime e^{i\omega t^\prime}\mathcal{B}_{\alpha\beta}(t^\prime).
\end{equation}
 $  \mathcal{B}_{\alpha\beta}(t^\prime)=\langle \hat{B}_\alpha(t^\prime)\hat{B}_{\beta}(0)\rangle=\text{Tr}_B[\hat{B}_\alpha(t^\prime)\hat{B}_{\beta}(0)\rho_B]$ are the bath correlation functions and $\text{Tr}_B$ denotes the partial trace over
degrees of freedom of the baths. We assume that the bath is stationary so that $[\hat{\rho}_B,\hat{H}_B]=0$. 

Neglecting the rapidly oscillating terms in the interaction picture is commonly known as the \textit{secular approximation}. In the so-called \textit{full secular approximation}, all the terms in Eq.~(\ref{GKLS}) with $\omega^\prime \neq \omega$ are neglected. In contrast, in our setup we retain the slowly rotating terms with $\omega^\prime \neq \omega$. This approach is referred to as the \textit{partial secular approximation} \cite{Cattaneo_2019,Jeske2015,Farina2019}. Specifically, we eliminate all the terms for which we can find a coarse-graining time $t^*$ such that
\begin{equation}\label{coarseGrainingTime}
\exists t^* \text{such that} \quad |\omega_1-\omega_2|^{-1} \ll t^*\ll\tau_R,
\end{equation}
where $\tau_R$ represents the relaxation time of the system—the timescale over which the system density matrix $\rho_S$ evolves toward dynamical equilibrium~\cite{Breuer, Cattaneo_2019}. 

As a consequence, we can rewrite the final Bloch-Redfield master equation in partial secular approximation, coming back to Schr\"odinger's picture, such as~\cite{Cattaneo_2019}
\begin{equation}\label{me}
     \Dot{\hat{\rho}}_S(t)=-i[\hat{H}_S+\hat{H}_{LS},\rho_S(t)]+\mathcal{D}[\rho_S(t)],
\end{equation}
where the Lamb-Shift Hamiltonian is given by
\begin{equation}
    \hat{H}_{LS}=\sum_{\omega,\omega^\prime}\sum_{\alpha,\beta}S_{\alpha\beta}(\omega,\omega^\prime)\hat{A}^\dagger_\alpha(\omega^\prime)\hat{A}_{\beta}(\omega)
\end{equation}
and the dissipator in the above master equation is
\begin{equation}
\begin{aligned}
    \mathcal{D}[\rho_S(t)]=&\sum_{\omega,\omega^\prime}\sum_{\alpha,\beta}\gamma_{\alpha\beta}(\omega,\omega^\prime)
   \times\big[\hat{A}_{\beta}(\omega)\rho_S(t)\hat{A}_{\alpha}^\dagger(\omega^\prime)\\ 
    &- \frac{1}{2}\{\hat{A}_{\alpha}^\dagger(\omega^\prime)\hat{A}_{\beta}(\omega),\rho_S(t)\}\big],
    \end{aligned}
\end{equation}
where $S_{\alpha\beta}(\omega,\omega^\prime)$ and $\gamma_{\alpha\beta}(\omega,\omega^\prime)$ are functions of the autocorrelation
functions of the bath operators $\hat{B}_\alpha$, and are given by
\begin{align}
S_{\alpha\beta}(\omega, \omega^\prime) &=\frac{1}{{2i}}\big[\Gamma_{\alpha\beta}(\omega) - \Gamma_{\beta\alpha}^*(\omega^\prime)\big], \\
\gamma_{\alpha\beta}(\omega, \omega^\prime) &= \Gamma_{\alpha\beta}(\omega) + \Gamma_{\beta\alpha}^*(\omega^\prime).\label{coeff}
\end{align}

While the master equation~\eqref{me} in partial secular approximation may not always be written in the Gorini-Kossakowski-Lindblad-Sudarshan (GKLS) form, it can indeed be brought into this form in many physical scenarios if the Born-Markov approximations are applied consistently \cite{Farina2019}. Moreover, this equation can be slightly modified to obtain a mathematically well-defined GKLS form by replacing the coefficients \(\gamma_{\alpha\beta}(\omega, \omega')\) with \(\gamma_{\alpha\beta}(\overline{\omega}) = 2 \mathrm{Re}[\Gamma_{\alpha\beta}(\overline{\omega})]\), where \(\overline{\omega} = (\omega + \omega')/2\). This procedure has been proposed and formalized in slightly different ways in a number of recent papers \cite{PhysRevLett.122.150603,McCauley2020a,Nathan2020a,PhysRevA.103.062226}. In our work, for the sake of numerical simulations for a two-qubit system we have employed the master equation~\eqref{me} and verified that the same results can be obtained through the ``unified'' master equation \cite{PhysRevA.103.062226} with $\gamma_{\alpha\beta}(\omega, \omega')\rightarrow \gamma_{\alpha\beta}(\overline{\omega})$.

\subsection{Single-qubit and two-qubit master equations}

\subsubsection{Single qubit}
The Hamiltonian of a qubit interacting with a bosonic bath is defined as
\begin{equation}
\hat{H}=\hat{H}_S+\hat{H}_B+\sum_k f_k(\mu_z\hat{\sigma}_z+\mu_x\hat{\sigma}_x)(\hat{b}^\dagger_k+\hat{b}_k),
\end{equation}
where $\mu_x$ and $\mu_x$ describe the dephasing and dissipative couplings of the qubit with the bath, respectively. We assume a weak system-bath coupling such that $\mu_x,\mu_z/\omega_0\ll1$. The system and the bath Hamiltonians are given as $\hat{H}_S=\frac{1}{2}\omega_0\hat{\sigma}_z$, $\hat{H}_B=\sum_k\omega_k\hat{b}^\dagger_k\hat{b}_k$, respectively.  The bath is assumed to be in a thermal state $\rho_B$ with inverse temperature $\beta=1/k_BT$. $f_k$ are real numbers that define the spectral density of the bath, as explained in Appendix~\ref{apA}. In this work, we always use an Ohmic spectral density with cutoff frequency $\omega_c=20$, for both single-qubit and two-qubit systems, \textcolor{black}{in units of $\omega_0$ and $\omega_1$, respectively}.
 
The GKLS master equation for the qubit dynamics can be written as~\cite{Breuer}
\begin{equation}\label{meQ}
\begin{aligned}
\Dot{\rho}_S(t)= &-i \big[\hat{H}_S+\hat{H}_{LS}, \rho_S(t)\big] + \gamma_{\downarrow} \mathcal{D}[\hat{\sigma}_-]\rho_S(t) \\
&+ \gamma_{\uparrow} \mathcal{D}[\hat{\sigma}_+]\rho_S(t) + \gamma_{0} \mathcal{D}[\hat{\sigma}_z]\rho_S(t),
\end{aligned}
\end{equation}
where the Lindblad dissipator for a jump operator $\hat{a}$ is defined as $\mathcal{D}[\hat{a}]\rho_S(t)=\hat{a}\rho_S(t)\hat{a}^\dagger-\{\hat{a}^\dagger\hat{a},\rho_S(t)\}/2$. Furthermore, \(\gamma_{\uparrow}\), \(\gamma_{\downarrow}\), and \(\gamma_{0}\) correspond to the absorption, emission, and dephasing rates, respectively, and are given in Appendix~\ref{apA}.
The \textit{Lamb-shift} Hamiltonian for a single qubit reads $\hat{H}_{LS}=s_0\hat{\sigma}_z/2$, where the coefficient $s_0$ can be found in Appendix~\ref{apA}. %It is important to note that the distinction between the partial secular and full secular approximations becomes noticeable only when the energy differences in the spectrum of the system Hamiltonian are small. This scenario arises, for instance, in the case of two slightly detuned qubits: if the qubit frequencies are \(\omega_1\) and \(\omega_2\), which are close but not identical, then the difference \(\omega_1 - \omega_2\) is small and cannot be neglected under the full secular approximation. However, for a single qubit, the system is characterized by a single transition frequency \(\omega_0\), corresponding to a single energy difference. In this case, the full secular and partial secular approximations fully coincide, ensuring that the results derived in this context are always consistent. 

\subsubsection{Two qubits}\label{psa_twoQ}
The Hamiltonian of two interacting qubits with transition frequencies $\omega_1$ and $\omega_2$ is given by
\begin{equation}\label{2qubitH}
     \hat{H}_S=\frac{\omega_1}{2}\hat{\sigma}_1^z+\frac{\omega_2}{2}\hat{\sigma}_2^z+k\hat{\sigma}^x_1\hat{\sigma}^x_2,
\end{equation}
 where $k$ represents the qubit-qubit coupling strength. In this work, we also explore the scenario where there is no direct qubit-qubit interaction, meaning $k=0$.
 
The interaction Hamiltonian of two qubits coupled to local baths and a common bath is given by
\begin{equation}
\begin{aligned}
\hat{H}_I = (\mu_x^{l_1} \hat{\sigma}_1^x+ \mu_z^{l_1} \hat{\sigma}_1^z)\hat{B}^{l_1} 
    + (\mu_x^{l_2} \hat{\sigma}_2^x+\mu_z^{l_2} \hat{\sigma}_2^z ) \hat{B}^{l_2} \\
    + \big(\mu_x^{c_1}\hat{\sigma}_1^x + \mu_x^{c_2}\hat{\sigma}_2^x+\mu_z^{c_1}\hat{\sigma}_1^z + \mu_z^{c_2}\hat{\sigma}_2^z\big) \hat{B}^{c}.
    \end{aligned}
\end{equation}
The bath operators are given as $\hat{B}^i=\sum_kf_{k,i}(\hat{a}^\dagger_{k,i}+\hat{a}_{k,i})$, where $i=l_1,l_2,c$ specifies the baths.
$\mu_{x,z}^{(i)}$ are the dissipative and dephasing coupling strengths. For simplicity, if not stated otherwise, we assume  $\mu_{x}^{l_1}=\mu_{x}^{l_2}=\mu_{x}^{c}=\mu_x$ and $\mu_{z}^{l_1}=\mu_{z}^{l_2}=\mu_{z}^{c}=\mu_z$ (when all the qubit-bath couplings are equal), where $\mu_{x,z}\ll\omega_{1,2}$ are weak dissipative and dephasing coupling constants. In the scenarios where the local baths are switched off, we implicitly assume $\mu_{x}^{l_1}=\mu_{x}^{l_2}=0$ and analogously for dephasing.
The jump operators that describe the emission and absorption processes of the system can be found by using Eq.~(\ref{jump}), and they correspond to single-qubit raising and lowering operators. %These are $\sigma^x_j(\omega_j)=\sigma^{-}_j$, $\sigma^x_j(-\omega_j)=\sigma^{+}_j$ and $\sigma^z_j(0)=\sigma^{z}_j$ with $j=1,2$.  

For two uncoupled qubits or for $k\ll \omega_1,\omega_2$, the master equation governing the two-qubit dynamics is given by~\cite{Cattaneo_2019}
\begin{equation}\label{fme}
\begin{aligned}
    \Dot{\hat{\rho}}_S(t)=&-i[\hat{H}_S+\hat{H}_{LS},\rho_S(t)]+\sum_{i,j=1,2}\Big(\gamma_{ij}^\downarrow\mathcal{D}[\hat{\sigma}^-_i,\hat{\sigma}^-_j]\rho_S(t)\\&+\gamma^\uparrow_{ij}\mathcal{D}[\hat{\sigma}^+_i,\hat{\sigma}^+_j]\rho_S(t)+\eta_{ij}\mathcal{D}[\hat{\sigma}^z_i,\hat{\sigma}^z_j]\rho_S(t)\Big),
\end{aligned}
\end{equation}
where we have introduced the notation $D[a,b]\rho_S(t)=a\rho_S(t)b^\dagger-\{b^\dagger a,\rho_S(t)\}/2$. The Lamb-shift Hamiltonian has the following form:
\begin{equation}
    \hat{H}_{LS}=\sum_{ij}\Big(s_{ij}\hat{\sigma}_j^+\hat{\sigma}_i^-+\tilde{s}_{ij}\hat{\sigma}_j^-\hat{\sigma}_i^++2s_0\hat{\sigma}^z_1\hat{\sigma}^z_2\Big).
\end{equation}
The coefficients of the master equation can be calculated in a manner similar to the calculations provided in Appendix~\ref{apA}, see for instance~\cite{Cattaneo_2019}, and can be easily tuned to reproduce the unified master equation \cite{PhysRevA.103.062226}. Note that in the master equation~\eqref{fme} there are both dissipative and unitary (i.e., in the Lamb shift) cross-terms coupling qubit 1 and 2, even if $k=0$. These cross-terms appear under the partial secular approximation in the limit of small detuning, $\omega_-=\abs{\omega_1-\omega_2}$ and $\omega_-/\omega_1\ll 1$ (compare this expression with Eq.~\eqref{coarseGrainingTime}). If the detuning is large and $k=0$, then the two qubits are effectively decoupled in the open dynamics, and the quantum correlations between them are negligible.

If the qubit-qubit coupling $k$ is not weak, then the expression of the master equation is more involved. In particular, it is not ``local'' anymore, meaning that the jump operators are not single-qubit operators anymore. We provide more details about this ``global'' master equation in Appendix~\ref{sec:appendixGlobalME}. Moreover, we refer the interested readers to the derivation of the global master equation for two strongly interacting qubits in Ref.~\cite{Cattaneo_2019}.
%==================================================================================%

\section{Results}\label{res}
In this section, we work with dimensionless units and scale all the parameters in this work with the frequency $\omega_1$ of the first qubit ($\omega_0$ in the single-qubit case), unless otherwise specified. In addition, in the two-qubit scenario, we set $\omega_1>\omega_2$, without loss of generality. We recall that we fix $\hbar=1$ and $k_B=1$.
\subsection{Single qubit}

%-------------------------------------------------------------------------%
\begin{figure}[t!]
    \centering
\subfloat[]{\includegraphics[scale=0.62]{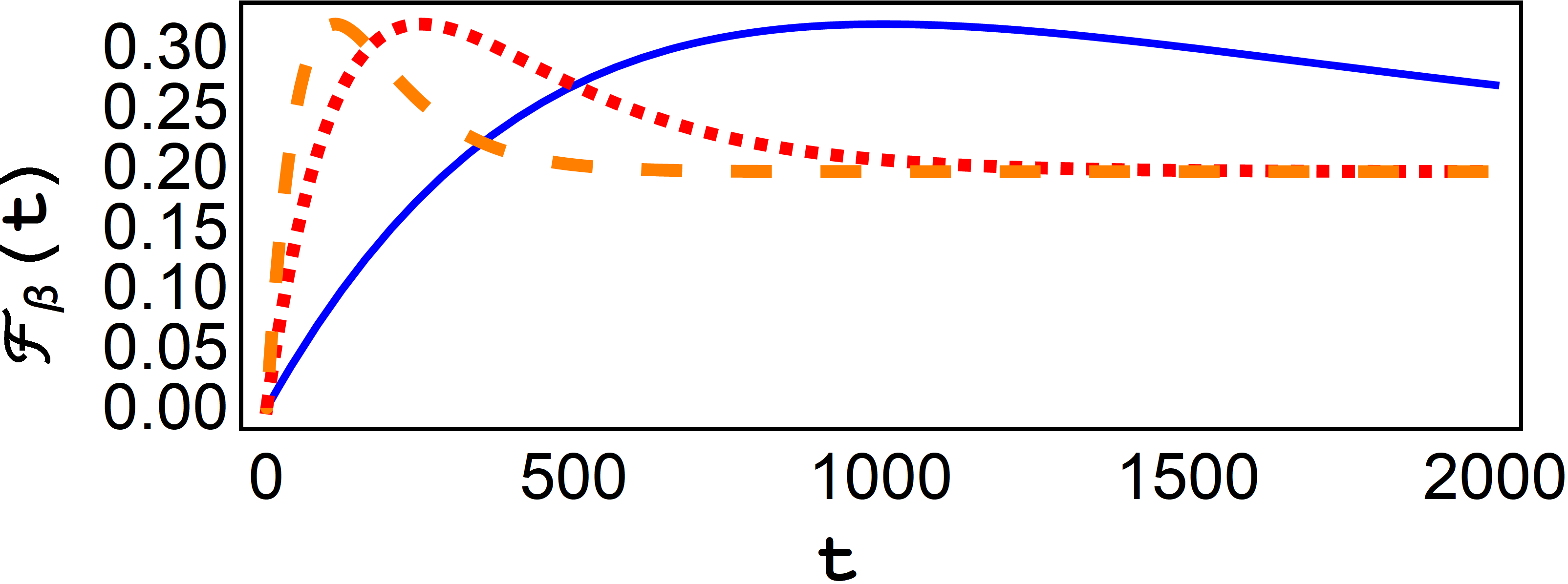}}\\
\subfloat[]{\includegraphics[scale=0.62]{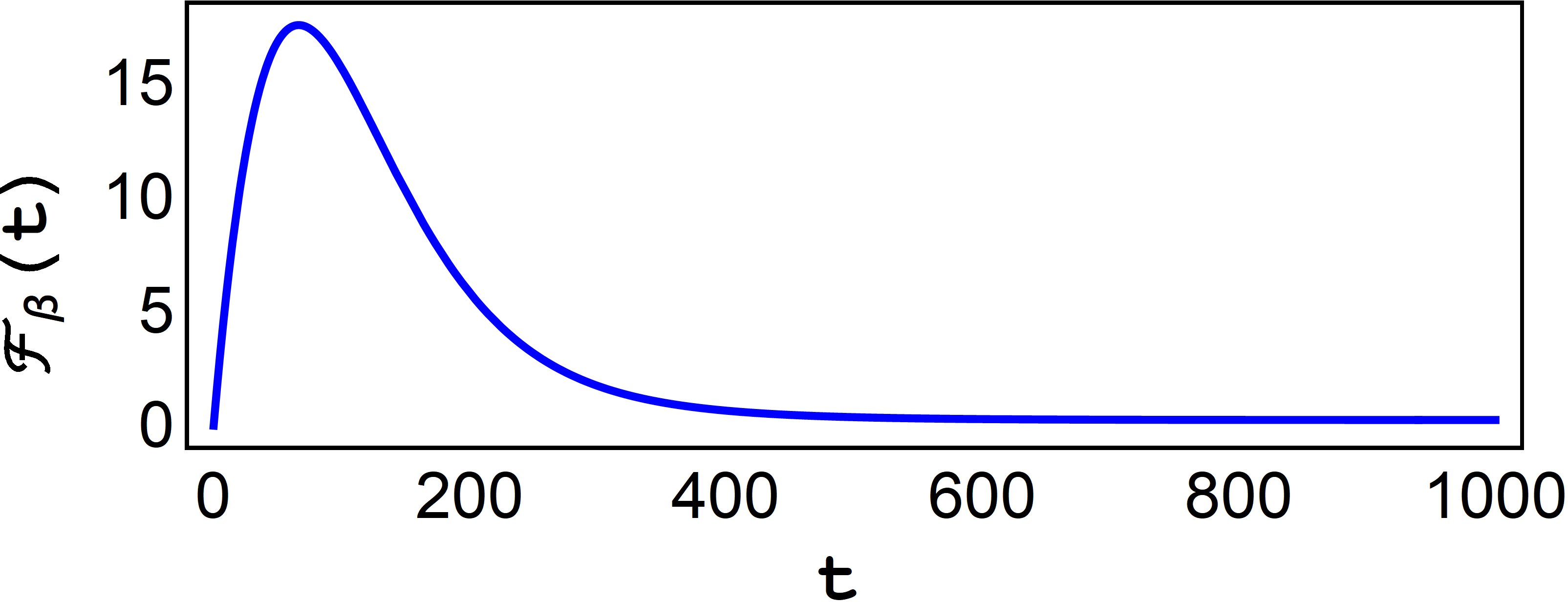}}\\
\subfloat[]{\includegraphics[scale=0.62]{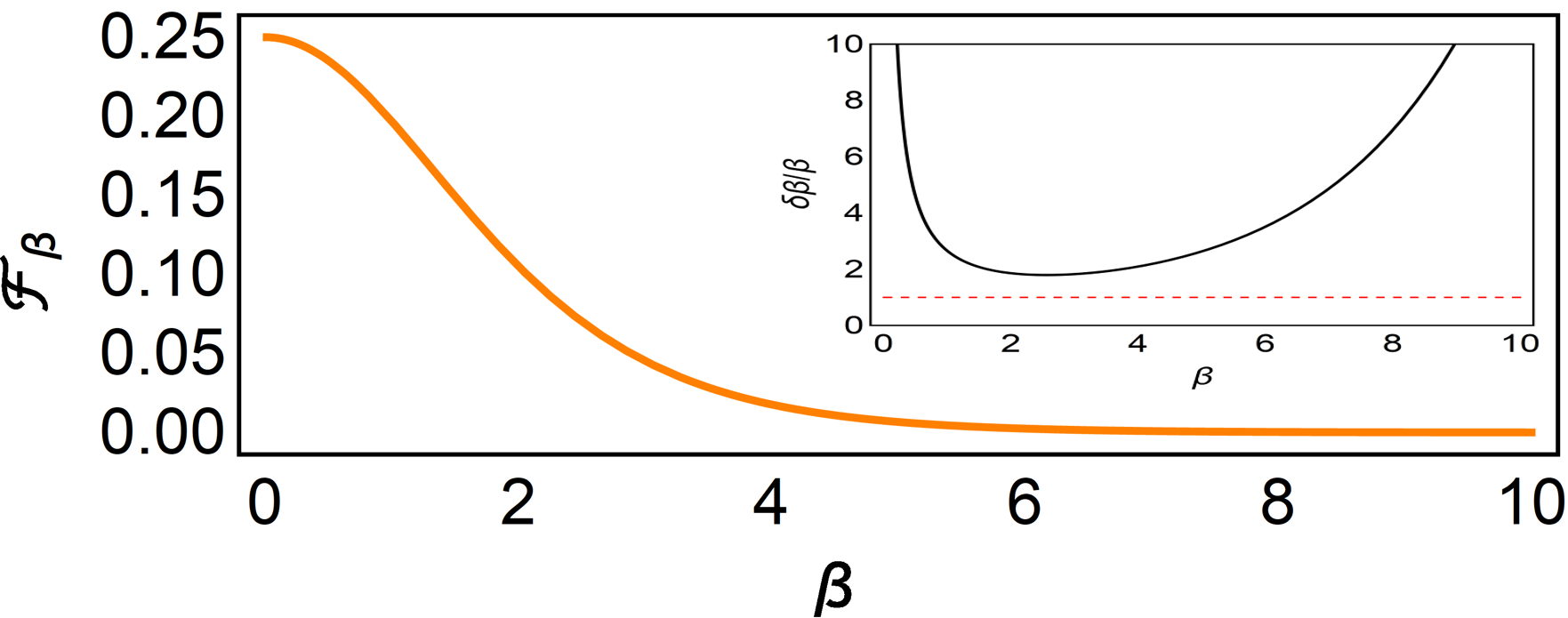}}
    \caption{\textcolor{black}{Scenario: Single qubit coupled to a thermal bath. \textbf{(a)} The QFI $\mathcal{F}_\beta(t)$ as a function of time for the estimation of the inverse temperature $\beta$ of the bath for different values of dissipation rate $\mu_x$. The solid blue, red dotted, and orange dashed curves represent $\mu_x = 0.01$, $\mu_x = 0.02$, and $\mu_x = 0.03$, respectively. The evolution begins in the excited state $|1\rangle$ of the qubit, and the dynamics is described by the master equation~(\ref{meQ}). The parameters are set to $\omega_0=1$ and  $\beta=1$. \textbf{(b)} Same as \textbf{(a)}, but for inverse temperature $\beta=0.1$, $\mu_x=0.01$.
    \textbf{(c)} QFI $\mathcal{F}_\beta$ as a function of bath inverse temperature $\beta$ for the steady solution of the master equation~\eqref{meQ}. The inset shows the relative error \(\delta\beta/\beta\) of inverse temperature $\beta$ as a function of inverse temperature $\beta$.}
    }
    \label{figsb}
\end{figure}
%----------------------------------------------------------------------%

For simplicity, we provide the analytical solution of the master equation~\eqref{meQ} at any time $t$ when the qubit is prepared in the excited state $|1\rangle$, so that dephasing does not play a role (we may set $\mu_z=0$):
    \begin{equation}\label{sol}
       \rho_S(t) = \left(
\begin{array}{cc}
 \frac{\gamma_{\uparrow}(1-e^{-(\gamma_{\downarrow}+\gamma_{\uparrow})t} )}{\gamma_{\downarrow}+\gamma_{\uparrow}} & 0 \\
 0 & \frac{\gamma_{\downarrow}+\gamma_{\uparrow}e^{- (\gamma_{\downarrow}+\gamma_{\uparrow})t}}{\gamma_{\downarrow}+\gamma_{\uparrow}} \\
\end{array}
\right).
\end{equation}
The above solution is derived by neglecting the Lamb-shift term, as we have verified that its contribution does not significantly affect the dynamics of the system and of the QFI.

The QFI for the estimation of the inverse temperature $\beta$ for the qubit dynamics in Eq.~\eqref{sol} can be easily calculated using Eq.~\eqref{qfi}, and it reads 
\begin{equation}
    \mathcal{F}_\beta(t)=\frac{\eta(t) ^2 \Bar{n}^4 \omega_0^2 e^{2\beta \omega_0} \left(\coth \left(\frac{\chi(t) }{2}\right)-1\right)}{2  \left(e^{\beta\omega_0}+1\right)^2 \left(1+e^{\beta\omega_0+\chi(t)}\right)}.
\end{equation}
The parameters in the above equation are defined as follows,
\begin{equation}
\begin{aligned}
    \eta(t) &=e^{\omega_0\beta} \left(2+4 c t-e^{\beta\omega_0}\right)+4 c t+\frac{e^{\chi(t) }}{\Bar{n}^2}-1,\\
    \chi(t) &=2 c t \coth \left(\frac{\beta\omega_0}{2}\right),
    \end{aligned}
\end{equation}
where $c=\pi J(\omega_0) \mu_x^2$ with $J(\omega_0)$ is the Ohmic spectral density of the bath and $\Bar{n}=(1-e^{\beta\omega_0})^{-1}$ is the mean photon number.
In Fig.~\ref{figsb}(a), we plot the QFI, $\mathcal{F}_\beta(t)$, as a function of time $t$ for the estimation of bath temperature, $\beta$, which is fixed to $\beta=1$. However, for high temperatures, such as $\beta=0.1$, the QFI in the transient regime is notably very high, as illustrated in Fig.~\ref{figsb}(b). This occurs during a short period at the initial stage of the evolution, after which the QFI decays and eventually converges to a steady-state value of $\sim0.25$, as shown in Fig.~\ref{figsb}(c).
The results in Fig.~\ref{figsb}(a) show that the dissipation constant $\mu_x$ significantly influences the QFI dynamics. For smaller values of $\mu_x$, the QFI takes longer to reach its steady-state value, indicating slower thermalization. In contrast, for larger values of $\mu_x$, the stronger dissipative coupling accelerates the process, enabling the QFI to reach its steady-state value more quickly, given by
\begin{equation}
    \mathcal{F}_\beta(t\rightarrow\infty)=\frac{\omega_0^2}{2+2 \cosh (\beta \omega_0)}.
\end{equation}
  This is a well-known result; see, for example, Refs.~\cite{PhysRevResearch.3.043039,Liu2020}. The above QFI for the steady state is plotted in Fig.~\ref{figsb}(c) as a function of inverse temperature $\beta$. We note that $\mathcal{F}_\beta$ is higher for very high temperatures ($\beta\ll 1$) and then decreases towards 0 for very low temperatures ($\beta \approx 10$). We remark that the behavior of $\mathcal{F}_\beta$ differs from that of $\mathcal{F}_T$ at large temperatures, for which $\mathcal{F}_T$ decays towards zero due to the $1/T^4$ factor in Eq.~\eqref{qfi_wrt_T}. \textcolor{black}{
The inset in Fig.~\ref{figsb}(c) illustrates the relative error \(\delta\beta/\beta\) as a function of the inverse temperature for the steady state of a single qubit. We observe that \(\delta\beta/\beta\) is significantly high at both extremes—very high temperatures (\(\beta \to 0\)) and very low temperatures (\(\beta > 10\)). However, within an intermediate range of \(\beta\), the relative error remains notably low (with the red dotted line marking \(\delta\beta/\beta = 1\)), indicating a more reliable estimation in this regime.
}

Building on these insights from the single-qubit case, we now analyze the two-qubit scenario, aiming to explore temperature sensing using single-qubit probes in the presence of different baths, with the help of correlations generated by the common bath.

\subsection{Two qubits}\label{NoDir}
For simplicity, we study the two-qubit dynamics when the initial state is separable (no initial quantum correlations) and maximally coherent with respect to the canonical basis. More precisely, we assume
\begin{equation}
    \rho_S(0) = \ket{+}\bra{+}\otimes\ket{+}\bra{+},
\end{equation}
where $\ket{+}=(\ket{0}+\ket{1})/\sqrt{2}$.

In what follows, we explore different setups for two-qubit thermometry. First, we switch off the local baths and consider a scenario of two uncoupled qubits in a single common bath at inverse temperature $\beta^c$. To test the sensitivity of single-qubit thermometry in this scenario, we compute the QFI with respect to $\beta^c$ of the reduced state of a single qubit only as a function of time. This model is similar to the one studied in Ref.~\cite{PhysRevLett.128.040502}, but it presents a few crucial differences. However, it is worth noting that joint measurements on both qubits can, in general, lead to a higher QFI than local measurements on a single qubit~\cite{PhysRevA.109.023309}.
Then, we switch on the local baths and investigate whether their effect is detrimental to the estimation of the common bath temperature through a single-qubit probe.

Next, we explore the possibility of remote temperature sensing in the scenario with both a common bath and local baths.
We compute the QFI of the reduced state of qubit 2, which acts as the probe, with respect to the inverse temperature $\beta^{l_1}$ of the local bath coupled to qubit 1, which serves as the sample in this configuration. We show that the correlations induced by the common bath enable remote temperature sensing even when the probe is not coupled to the sample qubit, if the temperature of the local bath 1 is sufficiently high. Then, we study the same model in the presence of a direct qubit-qubit coupling, showing that the effects of this coupling are similar to those induced by the common bath. In addition, we investigate the relevance of the Lamb-shift term for the emergence of correlations (see Sec.~\ref{lambshift}), the effects of dephasing (see Appendix~\ref{deph}).
%-------------------------------------------------------------------------%
\begin{figure}[t!]
    \centering
    \subfloat[]{\includegraphics[scale=0.6]{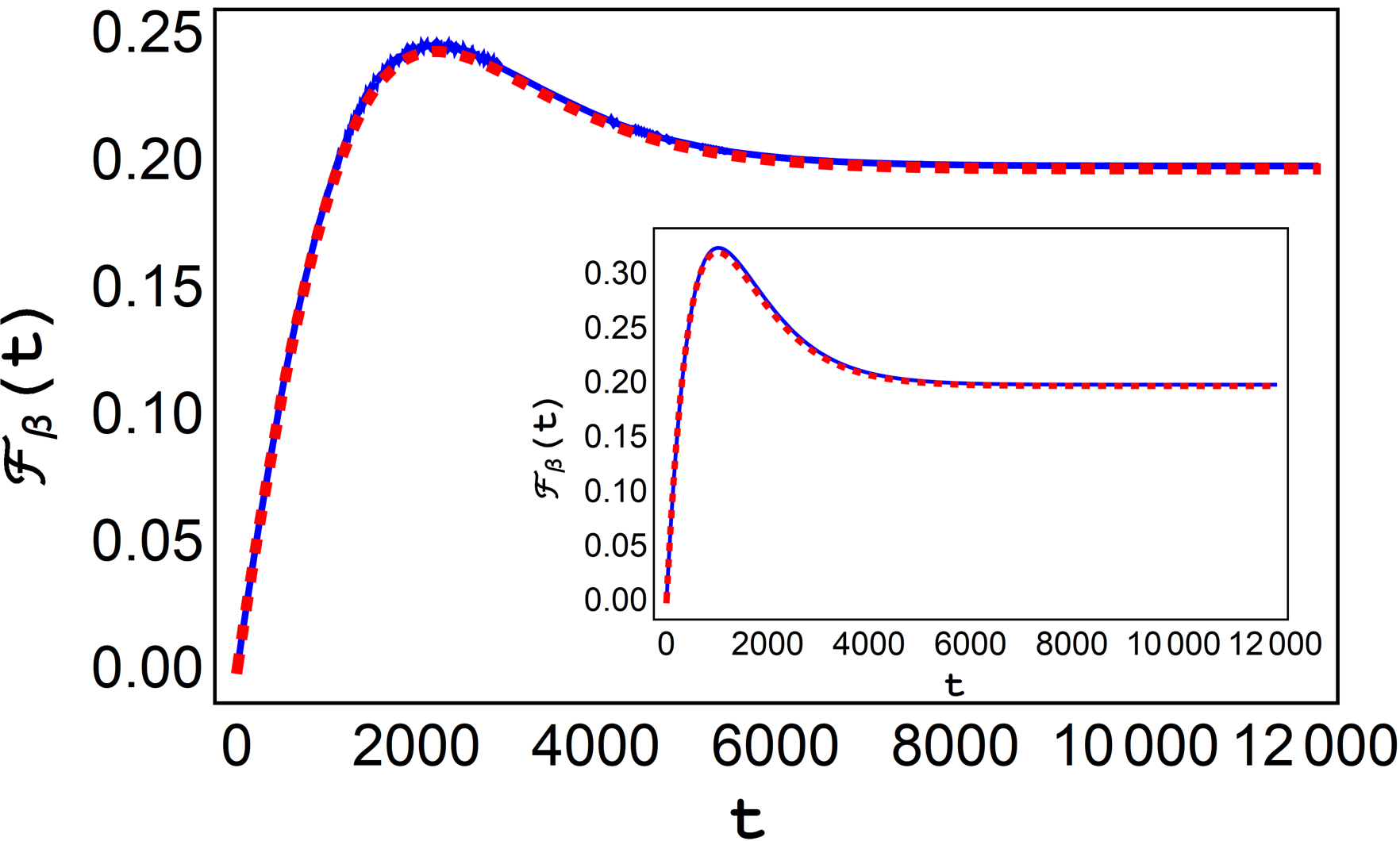}}\\
    \subfloat[]{\includegraphics[scale=0.6]{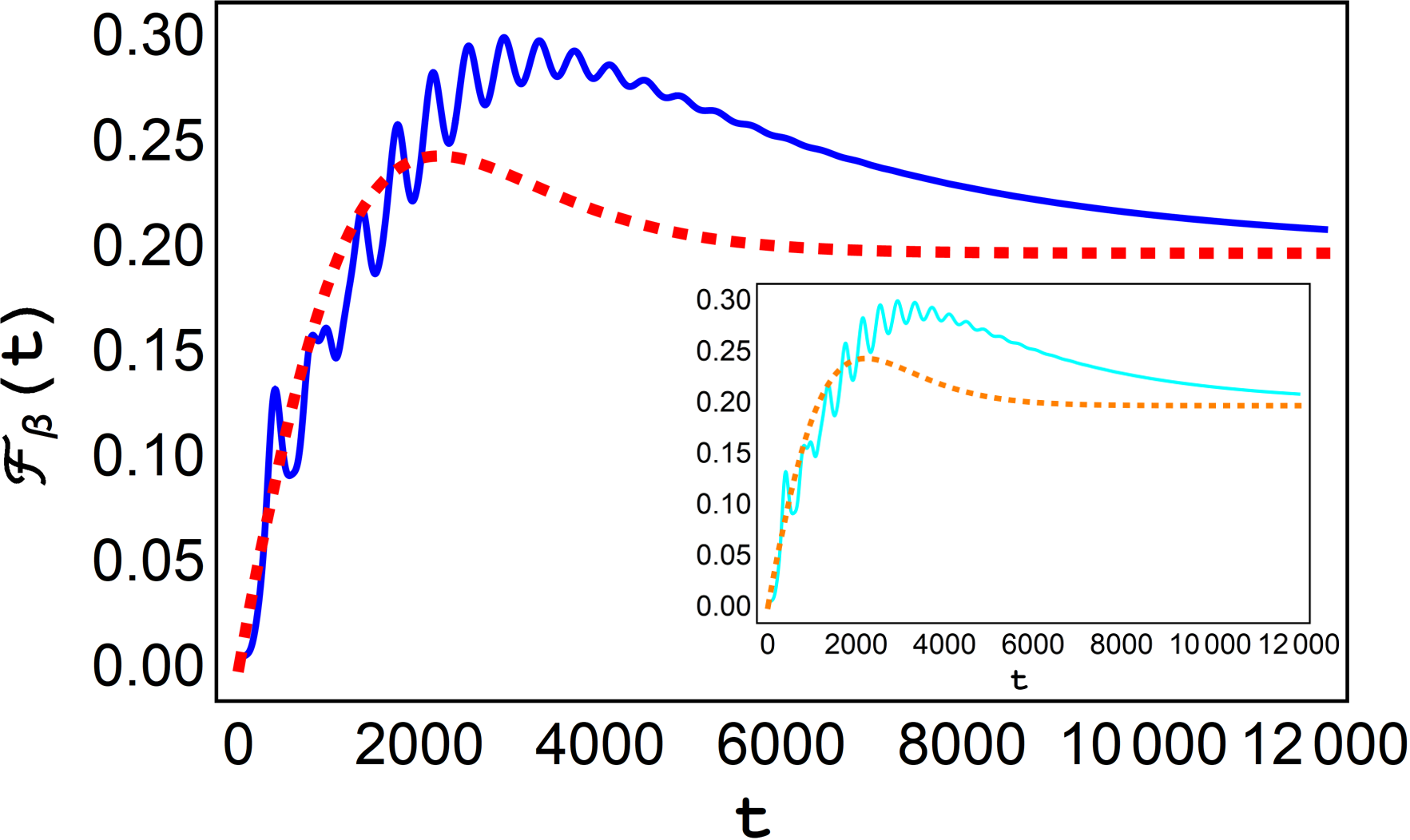}}
    \caption{Scenario with two uncoupled qubits ($k=0$) immersed in a common bath and in the absence of local baths. The interaction is purely dissipative: $\mu^c_x = 10^{-2}$, $\mu^c_z =0$. We plot the QFI as a function of time for the estimation of common bath temperature $\beta^c$ using the first qubit as a probe. The inverse temperature of the dissipative common bath is set to $\beta^c=1$. The solid blue curve corresponds to partial secular approximation, while the red dashed curve is for full secular approximation. \textbf{(a)}: Large detuning, $\omega_-=0.5$. The inset shows the same dynamics when both qubits start from the excited state, analogously to Fig.~\ref{figsb}(a). \textbf{(b)}: Small detuning, $\omega_-=0.01$. The inset shows the QFI for a unified GKLS master equation~\cite{PhysRevA.103.062226}, as discussed in the main text.}
    \label{fig3}
\end{figure}

\begin{figure*}
    \centering
    \subfloat[]{\includegraphics[scale=0.35]{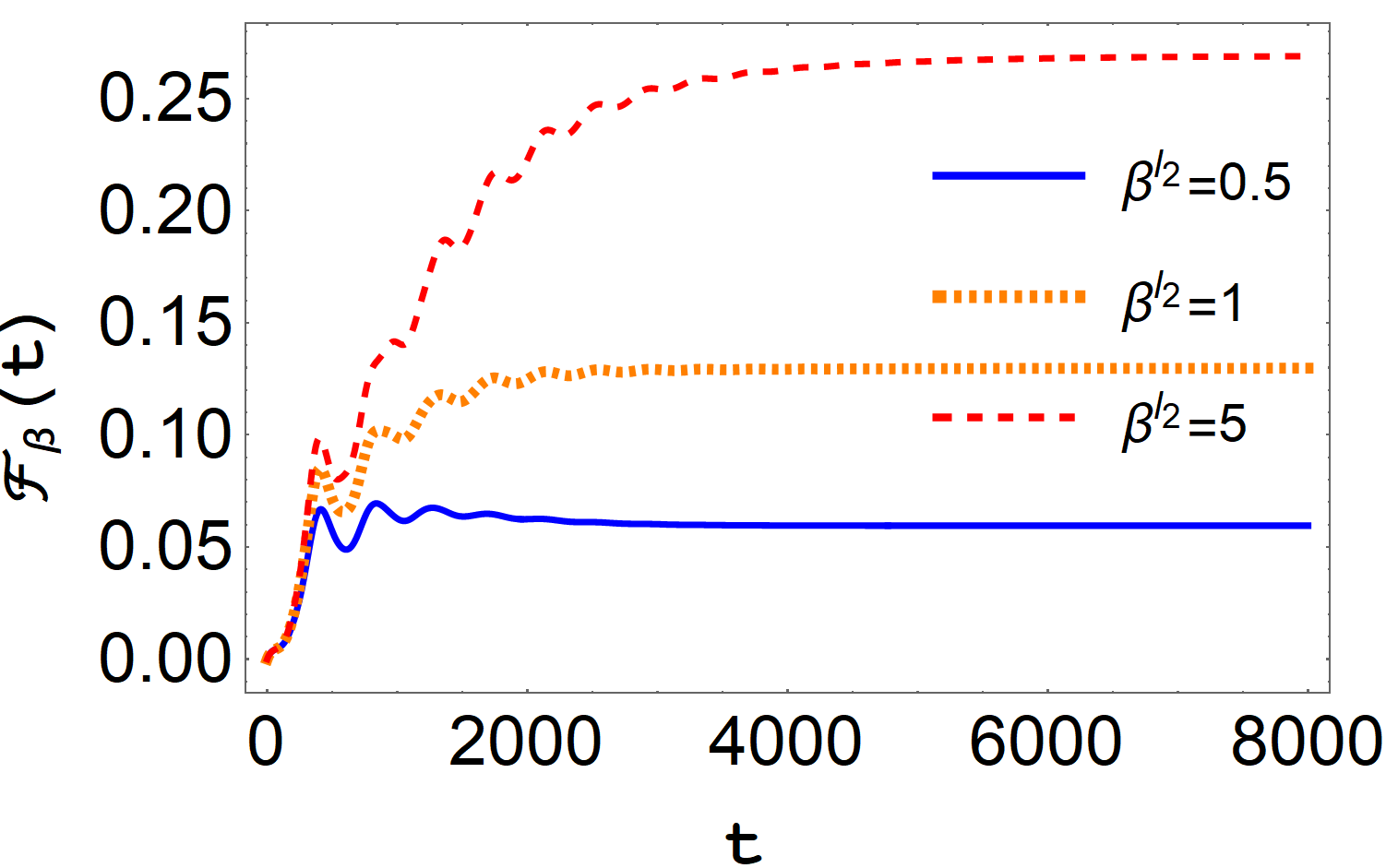}}
    \subfloat[]{\includegraphics[scale=0.35]{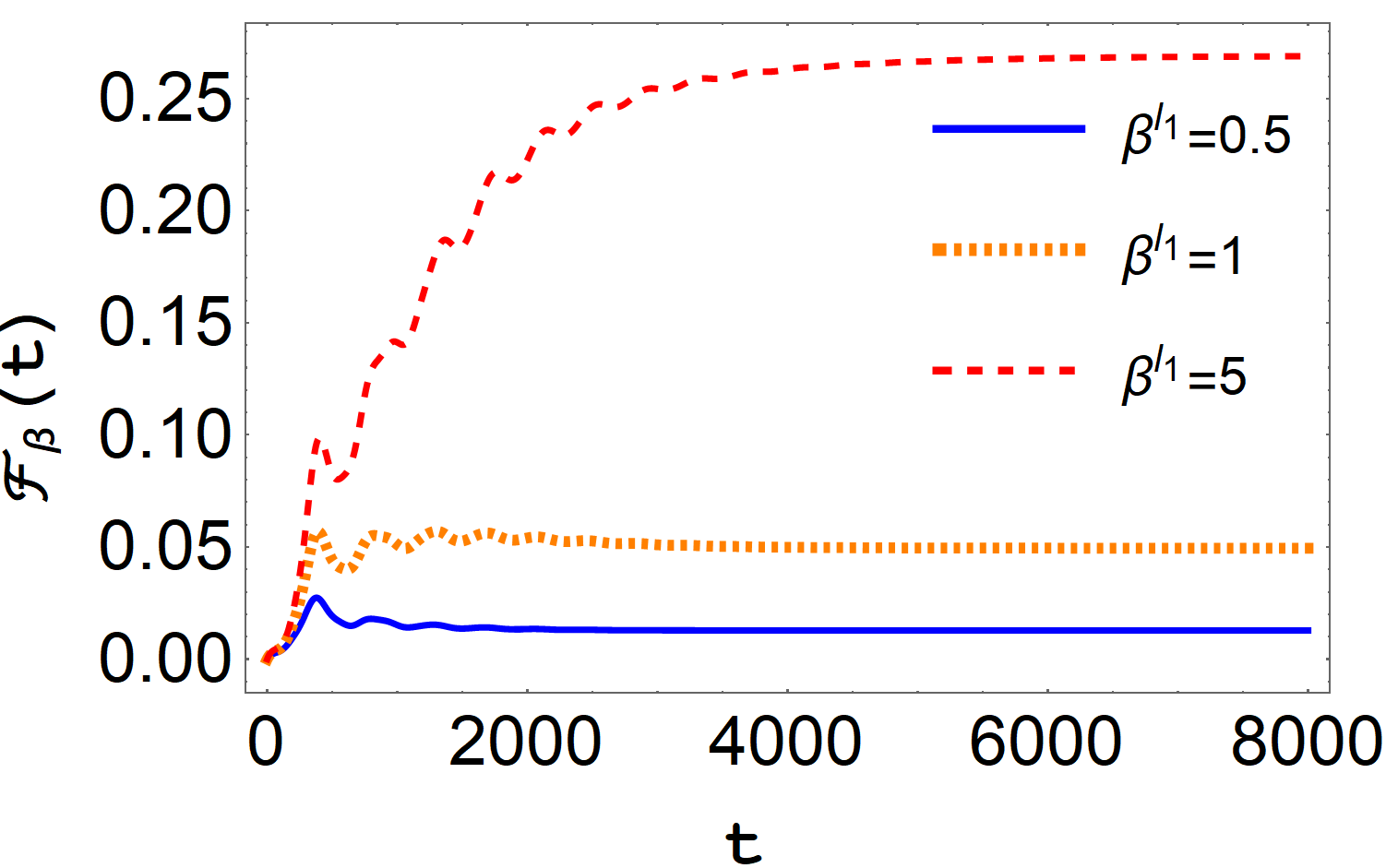}}
    \subfloat[]{\includegraphics[scale=0.35]{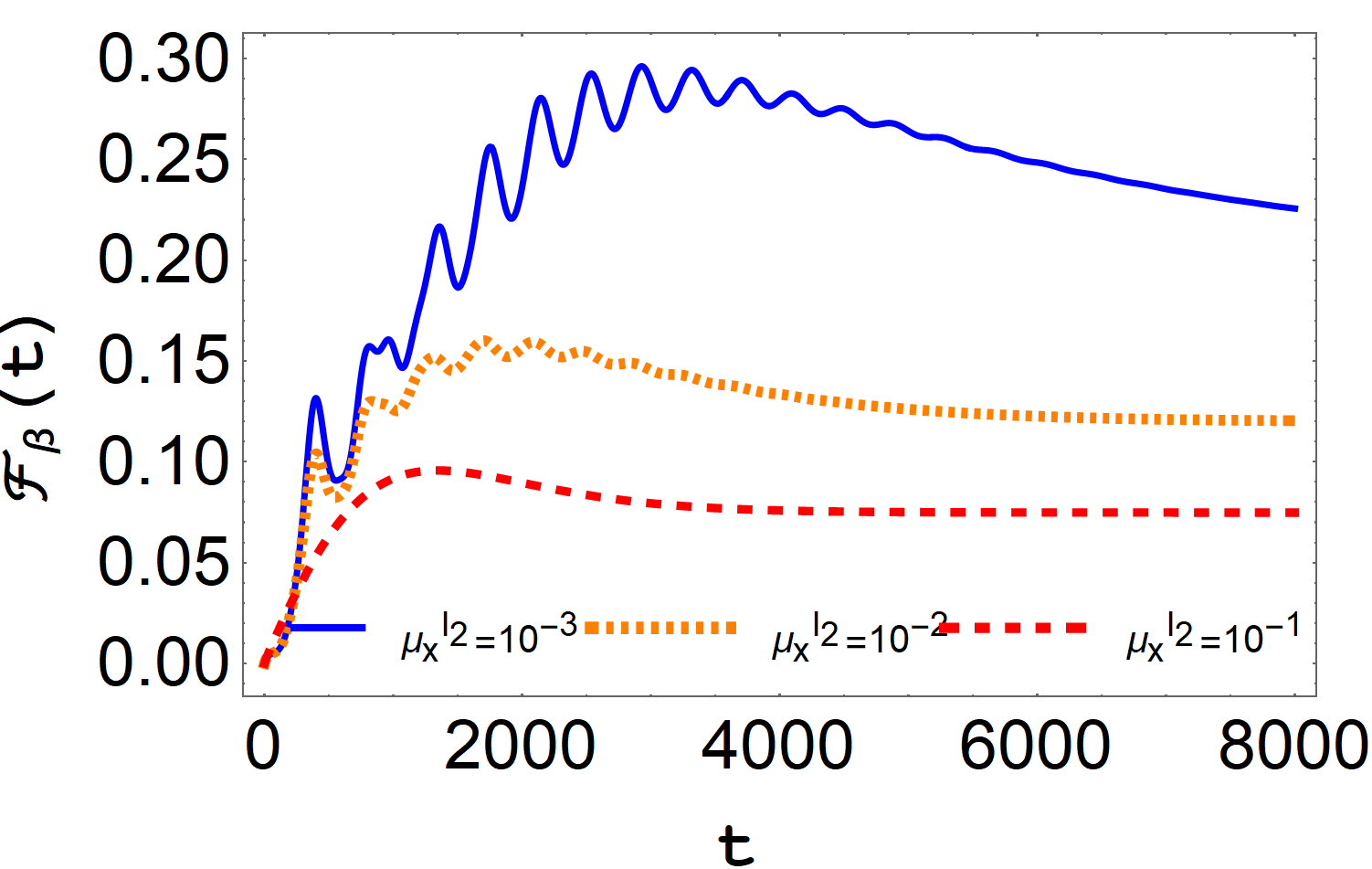}}
    \subfloat[]{\includegraphics[scale=0.35]{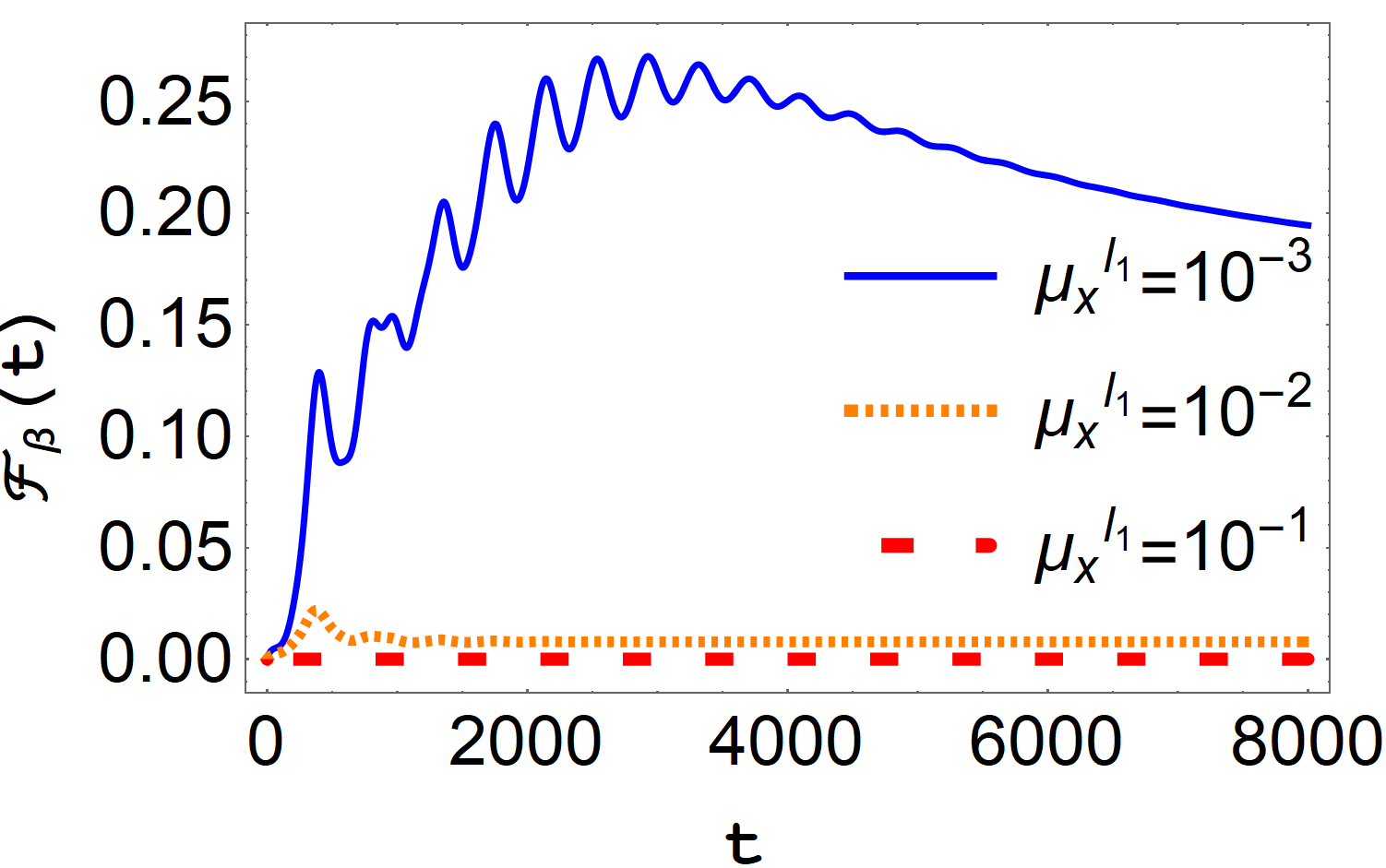}}
    \caption{QFI as a function of time for estimation of common bath temperature using the qubit 1 as a probe, and in the presence of local baths. The qubits are decoupled ($k=0$) and the detuning is small, $\omega_-=0.01$. All the interactions are purely dissipative. \textbf{(a)} We fix the temperatures of the local bath of qubit 1 and of the common baths at $\beta^{l_1}=5$ and $\beta^c=1$ respectively, while we vary the temperature of the local bath on qubit 2, $\beta^{l_2}$. The solid blue, orange dotted, and red dashed curves correspond to $\beta^{l_2}=0.5,1,5$, respectively. $\mu_x^c=\mu_x^{l_1}=\mu_x^{l_2}=10^{-2}$. \textbf{(b)} Same as panel \textbf{(a)}, but we vary the local bath temperature $\beta^{l_1}$, while the temperatures of the common bath and local bath 2 are set to $\beta^c=1$ and $\beta^{l_2}=5$, respectively. Here, the solid blue, orange dotted, and red dashed curves correspond to $\beta^{l_1}=0.5,1,5$, respectively.
    \textbf{(c)} We fix the bath temperatures, $\beta^c=1$, $\beta^{l_1}=0.4$, and $\beta^{l_2}=1$, while we vary the dissipative couplings of the local baths. The dissipative coupling to the common bath and local bath 1 are fixed at $\mu_x^{c}=10^{-2}$ and $\mu_x^{l_1}=10^{-4}$, respectively. The solid blue, orange dotted, and red dashed curves correspond to $\mu_x^{l_2}=10^{-3}, 10^{-2}, 10^{-1}$, respectively. \textbf{(c)} The dissipative couplings to the common bath and local bath 2 are fixed at $\mu_x^{c}=10^{-2}$ and $\mu_x^{l_2}=10^{-4}$, respectively. The solid blue, orange dotted, and red dashed curves correspond to $\mu_x^{l_1}=10^{-3}, 10^{-2}, 10^{-1}$, respectively. }
    \label{figCs}
\end{figure*}
%-------------------------------------------------------------------------%
\subsubsection{No local baths, estimation of the common bath temperature}
In this section, we fix $\mu_{x}^{l_1}=\mu_{x}^{l_2}=\mu_{z}^{l_1}=\mu_{z}^{l_1}=0$ (no local baths), and $k=0$ (no qubit-qubit interaction). Only the common bath at inverse temperature $\beta^c=1$ is present.
The dissipative  qubit-bath coupling is set to $\mu^c_{x}=10^{-2}$, while we assume that there is no dephasing, $\mu^c_z=0$.
We numerically calculate the QFI for the estimation of $\beta^c$~\cite{PhysRevResearch.4.043057}. The QFI is computed for the reduced state of qubit 1 at time $t$, which acts as the probe.

The results are shown in Fig.~\ref{fig3}. We plot the QFI calculated both through the partial secular master equation introduced in Sec.~\ref{psa_twoQ}, and through a master equation in full secular approximation in which we remove all the rotating terms (see Sec.~\ref{generalFormalism} for more details). The latter master equation is valid only in the limit of large detuning. Indeed, we observe that both the partial and full secular approximations yield identical results when the detuning is large, such as $\omega_-=0.5$, shown in Fig~\ref{fig3}(a). In this case, the two qubits are effectively decoupled in the open dynamics, and no correlations are built. Therefore, this scenario is analogous to the single qubit case discussed in the previous section (compare the results in the inset of Fig.~\ref{fig3}(a) with Fig.~\ref{figsb}).
%----------------------------------------------------%
\begin{figure}[b!]
    \centering
    \includegraphics[scale=0.3]{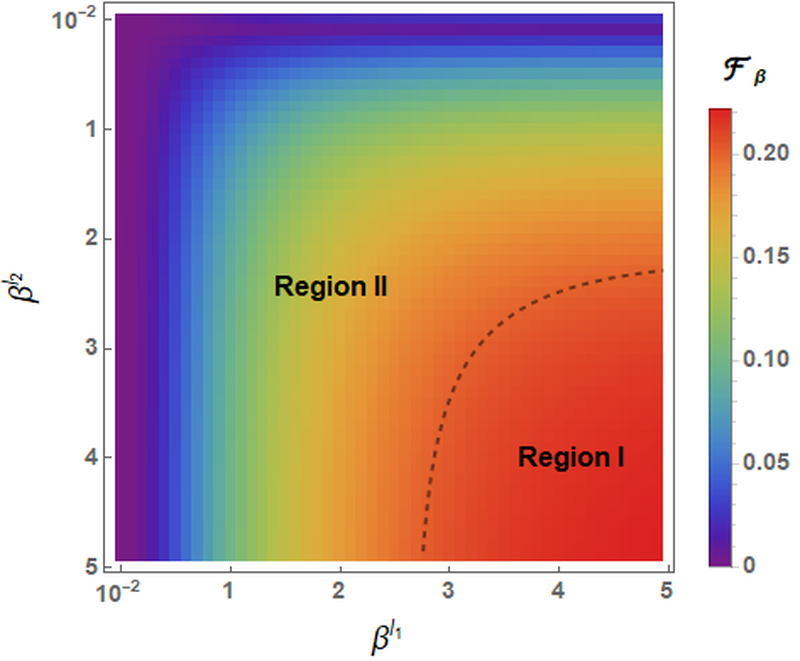}
      \caption{Heatmap of the QFI as a function of inverse temperatures $\beta^{l_1}$ and $\beta^{l_2}$ for the estimation of common bath temperature $\beta^c$ when $k=0$. The color scale indicates the magnitude of the QFI, with warmer colors representing higher values. The black dashed contour corresponds to the level set where $\mathcal{F}_\beta = 0.2$ (steady state QFI in the absence of local baths, as shown in Fig.~3(b)), separating two regions with distinct sensitivity to inverse temperature changes. Regions I and II, divided by the contour line, correspond to parameter regimes with higher and lower thermometric precision than in the absence of local baths, respectively. The rest of the parameters are $\mu_x=10^{-2}$, $\mu_z=0$, and $\omega_-=0.01$.}
    \label{fig:arrayQFI}
\end{figure}
%----------------------------------------------------%
In contrast, for small detuning  (e.g., $\omega_- = 0.01$), the results are now different, with the QFI being higher compared with the large detuning (i.e., single qubit) case in the transient dynamics, while the reduced steady state of qubit 1 is identical to that of the single-qubit case. The master equation with the partial secular approximation predicts a higher QFI compared with the full secular approximation, which is wrong in this regime and mimics the single-qubit case, as shown in Fig.~\ref{fig3}(b) (solid blue curve). For the master equation with full secular approximation, the QFI smoothly reaches its saturation value over time (red dashed curve). In contrast, under the partial secular approximation, the QFI exhibits oscillations over a short time interval. These oscillations can be interpreted as ``quantum beats''---a typical quantum effect in the presence of superradiance \cite{Ficek1987}---in the dynamics of the QFI. This kind of collective effect in the qubit dynamics enhances the sensitivity of single-qubit thermometry in the transient regime. Nevertheless, for longer time intervals, both approximations converge to the same value of the QFI, as expected. We note that the same results remain valid when employing the unified GKLS master equation, as proposed in Ref.~\cite{PhysRevA.103.062226} (see the inset of Fig.~\ref{fig3}(b)).  

We point out that our findings are consistent with the predictions of Ref.~\cite{PhysRevA.101.032112, PhysRevA.109.023309}, where it is shown that bath-induced correlations can be exploited to enhance thermometry precision in transient scenarios. Beyond such dynamical regimes, other studies have demonstrated that steady-state bath-induced correlations can also serve as a resource for quantum thermometry. For instance, in Ref.~\cite{PhysRevLett.128.040502}, a bosonic bath drives a multipartite, yet non-interacting system of quantum harmonic oscillators toward an entangled steady state, which enhances temperature estimation (see also a related scenario in Ref.~\cite{segalMultispin}). In contrast, in our model, the steady state of the system is thermal, thus, it cannot display any correlations between the non-interacting qubits. Therefore, the enhancement in temperature sensing is observed in the transient scenario. We also remark that our model is of practical relevance for platforms that are widely employed for quantum technologies, such as the system of superconducting qubits coupled to a common bath~\cite{Cattaneo2021engineering}.

In addition, we believe that it is important to study the effects of additional noise for temperature estimation in practical scenarios where qubits can couple to multiple local or shared baths. In the next section, we explore whether the enhancement in the QFI is robust with respect to the presence of additional sources of dissipation, which we assume to be known, acting locally on each qubit.

\subsubsection{Estimation of common bath temperature in the presence of local baths} \label{sec:common+Locals}
In this section, we switch on the local baths and study their effect on the temperature estimation of the common bath. We fix $\mu_x^c=10^{-2}$, and no dephasing ($\mu_{z}^{l_1}=\mu_{z}^{l_2}=\mu_z^c=0$) nor direct qubit-qubit coupling, $k=0$. We introduce a small detuning, $\omega_-=0.01$, and calculate the QFI for estimating the temperature of the common bath, $\beta^c$, using the first qubit as a probe. 

In Fig.~\ref{figCs}(a), we plot the QFI as a function of time in a scenario where we fix the temperatures of the common bath and of the local bath on the probe qubit 1, respectively $\beta^{c}=1$ and $\beta^{l_1}=5$, and vary the temperature of the local bath on qubit 2 ($\beta^{l_2}$). In Fig.~\ref{figCs}(b) we analyze the same scenario but we fix $\beta^{l_2}$ while varying $\beta^{l_1}$. In both figures, we choose $\mu_x^{l_1}=\mu_x^{l_2}=\mu_x^c$.

We observe that, in this scenario, the presence of hot local baths is highly detrimental for single-qubit temperature sensing. In fact, the QFI in Figs.~\ref{figCs}(a) and (b) is significantly reduced compared to Fig.~\ref{fig3} if any of the local baths is not at a very cold temperature ($\beta=5$). We note that the effects of the temperature of the local bath 2 are less prominent than those of the local 1, which is directly acting on the probe. However, even a local bath 2 at temperature $\beta^{l_2}=1$ is sufficient to significantly reduce the QFI, as observed in Fig.~\ref{figCs}(b).

That being said, we also observe a peculiar effect due to the local baths. If \textit{both} of them are very cold, $\beta^{l_j}=5$, then the steady-state value of the QFI is actually larger than in the absence of local baths. In other words, the two-qubit system is driven towards a non-thermal steady state for which the estimation of the temperature of the common bath is more precise than for a single-qubit thermal state. This behavior is illustrated in Fig.~\ref{fig:arrayQFI}, where we identify a threshold line in the parameter space of local inverse temperatures ($\beta^{l_1}$, $\beta^{l_2}$). Below this line—i.e., when both local baths are very cold—the QFI increases and eventually surpasses the steady-state value of QFI shown in Fig.~\ref{fig3}(b) in the absence of local baths. This reveals a transition between regions of degraded (Region II) and enhanced thermometric precision (Region I), highlighting the critical role of local bath temperatures in the estimation of common bath temperature $\beta^c$ in the steady state regime.

Moreover, in Fig.~\ref{figCs}(c), we investigate the effect of different values of qubit-bath coupling on the estimation of common bath temperature. We fix the temperatures of the baths $\beta^{c}=1$, $\beta^{l_1}=0.4$, and $\beta^{l_2}=1$. The dissipative couplings for the common bath and the hot bath are set to $\mu_x^c=10^{-2}$ and $\mu_x^{l_1}=10^{-4}$, respectively. We then investigate the influence of varying the cold qubit-bath coupling strength, $\mu_x^{l_2}$, on the estimation of the common bath temperature using the first qubit as a probe. 

When the coupling strength between the second qubit and its local bath is weak (e.g., $\mu_x^{l_2}=10^{-3}$), the QFI as a function of time closely resembles the QFI for common bath estimation in the absence of local baths (represented by the solid blue curve). In contrast, stronger qubit-bath coupling values, $\mu_x^{l_2}$, significantly reduce the QFI. This highlights the critical role of qubit-bath coupling in the presence of local baths on the estimation of common bath temperature.

Similarly, the coupling between the first qubit (which is a probe) and its local bath also plays a pivotal role (see Fig.~\ref{figCs}(d)). Since the local bath of the first qubit has a higher temperature compared with the other two baths, it exerts a strong negative effect on the QFI. Even for relatively strong qubit-bath couplings, such as $\mu_x^{l_1}=10^{-1}$, the QFI remains negligible compared to the values obtained for $\mu_x^{l_2}=10^{-1}$ when $\mu_x^{l_1}=10^{-4}$, as shown in Fig.~\ref{figCs}(d). In particular, the probe qubit must be weakly coupled to its local bath to get enhanced sensitivity for the estimation of common bath temperature. The key takeaway is that local bath coupling strengths significantly impact the estimation of common bath temperature, and they should ideally be kept weak to minimize their adverse effects.

\subsubsection{Estimation of local baths temperature in the presence of a common bath and no qubit-qubit interaction} \label{sec:localsNoDir}
Here, we consider the same model as in the previous section, with both a common bath and two separate local baths. However, we focus on the estimation of the temperature of a local bath instead of the common bath, using a qubit that is not directly coupled to this local bath. For now, we set $k=0$, i.e., no qubit-qubit coupling. 
Although the two qubits do not interact, it is still possible to remotely sense the temperatures of their local baths through correlations induced by the common bath.
%------------------------------------------------------------------%
\begin{figure}[t!]
    \centering
    \subfloat[]{\includegraphics[scale=0.6]{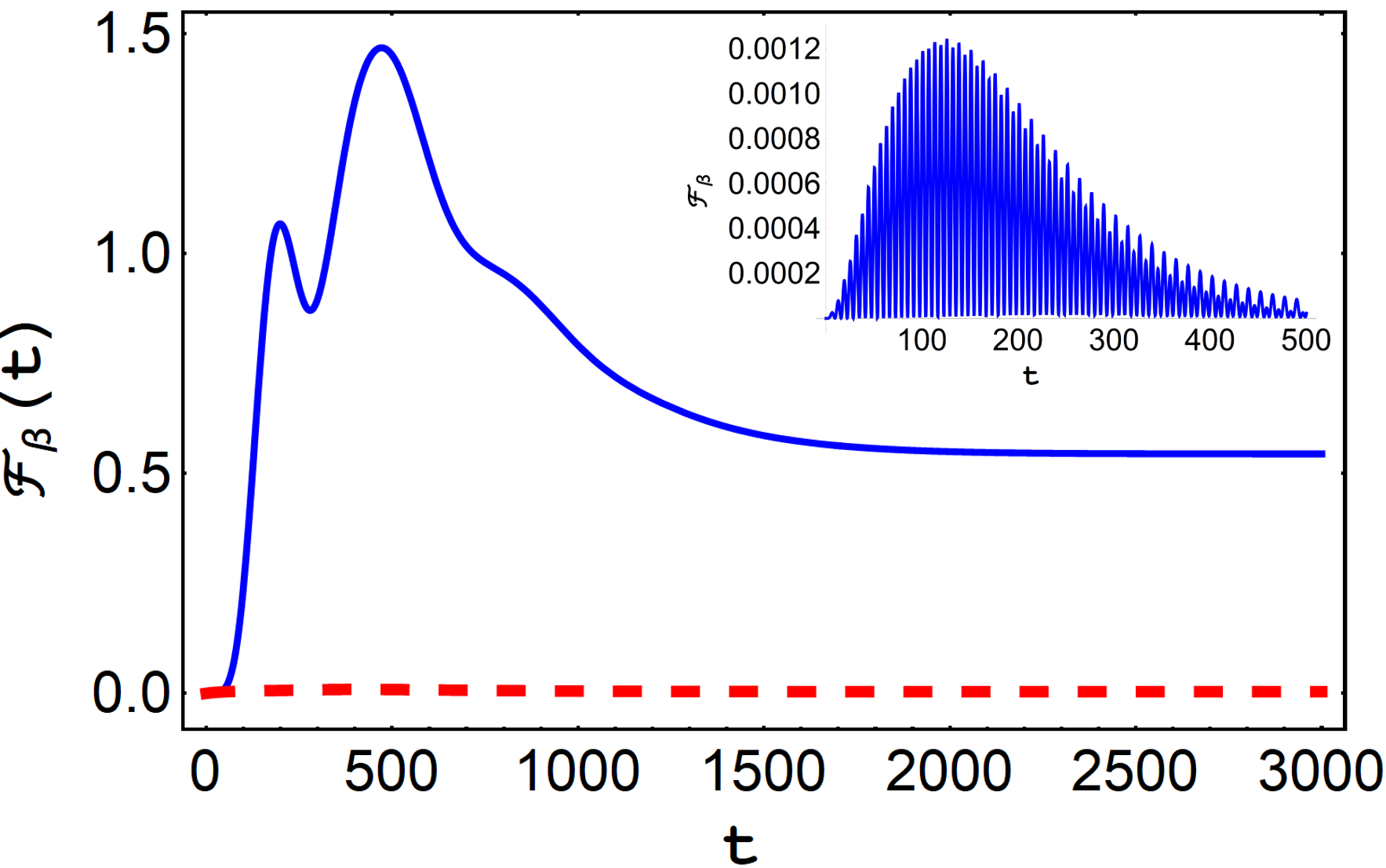}}\\
  \subfloat[]{ \includegraphics[scale=0.62]{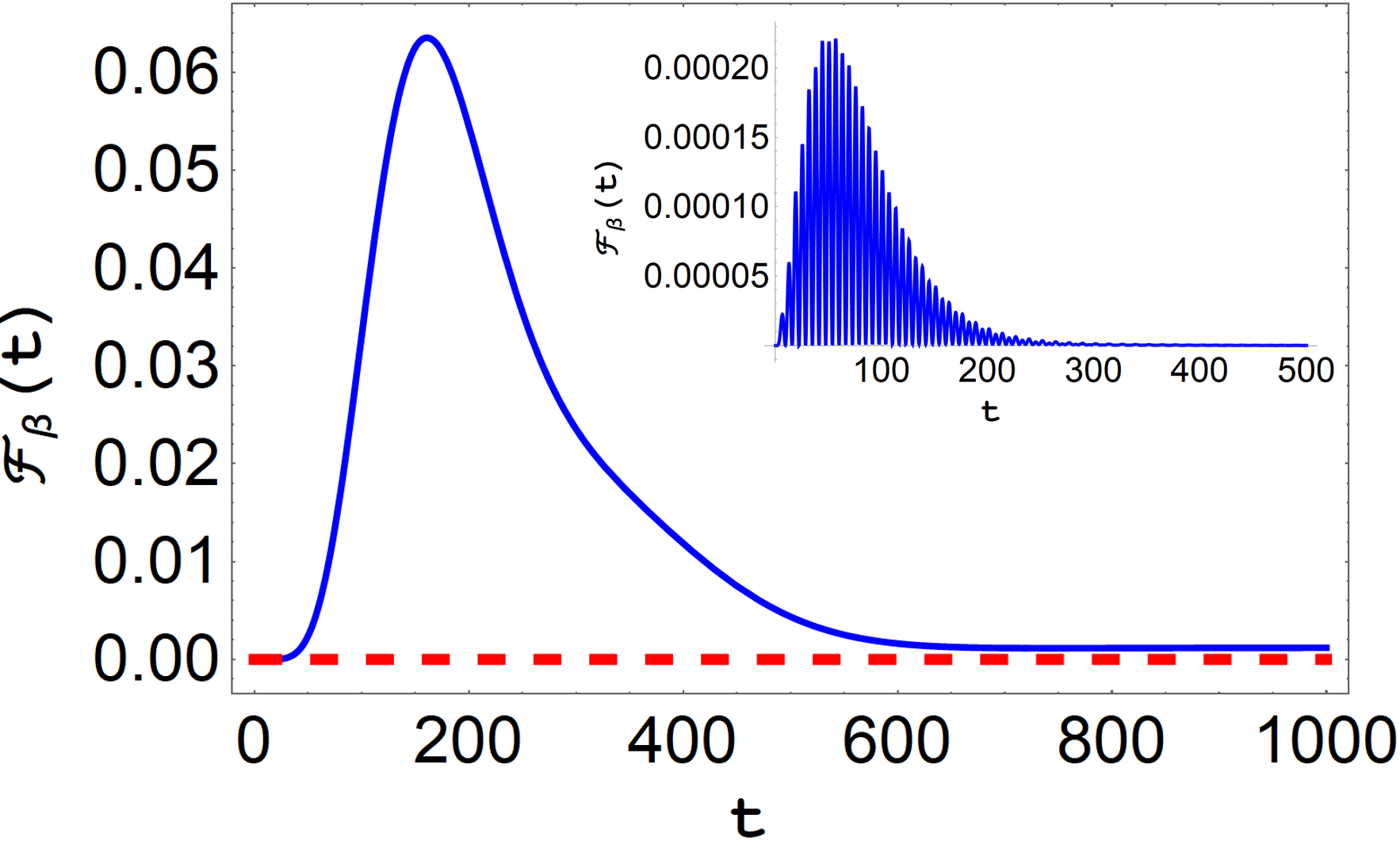}}
    \caption{Scenario with two uncoupled qubits ($k=0$) interacting with both a common bath and two local baths. The interaction is purely dissipative: $\mu_x = 10^{-2}$, $\mu_z =0$. \textbf{(a)} QFI  as a function of time $t$ for the estimation of the local bath temperature $\beta^{l_1}$ (blue curve) using qubit 2 as a probe and for the estimation of the temperature $\beta^{l_2}$ (red curve) using qubit 1 as a probe, for small detuning ($\omega_-=0.01$). The inset illustrates the QFI as a function of time for the estimation of $\beta^{l_1}$ and large detuning ($\omega_- = 0.5$) . The inverse temperatures are $\beta^c = 1$, $\beta^{l_1} = 0.1$, and $\beta^{l_2} = 1$. 
    \textbf{(b)} QFI for the same set of parameters but with $\beta^c=0.1$ for small detuning such as $\omega_-=0.01$. The inset shows the QFI as a function of time for big detuning such as $\omega_-=0.5$. All the other parameters are the same as in panel (a).
    }
    \label{fig4}
\end{figure}
%-----------------------------------------------------------------%
We fix $\beta^{l_1}=0.1$ and $\beta^{l_2}=1$. We see that the qubit 2 can estimate the temperature of qubit 1, enabling remote sensing when direct access to the bath is not feasible. At the same time, the sensitivity of the converse operation is very low.

We calculate the QFI for each local bath using Eq.~\eqref{qfi} and assume that we have already measured $\beta^c=1$. When calculating the QFI for the estimation of temperature \( \beta^{l_2} \), we use the reduced density matrix of qubit 1, we keep \( \beta^{l_1} \) fixed and exploit qubit 1 as a probe, and vice versa. We analyze two scenarios: large detuning (\(\omega_- = 0.5\)) and  small detuning (\(\omega_- = 0.01\)).  %To measure \(\beta^{l_2}\), the hot qubit is used as a probe, while the cold qubit serves as the probe for measuring \(\beta^{l_1}\).  
%-------------------------------------------------------------------------%
\begin{figure}[t!]
    \centering
  \subfloat[]{\includegraphics[scale=0.61]{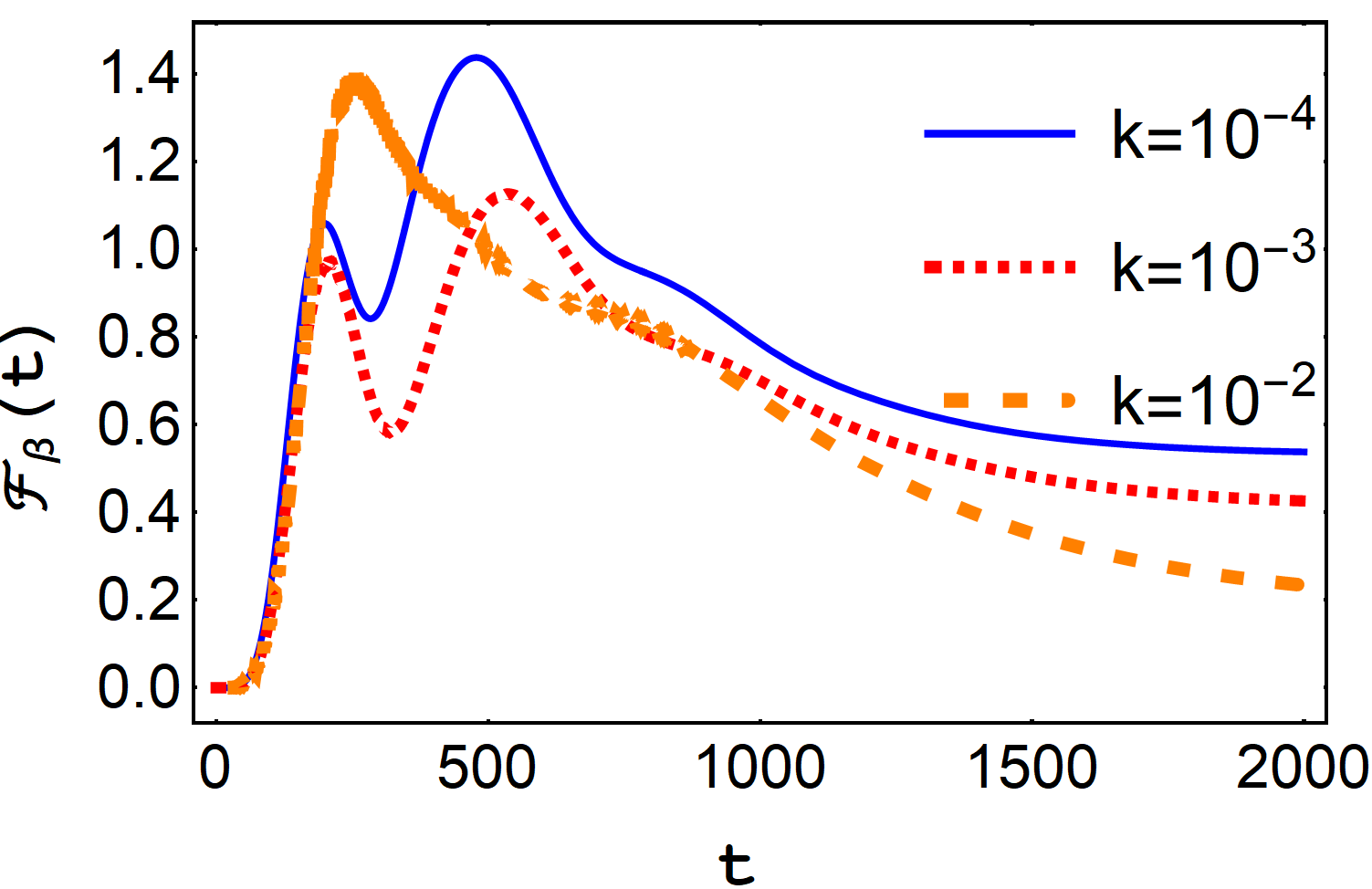}}\\
        \subfloat[]{\includegraphics[scale=0.61]{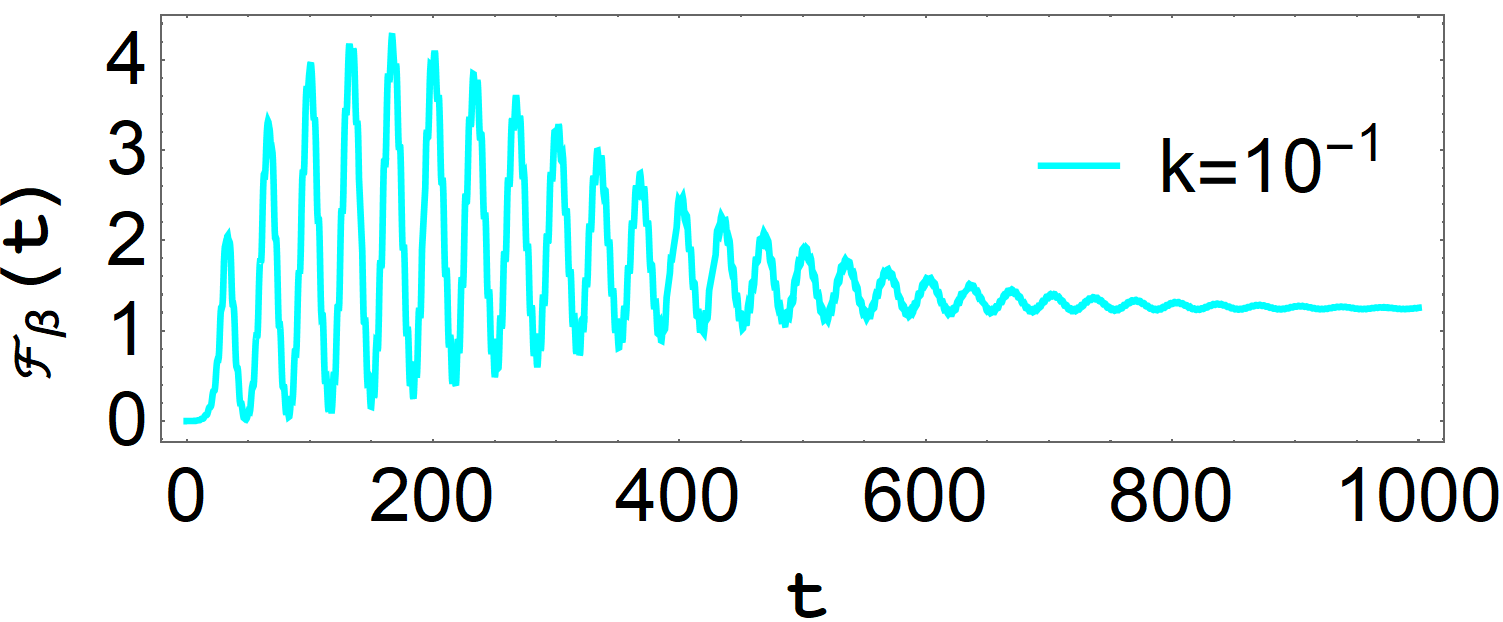}}
    \caption{Scenario with two coupled qubits interacting with both a common bath and two local baths. The interaction is purely dissipative: $\mu_x = 10^{-2}$, $\mu_z =0$. We plot the  QFI  as a function of time $t$ for the estimation of $\beta^{l_1}$ using qubit 2 as a probe. We consider small detuning ($\omega_-=0.01$). The inverse temperatures are $\beta^c = 1$, $\beta^{l_1} = 0.1$, and $\beta^{l_2} = 1$. \textbf{(a)} Weaker values of the qubit-qubit coupling constant $k$: the blue solid, red dotted, and orange dashed curves correspond to $k=10^{-4}$, $k=10^{-3}$, and $k=10^{-2}$, respectively. $k$ is the qubit-qubit coupling constant. \textbf{(b)} Stronger qubit-qubit coupling constant, $k=10^{-1}$.}
    \label{fig9}
\end{figure}
%------------------------------------------------------------------------------%

Fig.~\ref{fig4}(a) shows the QFI for estimating the temperature of the local bath $\beta^{l_1}$ (blue curve) using qubit 2 as the probe, and $\beta^{l_2}$ (red curve) using qubit 1 as the probe, based on their respective reduced states. Focusing on the estimation of $\beta^{l_1}$, we observe that for small detuning the QFI   does not show rapid oscillations as it increases, reaching its maximum value and then decreasing, and eventually it attains a constant value over time. Over a longer timescale, the QFI stabilizes, reaching a steady-state value.  Note that the QFI is considerably enhanced with respect to the single-qubit case. This is often the case, as demonstrated in Ref.~\cite{PhysRevA.98.042124}, which shows that, due to coherence effects, measuring the probe qubit can yield more information than measuring the qubit directly coupled to the bath. In contrast, for large detuning, the QFI shows some rapid and negligible oscillations shown in the inset of Fig.~\ref{fig4}(a), because the qubits are effectively decoupled in this regime. The QFI reduces significantly if the common bath temperature has the same temperature as the hot bath temperature, such as $\beta^c=0.1$, as shown in Fig.~\ref{fig4}(b) for both small ($\omega_-=0.01$) and big detuning ($\omega_-=0.5$).

Focusing now on the estimation of $\beta^{l_2}$, we observe that the QFI in this case is very small. So, our protocol for remote sensing of the inverse temperature works only for enough high temperatures of the local bath.

The remote sensing of \( \beta^{l_1} \) using qubit 2 as a probe is enabled by the correlations induced by the common bath, allowing the probe to measure the temperature of a distant bath without direct access to it.

\subsubsection{Estimation of local baths temperature in the presence of a common bath and with directly coupled qubits}
We now consider the case in which there is a direct qubit-qubit coupling. 
We numerically solve the global master equation under partial secular approximation for the two coupled qubits~\cite{Cattaneo_2019} and calculate the QFI. In Fig.~\ref{fig9}(a), we present the QFI as a function of time $t$ for the estimation of $\beta^{l_1}$ when the two qubits are coupled. The qubit 2 is used as a probe, and a small detuning value of $\omega_- = 0.01$ is assumed. We use the reduced density matrix of qubit 2 to calculate the QFI for the estimation of $\beta^{l_1}$.

In the weak-coupling regime (e.g., $k = 10^{-4}$, represented by the solid blue curve in Fig.~\ref{fig9}(a)), the QFI is very similar to the case where there is no direct coupling between the qubits (solid blue curve in Fig.~\ref{fig4}(a)). For other values of the coupling strength in the weak-coupling regime, such as $k=10^{-3}$ and $k=10^{-2}$, the QFI does not change significantly, but the QFI reaches its maximum value more quickly, as shown in Fig.~\ref{fig9}(a). These results show that a direct qubit-qubit coupling generates correlations in a way similar to the action of a common bath in the non-interacting scenario. For weaker values of $k$, the magnitude of $k$ does not significantly affect how we may exploit these correlations for the sake of remote temperature sensing (i.e., the behavior of the QFI is similar).

Interestingly, if the coupling strength between the qubits is stronger, the maximum value of the QFI significantly improves, and it shows an oscillatory behavior as a function of time. For example, if $k = 10^{-1}$, the QFI is significantly larger compared with the scenario with uncoupled qubits, as shown in Fig.~\ref{fig9}(b). Another interesting factor is that the probing time for sensing the temperature is also reduced now. 
However, we have verified that the QFI for estimating the temperature of the cold bath is still nearly zero compared with that of the hot bath, even for strong coupling ($k=10^{-1}$) between the two qubits. This is because the temperature of the cold bath is significantly lower than that of the hot bath, making its measurement inherently more challenging, exactly as in the non-interacting scenario addressed in Fig.~\ref{fig4}(b).% Moreover, the quantum correlations induced by the common bath are insufficient to enable the temperature sensing of the cold bath.

%-------------------------------------------------------------------------%
\begin{figure}
    \centering
  \subfloat[]{\includegraphics[scale=0.63]{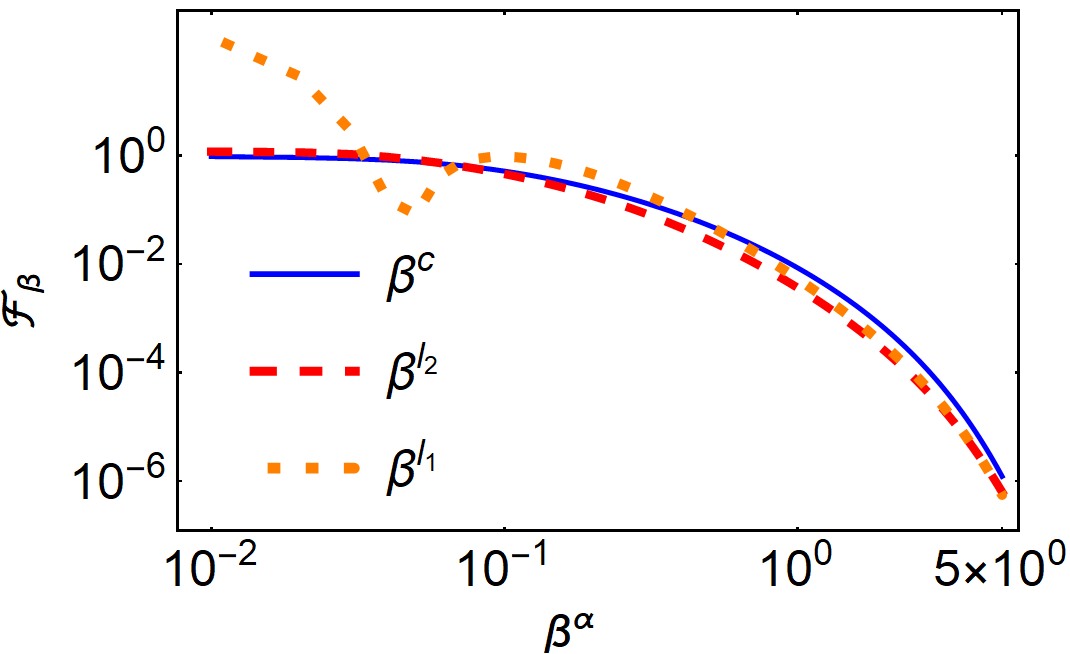}}\\
  \subfloat[]{\includegraphics[scale=0.63]{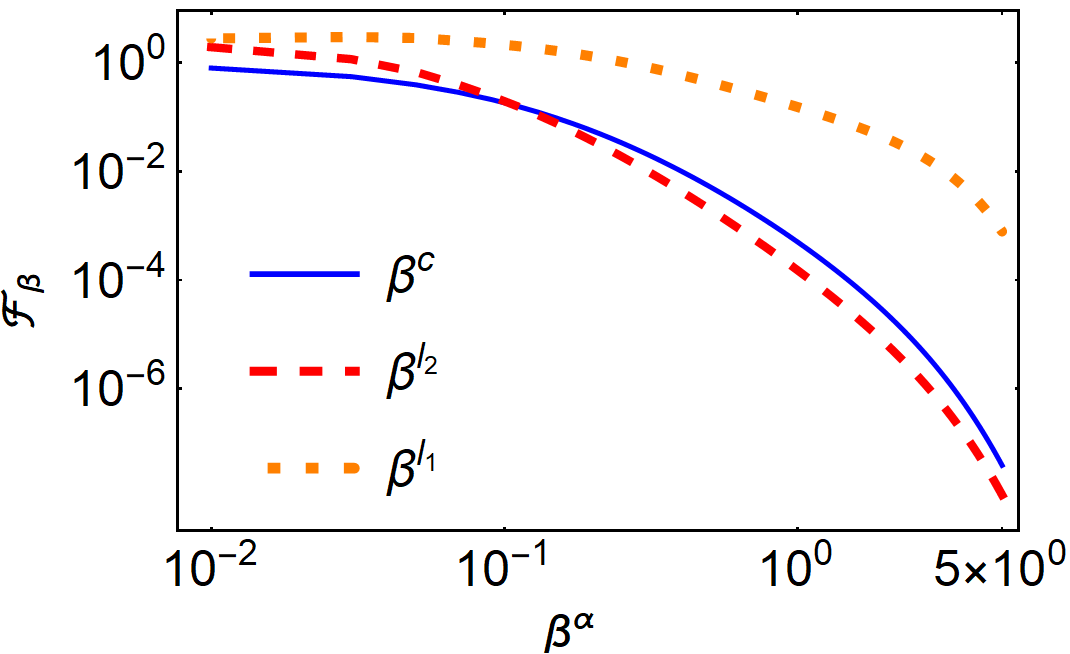}}
    \caption{\textbf{(a)} QFI in the steady state of a single-qubit reduced density matrix as a function of the inverse temperature $\beta^\alpha$, where $\alpha = l_1, l_2, c$ for two uncoupling qubits ($k=0$). The solid blue, red dashed, and orange dotted curves represent the estimation of the common bath, cold bath, and hot bath, respectively. For the estimation of the common bath, we vary $\beta^c$ while fixing the temperatures of the other two baths at $\beta^{l_1} = 0.1$ and $\beta^{l_2} = 5$. Similarly, for the estimation of $\beta^{l_1}$, we fix $\beta^{l_2} = 5$ and $\beta^c = 5$, and for the estimation of $\beta^{l_2}$, we set $\beta^{l_1} = 0.1$ and $\beta^c = 5$. The detuning is small, such as $\omega_-=0.01$. \textbf{(b)} QFI in the steady state for a single qubit reduced density matrix as a function of the inverse temperature $\beta^\alpha$, where $\alpha = l_1, l_2, c$ for two coupled qubits ($k=10^{-1}$). The solid blue, red dashed, and orange dotted curves represent the estimation of the common bath, cold bath, and hot bath, respectively. For the estimation of the common bath, we vary $\beta^c$ while fixing the temperatures of the other two baths at $\beta^{l_1} = 0.1$ and $\beta^{l_2} = 5$. Similarly, for the estimation of $\beta^{l_1}$, we fix $\beta^{l_2} = 5$ and $\beta^c = 5$, and for the estimation of $\beta^{l_2}$, we set $\beta^{l_1} = 0.1$ and $\beta^c = 5$.  The detuning is small, such as $\omega_-=0.01$. }
    \label{fig:steadystate}
\end{figure}
%------------------------------------------------------------------------------%

\subsubsection{Steady state QFI for estimation of common bath and local baths temperature with both uncoupled and directly coupled qubits}\label{steady-state}
We now consider the steady state of the system, i.e., the state obtained in the limit \( t \to \infty \), and investigate remote temperature estimation at equilibrium. We also study the temperature estimation of the common bath as a function of $\beta^c$ in the presence of one very hot and one very cold local bath. We assume a small detuning (\(\omega_- = 0.01\)) and consider a purely dissipative qubit-bath coupling.

First, the qubits are uncoupled ($k=0$), and we fix the other parameters as in Sec.~\ref{sec:localsNoDir}. We compute the QFI for the estimation of i) common bath temperature using qubit 1 as a probe; ii) ``cold'' local bath temperature using qubit 1 as a probe; iii) ``hot'' local bath temperature using qubit 2 as a probe. Within each scenario, we vary the temperature we aim to estimate, and fix the other two temperatures. The results are shown in Fig.~\ref{fig:steadystate}(a). We observe that we obtain a high value of the QFI only for low values of $\beta$, independently of the temperature we are estimating. In other words, our protocol for remote temperature sensing works only for sufficiently high temperatures. Moreover, the presence of a very hot local bath hinders the estimation of the temperature of the common bath, unless the common bath is also very hot (around $\beta^c=0.1$). This was expected from our analysis in Sec.~\ref{sec:common+Locals}.

Second, we compute the steady-state QFI for two directly interacting qubits. We consider the coupling strength between the qubits in the strong-coupling regime, specifically \( k = 10^{-1} \), while keeping all other parameters identical to the case of uncoupled qubits.
 As in Fig.~\ref{fig:steadystate}(b), we vary the temperature to be estimated while keeping the other two temperatures fixed. In this case, the temperatures of the local baths and the common bath are the same as in the case of uncoupled qubits, as the specific values of parameters are described in Fig.~\ref{fig:steadystate}(b).

For the estimation of all bath temperatures, the QFI attains higher values when the two qubits are strongly coupled (e.g., $k = 10^{-1}$). Notably, the QFI for estimating the common bath temperature remains nearly unchanged regardless of whether the qubits are coupled or not. Similarly, the QFI for the cold bath temperature shows a slight increase, whereas the QFI for the hot bath temperature exhibits a significant enhancement when the qubits are coupled. Interestingly, the QFI for the estimation of the inverse temperature of the ``hot bath'', $\beta^{l_1}$, is greatly enhanced in the strong-coupling regime also at high $\beta^{l_1}$, i.e. low temperature, provided all the other baths are very cold ($\beta^c=\beta^{l_2}=5$). For instance, if $\beta^{l_1}=1$, then $\mathcal{F}_{\beta^{l_1}}\approx 0.25$, which is higher than in the single-qubit case. This result shows that remote temperature sensing is also practical at lower temperatures if the qubits are directly coupled and all the other sources of dissipation are kept at very cold temperatures.

The value of the QFI for the estimation of $\beta$ in the steady state is a beneficial measure to characterize the measurement sensitivity in different situations. Moreover, the relative error on $\beta$, which can be obtained through the quantum Cramér-Rao bound, may also be an important indicator. The readers interested in the relative error on the steady-state values of $\beta$  as discussed in this section, can find more information and details in Appendix~\ref{relError}.
%-----------------------------------------------------------------------%
%-----------------------------------------------------------------------------%
\begin{figure}[t!]
    \centering
        \subfloat[]{\includegraphics[scale=0.65]{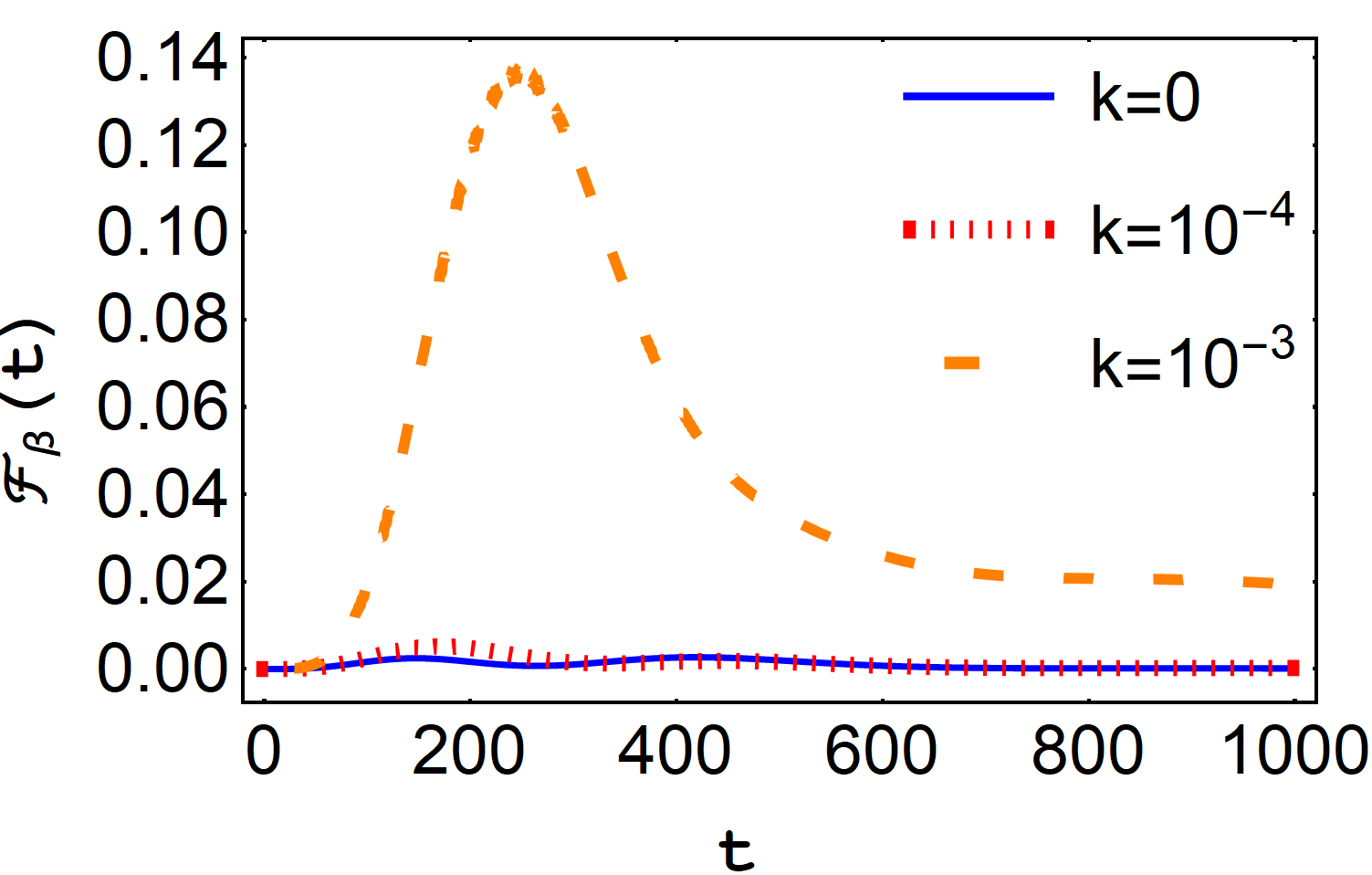}}\\
         \subfloat[]{\includegraphics[scale=0.64]{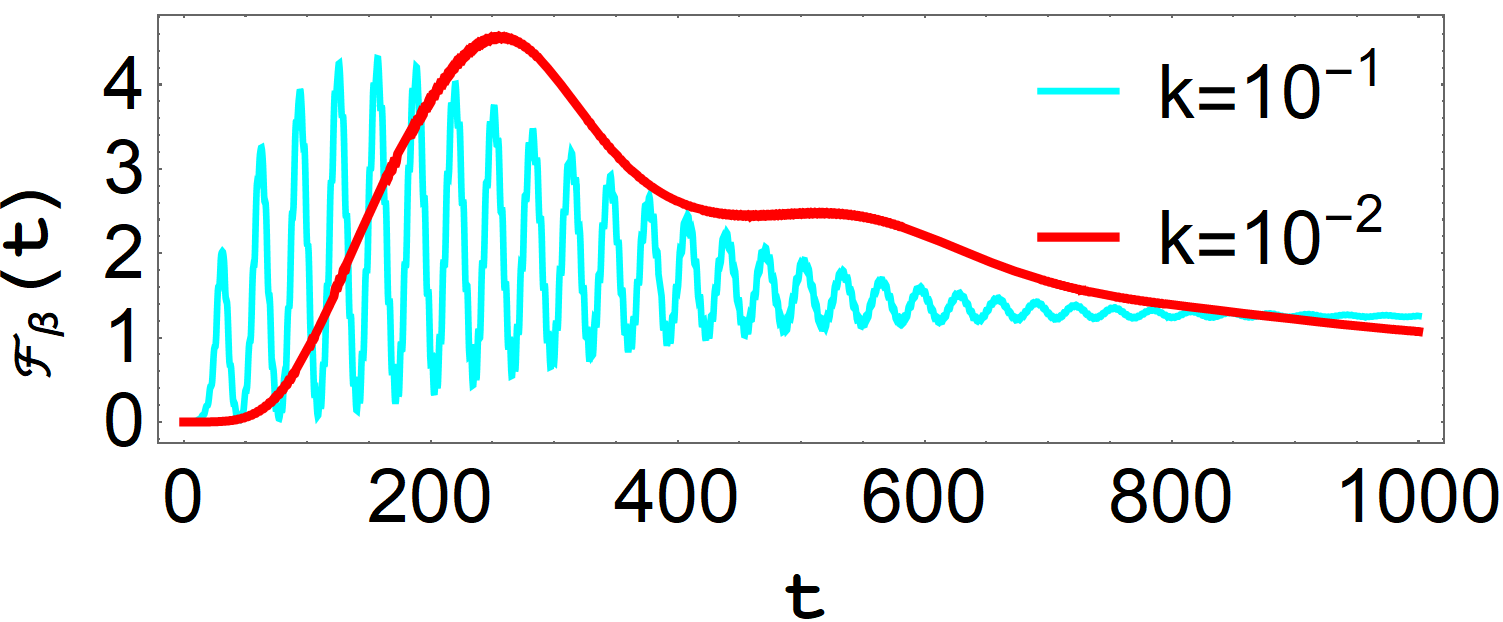}}
    \caption{Scenario with two interacting or non-interacting qubits, with both a common bath and two local baths, and in the absence of the Lamb shift. We plot the QFI as a function of time $t$ for the estimation of  $\beta^{l_1}$ using qubit 2 as a probe. The plots show the evolution for different values of the qubit-qubit coupling strength $k$. The parameters are set as follows. $\omega_- = 0.01$, $\mu_x = 10^{-2}$, $\mu_z=0$, $\beta^c = 1$, $\beta^{l_1} = 0.1$, and $\beta^{l_2} = 1$. }
    \label{fig:coupled}
\end{figure}
%--------------------------------------------------------------------------%
\subsubsection{Role of the Lamb-Shift}\label{lambshift}
In the scenario of non-interacting qubits, we have observed that the Lamb shift plays a significant role in remote temperature sensing. To check this, we remove the Lamb shift term from the system dynamics in both scenarios and examine whether the dissipation alone can generate a sufficient amount of common bath induced correlations to remotely estimate the temperature of the local baths. 

We consider the case of two non-interacting qubits and the QFI as a function of time when $k=0$ is plotted in Fig.~\ref{fig:coupled}(a) (blue curve) in the absence of the Lamb-shift term. The QFI is used to estimate the hot bath temperature $\beta^{l_1}$ using qubit 2 as a probe. We observe that, for the same set of parameters, the QFI is negligible compared with Fig.~\ref{fig4}(a) (blue curve). However, when the two qubits are weakly coupled ($k = 10^{-4}$), the QFI strength increases, as shown in Fig.~\ref{fig:coupled}(a) (red dotted curve). Nevertheless, its maximum value remains relatively small. 
For $k = 0$ and coupling strengths in the weak-coupling regime, such as $k = 10^{-4}$ ( the red dotted curve in Fig.~\ref{fig:coupled}(a)) and $k = 10^{-3}$ ( the orange dashed curve in Fig.~\ref{fig:coupled}(b)), the QFI exhibits small values and decays to approximately zero. In contrast, for stronger coupling (\( k = 10^{-2} \)), the QFI reaches a higher maximum and does not decay as quickly, as shown by the solid red curve in Fig.~\ref{fig:coupled}(b). This is due to the fact that the Lamb-shift coefficient for the cross-terms in our scenario is negative, so it effectively decreases the direct qubit-qubit coupling. For $k=0.1$ this effect is not felt anymore, and the dynamics is similar to that in the presence of the Lamb shift. Moreover, we have also verified that, if we choose a value of $k$ equal to the Lamb-shift coefficient for the cross-terms in the scenario of Fig.~\ref{fig4}(a), we obtain a very similar evolution as in Fig.~\ref{fig4}(a).
%Curiously enough, for strong couplings the QFI is higher than in the presence of Lamb shift: compare Fig.~\ref{fig5} with Fig.~\ref{fig9}(b). Moreover, the QFI exhibits rapid oscillations before eventually stabilizing at its saturation value.

These results show that the Lamb shift term is essential for the sake of remote temperature sensing with non-interacting qubits. In other words, if the qubits are uncoupled, we observe that the bath-induced correlations are mostly due to the collective Lamb shift generated by the common bath, which is consistent with previous related results in the literature~\cite{Solenov2007,Cattaneo2021collective}.

%Interestingly, the maximum value of the QFI in the absence of the Lamb shift for $k = 10^{-1}$ (Fig.~\ref{fig4}(c)) is approximately equal to the QFI in the presence of the Lamb shift for $k = 10^{-2}$ (solid red curve in Fig.~\ref{fig5}(b)). However, in the latter case, the QFI does not exhibit oscillatory behavior. An important observation is presented in Fig.~\ref{fig5}(b), where the QFI is plotted for \( k = 10^{-1} \) (a solid cyan curve). In the absence of direct qubit-qubit coupling, dissipation alone does not provide much information about local bath temperatures. The critical factor for this information is the Lamb shift, which introduces an effective qubit-qubit interaction due to quantum correlations created by this term. In the case of interacting qubits, the significant QFI arises from two factors: first, the direct coupling between the qubits, and second, the Lamb shift term, which induces quantum correlations.

%==================================================================================%
\section{Conclusions}\label{conc}
%==================================================================================%
\textcolor{black}{We investigated single-qubit probes for temperature sensing in the presence of local and collective baths}. Specifically, we considered two qubits that are coupled to a common bath and their respective local baths. We analyzed the open-system dynamics using a Lindblad master equation approach derived under partial secular approximation. 

 \mar{First, we explored the single-qubit scenario to measure the temperature of a single bath and understand the behavior of the QFI as a function of time and temperature. Next, we extended our analysis to two non-interacting qubits.}
 
 \mar{We began by examining the scenario in which there were no local baths, and a single qubit was employed as a probe to measure the temperature of the common bath. }
Our results showed that, if the detuning between the qubits is small such that correlations can be generated by the common bath, the sensitivity of non-equilibrium single-qubit thermometry is enhanced with respect to the scenario where a single qubit is present, or if there is large detuning (i.e., no collective effects) between the qubits. We have also observed \textit{quantum beats} in the dynamics of the QFI, that is, oscillations due to the interference effects generated by the common bath.

Moreover, we also explored the robustness of this result with respect to the presence of local baths acting on the qubits. We noticed that the local baths, in general, jeopardize the precision of temperature estimation, unless in the case where all of them are very cold ($\beta \approx 5$). In the latter case, the steady-state value of the QFI is actually enhanced by the presence of the local baths.

Next, considering both the common and the local baths, we studied whether qubit 1 could be employed as a probe of the temperature of the local bath on qubit 2 (or vice versa), enabling remote temperature sensing. 
 Although one qubit lacks direct access to the bath of the other qubit, the common bath induces correlations and enables remote sensing of the temperature of the distant local bath. We have observed that these correlations arise primarily due to the collective Lamb-shift term appearing in the master equation of the two-qubit system. Our findings suggest that collective dissipation alone is insufficient for precise temperature estimation; rather, the Lamb shift plays a key role in remote temperature sensing.

However, the sensitivity of remote temperature sensing is effective only for low values of the inverse temperature $\beta$, which corresponds to high temperatures. In particular, our results show that remote temperature sensing provides a great enhancement of the QFI for $\beta=0.1$ (renormalized by the qubit frequency), while it is not effective anymore for $\beta=1$. In systems of superconducting qubits, this value of $\beta$ may roughly correspond to $T=2$ K \cite{Cattaneo2021collective}, which is an extremely high temperature for superconducting circuits. The actual feasibility of our protocol for uncoupled qubits may therefore be limited on current quantum platforms.

Moreover, we have computed the QFI with respect to $\beta$ and noted that it would be less favorable if computed with respect to temperature $T$, due to the $1/T^4$ term in Eq.~\eqref{qfi_wrt_T}. Further studies are needed to explore factors such as temperature gradients or qubit-bath coupling strength that could improve the sensitivity to $T$. However, the sensitivity with respect to $\beta$ remains robust at high temperatures $(\beta=0.1)$, while the $1/T^4$ dependence indicates that QFI with respect to $T$ is significantly suppressed at $T=10$ (renormalized as before and with $k_B=1$), making $\beta$-based sensing more effective. Our analysis shows that while $\beta$-based estimation is more effective at high temperatures, $T$-based estimation becomes more favorable at lower temperatures. While \(T\) and \(\beta\) are physically interchangeable, the QFI is defined with respect to a specific parameter, and differences in how the state depends on \(T\) or \(\beta\) lead to distinct implications for error propagation and practical sensitivity.
In particular, we find that as temperature decreases, the QFI with respect to $T$ ($\mathcal{F}_T$) can surpass the QFI with respect to $\beta$ ($\mathcal{F}_\beta$)(see Eq.~\eqref{qfi_wrt_T}), indicating a regime where direct temperature estimation in terms of $T$ is more efficient. These findings highlight the complementary nature of $\beta$- and $T$-based sensing, with each providing advantages in different temperature ranges.

In addition, we have also considered the same model with a direct coupling between the qubits. For low values of the qubit-qubit coupling, we obtain similar results to those in the uncoupled scenario. This confirms that the Lamb-shift coupling plays the role of an effective direct qubit-qubit coupling that creates correlations. Quite intuitively, the QFI is instead highly enhanced in the strong-coupling regime and displays rapid oscillations over time that stabilize at a large steady-state value. This result has been obtained only at very high temperatures ($\beta \approx 0.1$). However, we have also observed that the steady-state value of the QFI for the inverse temperature of local bath 1 using the ``remote'' qubit 2 as a probe is higher than in the single-qubit scenario also at lower temperature values (such as $\beta^{l_1}=1$), provided that all other baths are very cold ($\beta^c=\beta^{l_2}=5$) and the coupling is sufficiently strong ($k=0.1$). Therefore, remote temperature sensing through an additional qubit-qubit coupling may be realized, for instance, on systems of coupled superconducting qubits immersed in both local and collective baths \cite{Cattaneo2021engineering}. In such systems, $\beta^{l_1}=1$ would correspond roughly to 200 mK \cite{Cattaneo2021collective}, which is a reasonable temperature for a superconducting chip.

 %\textcolor{blue}{We have also considered dephasing in our systems (see Appendix~\ref{deph}). Our results show that if dephasing is as strong as dissipation, then the sensitivity decreases, especially at late times. Furthermore, we investigate and highlight the importance of relative error bounds in assessing the performance of the temperature estimation protocol.} 
 
 The model we have considered in this work can be realized in different experimental platforms, including superconducting qubits in a common waveguide. Therefore, our results shed light on the practical possibilities and limitations of different protocols for temperature estimation in the presence of local and collective baths.
%==================================================================================%
\section*{Acknowledgments}
%==================================================================================%
M.C. would like to thank Matteo G. A. Paris for some interesting comments. This work is supported by the Scientific and Technological Research
Council of Türkiye (TÜBİTAK) under Project Grant Number 123F150. A.U. and \"O.M. thank TÜBİTAK for their financial support. M.C. acknowledges funding from the Research Council of Finland through the Centre of Excellence program grant 336810 and the Finnish Quantum Flagship project 358878 (UH), and from the COQUSY Project No. PID2022-140506NB-C21 funded by Grant No. MCIN/AEI/10.13039/501100011033.
%==================================================================================%
\appendix
\begin{widetext}
\section{Coefficients of the master equation for a single qubit}\label{apA}
In this appendix, we calculate the explicit expressions for the coefficients of the master equation for a single qubit attached to a thermal bath. 
Let us first calculate the bath correlation function, which is given by
\begin{equation}
\begin{aligned}
    \mathcal{B}(\tau)=&\langle B_\alpha(\tau)B_\alpha^\prime(0)\rangle\\
    &= \text{Tr}_B[B_\alpha(\tau)B_\alpha^\prime(0)\rho_B].
    \end{aligned}
\end{equation}
The general expression of $\Gamma(\omega)$ reads 
\begin{equation}
    \Gamma(\omega)=\int_0^\infty d\tau(\tau)e^{i\omega\tau}\mathcal{B}.
\end{equation}
The bath operators in the interaction Hamiltonian are given by
\begin{equation}
    B=\sum_k f_k(a_k+a_k^\dagger),
\end{equation}
where $f_k$ are real numbers.
The operators $a_k$ and $a^\dagger_k$ and the bath operators $B$ in the interaction picture can be written
\begin{equation}
    \begin{aligned}
        a_k(t)=&a_k e^{-i\omega_k t},\quad a_k^\dagger(t)=a^\dagger_k e^{i\omega_k t}\\
        B(t)&=\sum_kf_k(a_k e^{-i\omega_k t}+a^\dagger_k e^{i\omega_k t})
    \end{aligned}
\end{equation}
We can now calculate the expectation value $\langle B(t)B(0)\rangle$ as given below
\begin{equation}
    \begin{aligned}
        \langle B(t)B(0)\rangle&=\langle[\sum_kf_k(a_k e^{-i\omega_k t}+a^\dagger_k e^{i\omega_k t})\sum_k f_k(a_k+a_k^\dagger)]\rangle\\
        &=\langle\sum_kf_ka_ke^{-i\omega_k t}\sum_{k^\prime} f_{k^\prime}a_k^\prime\rangle
        +\langle\sum_kf_ka_ke^{-i\omega_k t}\sum_{k^\prime} f_{k^\prime}a^\dagger{_k^\prime}\rangle\\
        &+\langle\sum_ka^\dagger_k e^{i\omega_k t}\sum_{k^\prime} f_{k^\prime}a{_k^\prime}\rangle
        +\langle\sum_ka^\dagger_k e^{i\omega_k t}\sum_{k^\prime} f_{k^\prime}a^\dagger{_k^\prime}\rangle\\
        &=\sum_{kk^\prime} f_kf_{k^\prime}\langle a_ka^\dagger{_k^\prime}\rangle e^{-i\omega_k t}
        +\sum_{kk^\prime} f_kf_{k^\prime}\langle a^\dagger{_k^\prime}a_k\rangle e^{i\omega_k t}\\
        &=\sum_{kk^\prime} f_kf_{k^\prime}(n(\omega_k)+1)\delta_{kk^\prime} e^{-i\omega_k t} +\sum_{kk^\prime} f_kf_{k^\prime}n(\omega_k)\delta_{kk^\prime} e^{i\omega_k t}\\
        &=\sum_kf_k^2[n(\omega_k) e^{i\omega_k t}+(n(\omega_k)+1)e^{-i\omega_k t}]\\
        &=\int_0^\infty d\omega_kJ(\omega_k)[n(\omega_k) e^{i\omega_k t}+(n(\omega_k)+1)e^{-i\omega_k t}],
    \end{aligned}
\end{equation}
where $J(\omega)$ denotes the spectral density of the bath and is defined as
\begin{equation}
    J(\omega)=\sum_kf_k^2\delta(\omega-\omega_k).
\end{equation}
Typically, we assume an Ohmic bath spectral density for the bath such that
\begin{equation}
    J_{ohm}(\omega)=\omega\frac{\omega_c^2}{\omega_c^2+\omega^2},
\end{equation}
where $\omega_c$ is a cut-off frequency such that $\omega_c\gg\omega$.

Next, we calculate the function $\Gamma(\omega)$ using the formula
\begin{equation}
    \int_0^\infty d\tau e^{ib\tau}=\pi\delta(b)+\frac{iP}{b},
\end{equation}
where $P$ denotes the Cauchy principal value. We can now write $\Gamma(\omega)$ as
\begin{equation}
\begin{aligned}
    \Gamma(\omega)=&\int_0^\infty dt e^{i\omega t}\int_0^\infty d\omega_kJ(\omega_k)\left[n(\omega_k) e^{i\omega_k t}+(n(\omega_k)+1)e^{-i\omega_k t}\right]\\
    &=\int_0^\infty dt \int_0^\infty d\omega_kJ(\omega_k)\left[n(\omega_k) e^{i(\omega_k+\omega) t}+(n(\omega_k)+1)e^{i(\omega-\omega_k) t}\right]\\
    &=\int_0^\infty d\omega_kJ(\omega_k)\left[n(\omega_k)\Big(\delta(\omega+\omega_k)+\frac{iP}{\omega+\omega_k}\Big)+(n(\omega_k)+1)\Big(\delta(\omega-\omega_k)+\frac{iP}{\omega-\omega_k}\Big)\right].
    \end{aligned}
\end{equation}
We simplify the above equation and get the following form:
\begin{equation}
   \begin{aligned}
        \Gamma(\omega)=\int_0^\infty d\omega_kJ(\omega_k)\left[iP\left(\frac{n(\omega_k)+1}{\omega-\omega_k}+\frac{n(\omega_k)}{\omega+\omega_k}
        \right)+(n(\omega_k)+1)\pi\delta(\omega-\omega_k)+n(\omega_k)\pi\delta(\omega+\omega_k)
        \right].
   \end{aligned}
\end{equation}
We now focus on the tricky case with $\omega=0$, such that
\begin{equation}
    \begin{aligned}
        \Gamma(0)=&\int_0^\infty d\tau\mathcal{B}(\tau) \\
        &=\int_0^\infty d\tau\langle B(\tau)B(0)\rangle\\
        &=\int_0^\infty d\tau\int_0^\infty d\omega_k\left[n(\omega_k) e^{i\omega_k t}+(n(\omega_k)+1)e^{-i\omega_k t}\right]\\
        &=\int_0^\infty d\omega_kJ(\omega_k)\left[n(\omega_k)\left(\pi\delta(\omega_k)+\frac{iP}{\omega_k}\right)+(n(\omega_k)+1)\left(\pi\delta(-\omega_k)+\frac{iP}{-\omega_k)}\right)\right].
    \end{aligned}
\end{equation}
Using the property $\delta(-\omega_k)=\delta(\omega_k)$ and simplification gives us
\begin{equation}
    \begin{aligned}
   \Gamma(0)=&\pi\int_0^\infty d\omega_kJ(\omega_k) (2n(\omega_k)+1)\delta(\omega_k)-iP\int_0^\infty d\omega_k\frac{J(\omega_k)}{\omega_k} \\
   &=\frac{\pi}{2}\lim_{\omega_k{\rightarrow}0^+}J(\omega_k)\coth({\frac{\beta\omega_k}{2}})-iP\int_0^\infty d\omega_k\frac{J(\omega_k)}{\omega_k},
    \end{aligned}
\end{equation}
where we used the following relation,
\begin{equation}
    n(\omega_k)=\frac{1}{e^{\beta\omega_k}-1}=\frac{1}{2}\left[\left(\frac{2}{e^{\beta\omega_k}-1}\right)+1-1\right]=\frac{1}{2}\left(\frac{e^{\beta\omega_k}+1}{e^{\beta\omega_k}-1}-1\right)=\frac{1}{2}\left(\coth{(\frac{\beta\omega_k}{2})}-1\right)
\end{equation}
and similarly 
\begin{equation}
    n(\omega_k)+1=\frac{1}{2}\left(\coth{(\frac{\beta\omega_k}{2})}+1\right).
\end{equation}
For a single qubit case, we can write the coefficients as follows:
\begin{equation}
    \begin{aligned}
        \gamma(\omega)=\Gamma(\omega)+\Gamma^*(\omega)=2\text{Re}\{\Gamma(\omega)\}
        \quad,s(\omega)=\frac{\Gamma(\omega)-\Gamma^*(\omega)}{2i}=\text{Im}\{\Gamma(\omega)\}.
    \end{aligned}
\end{equation}
We can easily calculate $\gamma(0)$ when $\omega=0$, that is 
\begin{equation}
    \begin{aligned}
        \gamma(0)=&2\text{Re}\{\Gamma(0)\}=\pi\lim_{\omega_k{\rightarrow}0^+}J(\omega_k)\coth({\frac{\beta\omega_k}{2}}).
    \end{aligned}
\end{equation}
Similarly, we can obtain the general form of $\Gamma(\omega)$ based on the sign of $\omega$ which is given by
\begin{equation}
\Gamma(\omega) = \left\{
\begin{array}{ll}
\frac{\pi}{2} J(\omega) \left( \coth\left( \frac{\beta \hbar \omega}{2} \right) + 1 \right) 
+ i P \int_0^\infty d\omega_k J(\omega_k) \left[ \frac{n(\omega_k) + 1}{\omega - \omega_k} + \frac{n(\omega_k)}{\omega + \omega_k} \right], & \text{if } \omega > 0, \\
\frac{\pi}{2} J(-\omega) \left( \coth\left( -\frac{\beta \hbar \omega}{2} \right) - 1 \right) 
+ i P \int_0^\infty d\omega_k J(\omega_k) \left[ \frac{n(\omega_k) + 1}{\omega - \omega_k} + \frac{n(\omega_k)}{\omega + \omega_k} \right], & \text{if } \omega < 0, \\
\frac{\pi}{2} \lim_{\omega_k \to 0^+} J(\omega_k) \coth\left( \frac{\beta \hbar \omega_k}{2} \right) 
- i P \int_0^\infty d\omega_k \frac{J(\omega_k)}{\omega_k}, & \text{if } \omega = 0.
\end{array} \right.
\end{equation}

As a result:
\begin{equation}
\gamma(\omega) = \hbar^2 \left\{
\begin{array}{ll}
\pi J(\omega) \left( \coth\left( \frac{\beta \hbar \omega}{2} \right) + 1 \right), & \text{if } \omega > 0, \\
\pi J(-\omega) \left( \coth\left( -\frac{\beta \hbar \omega}{2} \right) - 1 \right), & \text{if } \omega < 0, \\
\pi \lim_{\omega_k \to 0^+} J(\omega_k) \coth\left( \frac{\beta \hbar \omega_k}{2} \right), & \text{if } \omega = 0.
\end{array} \right.
\end{equation}
\begin{equation}
s(\omega) = \hbar^2 \left\{
\begin{array}{ll}
P \int_0^\infty d\omega_k J(\omega_k) \left[ \frac{n(\omega_k) + 1}{\omega - \omega_k} + \frac{n(\omega_k)}{\omega + \omega_k} \right], & \text{if } \omega \neq 0, \\
- P \int_0^\infty d\omega_k \frac{J(\omega_k)}{\omega_k}, & \text{if } \omega = 0.
\end{array} \right.
\end{equation}
We can now easily calculate all the coefficients appearing in the master equation~(\ref{meQ}) and the final forms of these expressions are given as
\begin{equation}
    \begin{aligned}
        \gamma_{\uparrow}=&\pi\mu_x^2J(\omega_0)[\coth{(\frac{\beta\omega_0}{2})-1}],\quad
        \gamma_{\downarrow}=\pi\mu_x^2J(\omega_0)[\coth{(\frac{\beta\omega_0}{2})+1}],\\
        &\gamma_{0}=\pi\mu_z^2\lim_{\omega^\prime \longrightarrow 0^+} J(\omega^\prime)\coth{(\frac{\beta\omega^\prime}{2})},\quad
        s_0=\mu_x^2P\int_0^\infty J(\omega_k)\coth{(\frac{\beta\omega_k}{2})}\big[\frac{1}{\omega_0-\omega_k}+\frac{1}{\omega_0+\omega_k}\big].
    \end{aligned}
\end{equation}
%*******************************************************%
\section{Global master equation with partial secular approximation}\label{sec:appendixGlobalME}
In the basis $\{|00\rangle,|10\rangle,|01\rangle,|11\rangle\}$, the Hamiltonian $\hat{H}_S$ of two interacting qubits (Eq.~\eqref{2qubitH} of the main text) can be written in matrix form as
\begin{equation}
   H_S= \left(
\begin{array}{cccc}
 \frac{\omega_+}{2} & 0 & 0 & k  \\
 0 & \frac{\omega_-}{2} & k  & 0 \\
 0 & k  & -\frac{\omega_-}{2} & 0 \\
 k  & 0 & 0 &  -\frac{\omega_+}{2} \\
\end{array}
\right),
\end{equation}
where $\omega_+=\omega_1+\omega_2$ and $\omega_-=\omega_1-\omega_2$ are defined for simplicity. It is straightforward to show that the eigenvalues of $H_S$ are given by
\begin{equation}
    \begin{array}{c}
E_a=-\frac{1}{2} \sqrt{4 k ^2 + \omega_-^2 },\quad E_b=-\frac{1}{2} \sqrt{4 k ^2 + \omega_+^2 }\\
E_c=\frac{1}{2} \sqrt{4  ^2 k+ \omega_-^2},\quad E_d=\frac{1}{2} \sqrt{4 k ^2 + \omega_+^2 },
\end{array}
\end{equation}
while the corresponding eigenstates are
\begin{equation}
\begin{aligned}
    |a\rangle=&\cos{\theta}|00\rangle-\sin{\theta}|11\rangle,
    |b\rangle=&-\sin{\phi}|10\rangle+\cos{\phi}|01\rangle\\
     |c\rangle=&\cos{\phi}|10\rangle+\sin{\phi}|01\rangle, |d\rangle=&\sin{\theta}|00\rangle+\cos{\theta}|11\rangle,
\end{aligned}
\end{equation}
where the parameters $\theta$, and $\phi$ are given by
\begin{equation}
    \begin{aligned}
        \tan2\theta=\frac{2k}{\omega_+}, \quad \tan2\phi=\frac{2k}{\omega_-}.
    \end{aligned}
\end{equation}
Following the same notation as in Ref.~\cite{Cattaneo_2019}, the master equation for two interacting qubits reads
\begin{equation}\label{me2Q}
\begin{aligned}
\Dot{\hat{\rho}}_S(t) &= -i \left[ \hat{H}_S+\hat{H}_{LS}, \rho_S(t) \right] 
+ \sum_{j,k=\text{I,II}} \sum_{m,n=1,2} \gamma_{jk}^{mn} 
\left( \hat{\sigma}_m^x(\omega_j) \rho_S(t) \hat{\sigma}_n^{x}(-\omega_k) 
- \frac{1}{2} \left\{ \hat{\sigma}_n^{x}(-\omega_k) \hat{\sigma}_m^x(\omega_j), \rho_S(t) \right\} \right) \\
&\quad + \sum_{j,k=\text{I,II}} \sum_{m,n=1,2} \Tilde{\gamma}_{jk}^{mn}
\left( \hat{\sigma}_m^x(-\omega_j) \rho_S(t) \hat{\sigma}_n^{x}(\omega_k) 
- \frac{1}{2} \left\{ \hat{\sigma}_n^{x}(\omega_k) \hat{\sigma}_m^x(-\omega_j), \rho_S(t) \right\} \right) \\
&\quad + \sum_{j,k=0,\pm\text{III}} \sum_{m,n=1,2} \eta_{jk}^{mn} 
\left( \hat{\sigma}_m^z(\omega_j) \rho_S(t) \hat{\sigma}_n^{z}(-\omega_k) 
- \frac{1}{2} \left\{ \hat{\sigma}_n^{z}(-\omega_k) \hat{\sigma}_m^z(\omega_j), \rho_S(t) \right\} \right) \\
&\quad + \sum_{j,k=0,\pm\text{IV}} \sum_{m,n=1,2} \zeta_{jk}^{mn} 
\left( \hat{\sigma}_m^z(\omega_j) \rho_S(t) \hat{\sigma}_n^{z}(-\omega_j) 
- \frac{1}{2} \left\{ \sigma_n^{z}(-\omega_j) \hat{\sigma}_m^z(-\omega_j), \rho_S(t) \right\} \right),
\end{aligned}
\end{equation}
where the jump frequencies and the jump operators  are given respectively in equations (\ref{freq}) and (\ref{JO}). Moreover, we have used the notation $\omega_{-\text{IV}} = -\omega_\text{IV}$, $\omega_{-\text{III}} = -\omega_\text{III}$, and $\omega_0 = 0$.
The Lamb-shift Hamiltonian is given by
\begin{equation}\label{lshift}
\begin{aligned}
\hat{H}_{LS} &= \sum_{j,k=\text{I,II}} \sum_{m,n=1,2}\Big( s_{jk}^{mn} \hat{\sigma}_n^x(-\omega_k) \hat{\sigma}_m^{x}(\omega_j) +\Tilde{s}_{jk}^{mn}\hat{\sigma}_n^x(\omega_k)\hat{\sigma}_m^x(-\omega_j)\Big) +\sum_{j,k=0,\pm\text{ IV}} \sum_{m,n=1,2} r_{jk}^{mn} \hat{\sigma}_n^z(-\omega_k) \hat{\sigma}_m^{z}(\omega_j) \\
&\quad + \sum_{j,k=0,\pm\text{III}} \sum_{m,n=1,2} u_{j}^{mn} \hat{\sigma}_n^z(-\omega_j) \hat{\sigma}_m^{z}(\omega_j).
\end{aligned}
\end{equation}
The jump frequencies are
\begin{equation}\label{freq}
    \begin{aligned}
        \omega_I=&E_d-E_b=E_c-E_a=\frac{1}{2}\sqrt{4k^2+\omega_+^2}+\frac{1}{2}\sqrt{4k^2+\omega_-^2},\\
        \omega_{II}=&E_d-E_c=E_b-E_a=\frac{1}{2}\sqrt{4k^2+\omega_+^2}-\frac{1}{2}\sqrt{4k^2+\omega_-^2},\\
        \omega_{III}=&E_d-E_a=\sqrt{4k^2+\omega_+^2},\quad \omega_{IV}=E_c-E_b=\sqrt{4k^2+\omega_-^2}.
    \end{aligned}
\end{equation}
Moreover, the jump operators can be derived from Eq.~\eqref{jump} and they read
\begin{equation}\label{JO}
    \begin{aligned}
        \hat{\sigma}_1^x(\omega_{I})      &= \cos(\theta + \phi)(|a\rangle \langle c| + |b\rangle \langle d|),\\
        \hat{\sigma}_1^x(\omega_{II})     &= \sin(\theta + \phi)(-|a\rangle \langle b| + |c\rangle \langle d|),\\
        \hat{\sigma}_2^x(\omega_{I})      &= \sin(\theta - \phi)(-|a\rangle \langle c| + |b\rangle \langle d|),\\
        \hat{\sigma}_2^x(\omega_{II})     &= \cos(\theta - \phi)(|a\rangle \langle b| + |c\rangle \langle d|),\\
        \hat{\sigma}_1^z(\omega_{III})    &= -\sin(2\theta)(|a\rangle \langle d|),\\
        \hat{\sigma}_1^z(\omega_{IV})     &= -\sin(2\phi)(|b\rangle \langle c|),\\
        \hat{\sigma}_2^z(\omega_{III})    &= -\sin(2\theta)(|a\rangle \langle d|),\\
        \hat{\sigma}_2^z(\omega_{IV})     &=  \sin(2\phi)(|b\rangle \langle c|),\\
         \hat{\sigma}_1^z(0)&= \cos(2\theta)(|d\rangle \langle d| - |a\rangle \langle a|) + \cos(2\phi)(|c\rangle \langle c| - |b\rangle \langle b|),\\
        \hat{\sigma}_2^z(0) &= \cos(2\theta)(|d\rangle \langle d| - |a\rangle \langle a|) + \cos(2\phi)(|b\rangle \langle b| - |c\rangle \langle c|).
    \end{aligned}
\end{equation}
The explicit expressions for the coefficients appearing in the master equation, Eqs.~\eqref{me2Q} and \eqref{lshift}, can be derived following a procedure similar to that outlined in Appendix~\ref{apA}; see also Ref.~\cite{Cattaneo_2019} for further details.
\end{widetext}
%%%%%%%%%%%%%%%%%%%%%%%%%%%%%%%%%%%%%%%%%%%%
\section{Effects of dephasing}\label{deph}
Incorporating dephasing into our current model is important for making our temperature sensing scheme more realistic and applicable to practical scenarios. To address this, we investigate the combined effect of both dissipation and dephasing dynamics on the estimation of bath temperatures. Specifically, we turn on the dephasing coupling of the qubits to the baths, setting it to $\mu_z=10^{-2}$ for both non-interacting ($k=0$) and interacting ($k \neq 0$) qubits. The remaining parameters are kept the same as discussed earlier in the text. 

\begin{figure}[t!]
    \centering
    \subfloat[]{
    \includegraphics[scale=0.62]{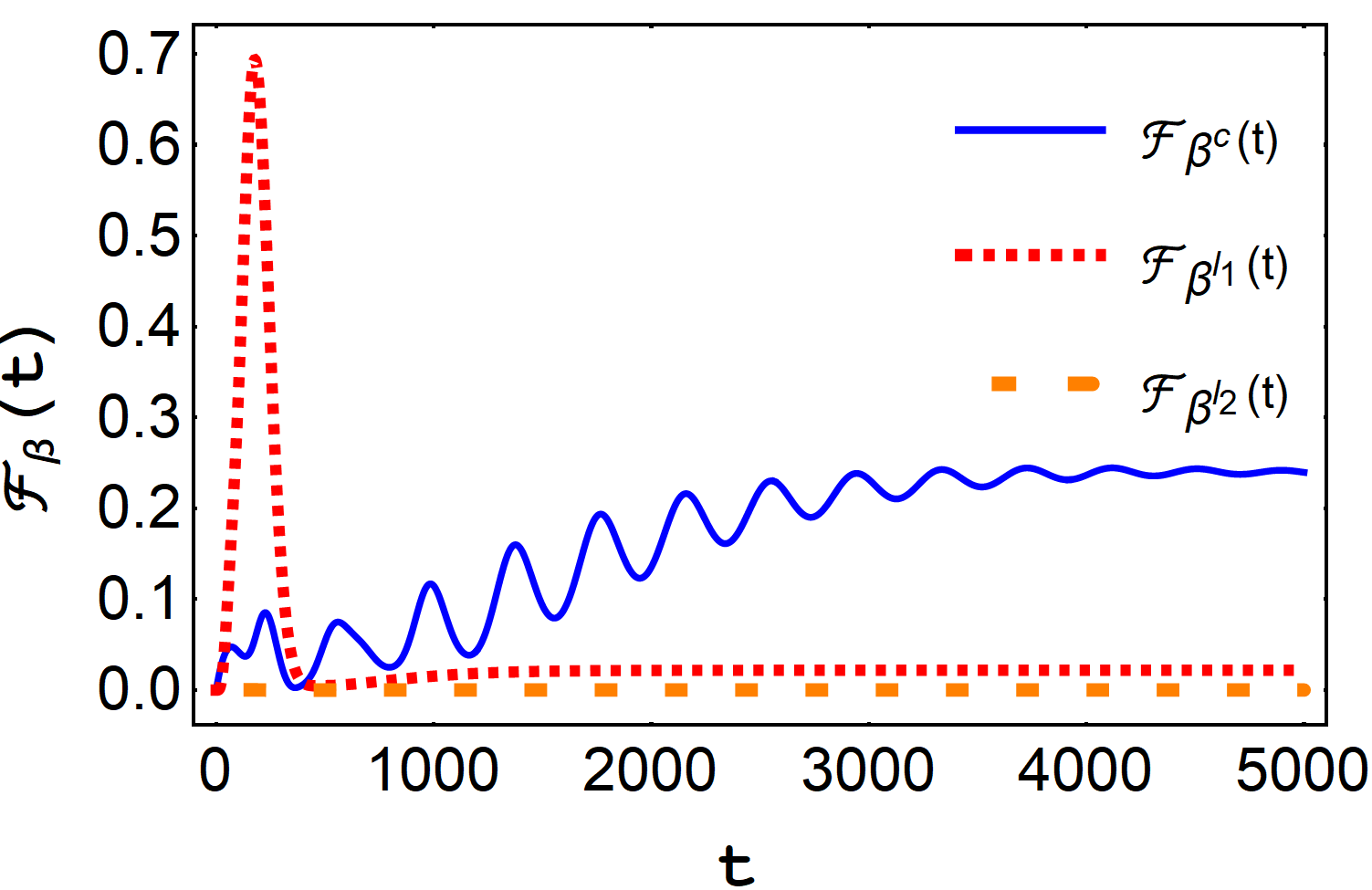}}\\
    \subfloat[]{\includegraphics[scale=0.62]{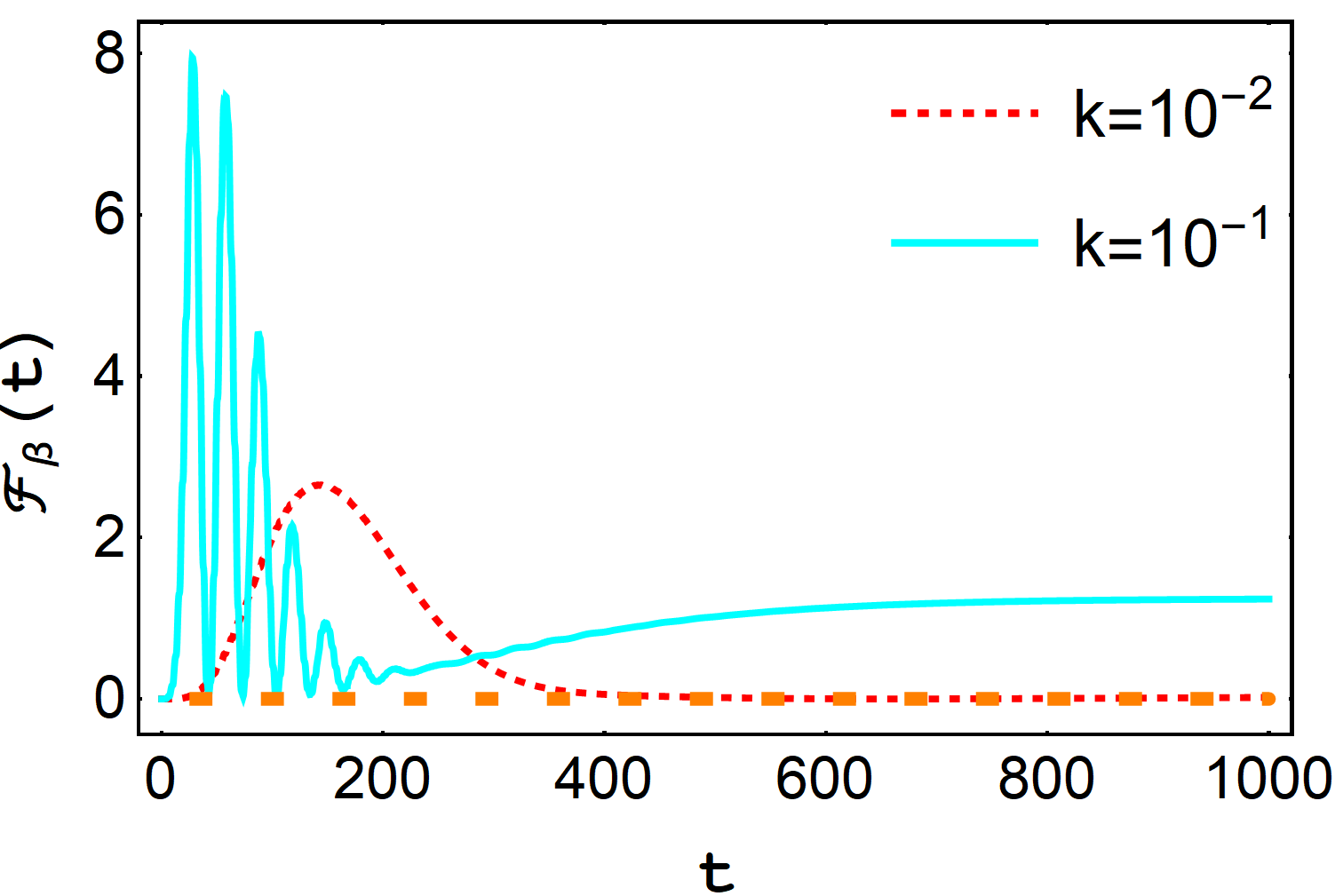}}
    \caption{Scenario in the presence of dephasing. \textbf{(a)} The QFI $\mathcal{F}_\beta(t)$ is plotted as a function of time $t$ for the estimation of the common bath temperature $\beta^c$ (solid blue curve) in the absence of local baths, the hot bath temperature $\beta^{l_1}$ (red dotted curve), and the cold bath temperature $\beta^{l_2}$ (orange dashed curve) when there is no direct coupling between the qubits ($k=0$). \textbf{(b)} The QFI $\mathcal{F}_\beta(t)$ as a function of time $t$ when there is a direct qubit-qubit coupling. Here, the orange dashed curve denotes the QFI for the estimation of cold bath and the red dotted and solid cyan curve represents the QFI for the estimation of hot bath for  $k=10^{-2}$ and  $k=10^{-1}$, respectively.
    The rest of the parameters are set to $\mu_x=10^{-2}$, $\mu_z=10^{-2}$, $\omega_-=0.01$, $\beta^{c}=1$, $\beta^{l_1}=0.1$, and $\beta^{l_2}=1$.}
    \label{fig:dd}
\end{figure}

Figure~\ref{fig:dd}(a) illustrates the QFI as a function of time for the estimation of the common bath temperature $\beta^c$ in the absence of local baths, as well as for the estimation of the hot bath temperature $\beta^{l_1}$ and the cold bath temperature $\beta^{l_2}$, for two non-interacting qubits and small detuning $\omega_-=0.01$. The QFI for all bath scenarios decreases when dephasing is introduced, compared with the case where only dissipation was present. This demonstrates that dephasing negatively impacts temperature estimation. In contrast, in the single-qubit case, dephasing can actually be exploited for the estimation of the bath temperature \cite{Razavian2019}.

%-----------------------------------------------------------------------%
%------------------------------------------------------------------------%
Figure~\ref{fig:dd}(b) illustrates the QFI for estimating the temperatures of the cold and hot baths in the presence of direct qubit-qubit coupling. The QFI for estimating the cold bath temperature is zero, as shown by the dashed orange curve in Fig.~\ref{fig:dd}(b) for \( k = 10^{-2} \). In contrast, the QFI for estimating the temperature \( \beta^{l_1} \) of the hot bath is depicted by the red dotted curve, where it varies as a function of time \( t \). Interestingly, even with dissipation and dephasing, the QFI still provides a significant level of precision.

However, a key observation is that while the QFI saturates in the absence of dephasing, it rapidly decays to zero when estimating the temperature of local baths under dephasing. In the strong-coupling regime (\( k = 10^{-1} \)), the QFI exhibits higher values with oscillatory behavior, as shown by the solid cyan curve in Fig.~\ref{fig:dd}(b). Unlike the weak-coupling case, the QFI does not decay to zero in this regime. The QFI in this scenario is actually enhanced. However, for such a strong value of the qubit-qubit coupling, the eigenstates of the system Hamiltonian are dressed, and the ``dephasing coupling'' to the bath (through $\sigma_z$) actually corresponds to pure dissipation, which may explain the enhancement of the sensitivity in this scenario.

In conclusion, a sufficiently strong dephasing limits the precision of remote temperature sensing. This effect disappears for stronger values of the qubit-qubit coupling. A possible solution may be to monitor the transient dynamics, as the detrimental effects of dephasing are mostly felt at the steady state.
%----------------------------------------------------------%
\begin{figure}[t!]
    \centering
     \subfloat[]{
    \includegraphics[scale=0.6]{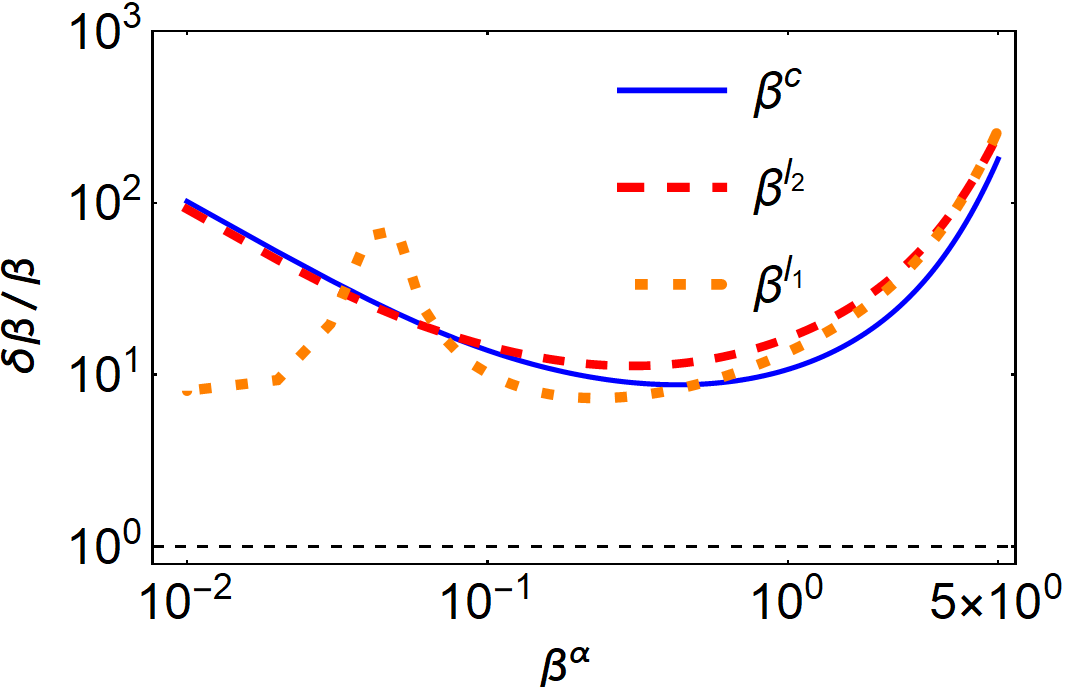}}\\
    \subfloat[]{
    \includegraphics[scale=0.6]{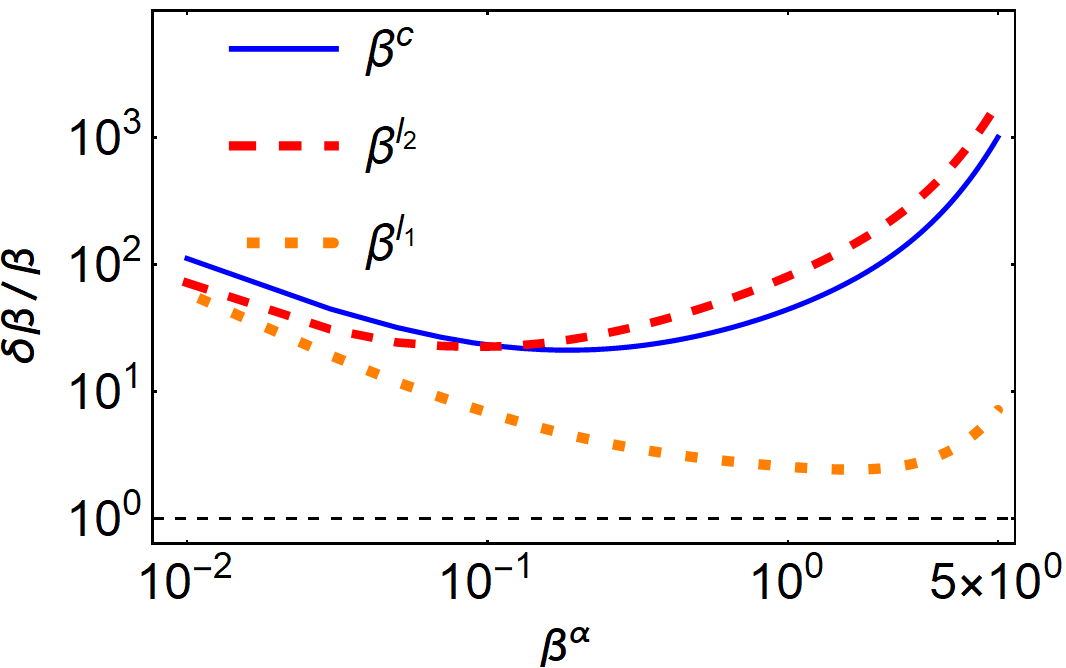}}
    \caption{Relative error $\delta\beta/\beta$ for the steady-state of a single-qubit reduced density matrix as a function of the inverse temperature $\beta^\alpha$, where $\alpha=l_1,l_2,c$ for \textbf{(a)} uncoupled qubits ($k=0$) and \textbf{(b)} coupled qubits ($k=10^{-1}$). All the parameters and their range are the same as discussed in the caption of Fig.~\ref{fig:steadystate}.}.
    \label{fig:error}
\end{figure}
%----------------------------------------------------------%
\section{Relative error bound in inverse temperature estimation}\label{relError}
In this section, we analyze the relative error in estimating the inverse temperatures of the common and local baths in the steady-state regime, given by~\cite{PhysRevLett.72.3439}
\begin{equation}
    \frac{\delta\beta}{\beta} = \frac{1}{\beta\sqrt{\mathcal{F}_\beta}}.
\end{equation}
We examine both uncoupled and coupled qubits, considering a small detuning (\(\omega_- = 0.01\)) in the presence of purely dissipative qubit-bath coupling. The parameters used are consistent with those investigated in Sec.~\ref{steady-state}. 

For uncoupled qubits ($k=0$), the relative error \(\delta\beta/\beta\) as a function of the inverse temperatures \(\beta^\alpha\) is illustrated in Fig.~\ref{fig:error}(a). We observe that the relative error for the common bath temperature \(\beta^c\) and the cold bath temperature \(\beta^{l_2}\) is nearly identical, both exhibiting lower values in the high-temperature regime. In contrast, the relative error in estimating the hot bath temperature \(\beta^{l_1}\) is significantly lower when the temperature is high.

Figure~\ref{fig:error}(b) presents the relative error \(\delta\beta/\beta\) as a function of inverse temperature \(\beta^\alpha\) for two strongly coupled qubits ($k=10^{-1}$). We find that the relative error is highest for the estimation of the cold bath temperature, while it is slightly lower for the common bath temperature. However, the relative error decreases substantially for the estimation of the hot bath temperature, particularly in the low-temperature regime where \(\beta^{l_1}\) takes large values. Notably, for the coupling strengths in the strong-coupling regime, the relative error for \(\beta^{l_1}\) is very close to the reference line (black dotted line) where \(\delta\beta/\beta = 1\), especially in the low-temperature regime. This result suggests that the estimation of the hot bath temperature is more reliable in this regime.
%==================================================================================%
\bibliography{PSA.bib}

\begin{thebibliography}{83}%
\makeatletter
\providecommand \@ifxundefined [1]{%
 \@ifx{#1\undefined}
}%
\providecommand \@ifnum [1]{%
 \ifnum #1\expandafter \@firstoftwo
 \else \expandafter \@secondoftwo
 \fi
}%
\providecommand \@ifx [1]{%
 \ifx #1\expandafter \@firstoftwo
 \else \expandafter \@secondoftwo
 \fi
}%
\providecommand \natexlab [1]{#1}%
\providecommand \enquote  [1]{``#1''}%
\providecommand \bibnamefont  [1]{#1}%
\providecommand \bibfnamefont [1]{#1}%
\providecommand \citenamefont [1]{#1}%
\providecommand \href@noop [0]{\@secondoftwo}%
\providecommand \href [0]{\begingroup \@sanitize@url \@href}%
\providecommand \@href[1]{\@@startlink{#1}\@@href}%
\providecommand \@@href[1]{\endgroup#1\@@endlink}%
\providecommand \@sanitize@url [0]{\catcode `\\12\catcode `\$12\catcode `\&12\catcode `\#12\catcode `\^12\catcode `\_12\catcode `\%12\relax}%
\providecommand \@@startlink[1]{}%
\providecommand \@@endlink[0]{}%
\providecommand \url  [0]{\begingroup\@sanitize@url \@url }%
\providecommand \@url [1]{\endgroup\@href {#1}{\urlprefix }}%
\providecommand \urlprefix  [0]{URL }%
\providecommand \Eprint [0]{\href }%
\providecommand \doibase [0]{http://dx.doi.org/}%
\providecommand \selectlanguage [0]{\@gobble}%
\providecommand \bibinfo  [0]{\@secondoftwo}%
\providecommand \bibfield  [0]{\@secondoftwo}%
\providecommand \translation [1]{[#1]}%
\providecommand \BibitemOpen [0]{}%
\providecommand \bibitemStop [0]{}%
\providecommand \bibitemNoStop [0]{.\EOS\space}%
\providecommand \EOS [0]{\spacefactor3000\relax}%
\providecommand \BibitemShut  [1]{\csname bibitem#1\endcsname}%
\let\auto@bib@innerbib\@empty
%</preamble>
\bibitem [{\citenamefont {Mehboudi}\ \emph {et~al.}(2019{\natexlab{a}})\citenamefont {Mehboudi}, \citenamefont {Sanpera},\ and\ \citenamefont {Correa}}]{Mehboudi_2019}%
  \BibitemOpen
  \bibfield  {author} {\bibinfo {author} {\bibfnamefont {Mohammad}\ \bibnamefont {Mehboudi}}, \bibinfo {author} {\bibfnamefont {Anna}\ \bibnamefont {Sanpera}}, \ and\ \bibinfo {author} {\bibfnamefont {Luis~A}\ \bibnamefont {Correa}},\ }\bibfield  {title} {\enquote {\bibinfo {title} {Thermometry in the quantum regime: recent theoretical progress},}\ }\href {\doibase 10.1088/1751-8121/ab2828} {\bibfield  {journal} {\bibinfo  {journal} {J. Phys. A: Math. Theor.}\ }\textbf {\bibinfo {volume} {52}},\ \bibinfo {pages} {303001} (\bibinfo {year} {2019}{\natexlab{a}})}\BibitemShut {NoStop}%
\bibitem [{\citenamefont {Giovannetti}\ \emph {et~al.}(2011)\citenamefont {Giovannetti}, \citenamefont {Lloyd},\ and\ \citenamefont {Maccone}}]{Giovannetti2011}%
  \BibitemOpen
  \bibfield  {author} {\bibinfo {author} {\bibfnamefont {Vittorio}\ \bibnamefont {Giovannetti}}, \bibinfo {author} {\bibfnamefont {Seth}\ \bibnamefont {Lloyd}}, \ and\ \bibinfo {author} {\bibfnamefont {Lorenzo}\ \bibnamefont {Maccone}},\ }\bibfield  {title} {\enquote {\bibinfo {title} {Advances in quantum metrology},}\ }\href {\doibase 10.1038/nphoton.2011.35} {\bibfield  {journal} {\bibinfo  {journal} {Nat. Photon.}\ }\textbf {\bibinfo {volume} {5}},\ \bibinfo {pages} {222--229} (\bibinfo {year} {2011})}\BibitemShut {NoStop}%
\bibitem [{\citenamefont {Dedyulin}\ \emph {et~al.}(2022)\citenamefont {Dedyulin}, \citenamefont {Ahmed},\ and\ \citenamefont {Machin}}]{Dedyulin_2022}%
  \BibitemOpen
  \bibfield  {author} {\bibinfo {author} {\bibfnamefont {S}~\bibnamefont {Dedyulin}}, \bibinfo {author} {\bibfnamefont {Z}~\bibnamefont {Ahmed}}, \ and\ \bibinfo {author} {\bibfnamefont {G}~\bibnamefont {Machin}},\ }\bibfield  {title} {\enquote {\bibinfo {title} {Emerging technologies in the field of thermometry},}\ }\href {\doibase 10.1088/1361-6501/ac75b1} {\bibfield  {journal} {\bibinfo  {journal} {Meas. Sci. Technol.}\ }\textbf {\bibinfo {volume} {33}},\ \bibinfo {pages} {092001} (\bibinfo {year} {2022})}\BibitemShut {NoStop}%
\bibitem [{\citenamefont {Albarelli}\ \emph {et~al.}(2023)\citenamefont {Albarelli}, \citenamefont {Paris}, \citenamefont {Vacchini},\ and\ \citenamefont {Smirne}}]{PhysRevA.108.062421}%
  \BibitemOpen
  \bibfield  {author} {\bibinfo {author} {\bibfnamefont {Francesco}\ \bibnamefont {Albarelli}}, \bibinfo {author} {\bibfnamefont {Matteo G.~A.}\ \bibnamefont {Paris}}, \bibinfo {author} {\bibfnamefont {Bassano}\ \bibnamefont {Vacchini}}, \ and\ \bibinfo {author} {\bibfnamefont {Andrea}\ \bibnamefont {Smirne}},\ }\bibfield  {title} {\enquote {\bibinfo {title} {Invasiveness of nonequilibrium pure-dephasing quantum thermometry},}\ }\href {\doibase 10.1103/PhysRevA.108.062421} {\bibfield  {journal} {\bibinfo  {journal} {Phys. Rev. A}\ }\textbf {\bibinfo {volume} {108}},\ \bibinfo {pages} {062421} (\bibinfo {year} {2023})}\BibitemShut {NoStop}%
\bibitem [{\citenamefont {Jacobs}(2014)}]{Jacobs_2014}%
  \BibitemOpen
  \bibfield  {author} {\bibinfo {author} {\bibfnamefont {Kurt}\ \bibnamefont {Jacobs}},\ }\href@noop {} {\emph {\bibinfo {title} {Quantum Measurement Theory and its Applications}}}\ (\bibinfo  {publisher} {Cambridge University Press},\ \bibinfo {year} {2014})\BibitemShut {NoStop}%
\bibitem [{\citenamefont {Correa}\ \emph {et~al.}(2015)\citenamefont {Correa}, \citenamefont {Mehboudi}, \citenamefont {Adesso},\ and\ \citenamefont {Sanpera}}]{PhysRevLett.114.220405}%
  \BibitemOpen
  \bibfield  {author} {\bibinfo {author} {\bibfnamefont {Luis~A.}\ \bibnamefont {Correa}}, \bibinfo {author} {\bibfnamefont {Mohammad}\ \bibnamefont {Mehboudi}}, \bibinfo {author} {\bibfnamefont {Gerardo}\ \bibnamefont {Adesso}}, \ and\ \bibinfo {author} {\bibfnamefont {Anna}\ \bibnamefont {Sanpera}},\ }\bibfield  {title} {\enquote {\bibinfo {title} {Individual quantum probes for optimal thermometry},}\ }\href {\doibase 10.1103/PhysRevLett.114.220405} {\bibfield  {journal} {\bibinfo  {journal} {Phys. Rev. Lett.}\ }\textbf {\bibinfo {volume} {114}},\ \bibinfo {pages} {220405} (\bibinfo {year} {2015})}\BibitemShut {NoStop}%
\bibitem [{\citenamefont {Mehboudi}\ \emph {et~al.}(2015)\citenamefont {Mehboudi}, \citenamefont {Moreno-Cardoner}, \citenamefont {Chiara},\ and\ \citenamefont {Sanpera}}]{Mehboudi_2015}%
  \BibitemOpen
  \bibfield  {author} {\bibinfo {author} {\bibfnamefont {M}~\bibnamefont {Mehboudi}}, \bibinfo {author} {\bibfnamefont {M}~\bibnamefont {Moreno-Cardoner}}, \bibinfo {author} {\bibfnamefont {G~De}\ \bibnamefont {Chiara}}, \ and\ \bibinfo {author} {\bibfnamefont {A}~\bibnamefont {Sanpera}},\ }\bibfield  {title} {\enquote {\bibinfo {title} {Thermometry precision in strongly correlated ultracold lattice gases},}\ }\href {\doibase 10.1088/1367-2630/17/5/055020} {\bibfield  {journal} {\bibinfo  {journal} {New J. Phys.}\ }\textbf {\bibinfo {volume} {17}},\ \bibinfo {pages} {055020} (\bibinfo {year} {2015})}\BibitemShut {NoStop}%
\bibitem [{\citenamefont {Brunelli}\ \emph {et~al.}(2012)\citenamefont {Brunelli}, \citenamefont {Olivares}, \citenamefont {Paternostro},\ and\ \citenamefont {Paris}}]{PhysRevA.86.012125}%
  \BibitemOpen
  \bibfield  {author} {\bibinfo {author} {\bibfnamefont {Matteo}\ \bibnamefont {Brunelli}}, \bibinfo {author} {\bibfnamefont {Stefano}\ \bibnamefont {Olivares}}, \bibinfo {author} {\bibfnamefont {Mauro}\ \bibnamefont {Paternostro}}, \ and\ \bibinfo {author} {\bibfnamefont {Matteo G.~A.}\ \bibnamefont {Paris}},\ }\bibfield  {title} {\enquote {\bibinfo {title} {Qubit-assisted thermometry of a quantum harmonic oscillator},}\ }\href {\doibase 10.1103/PhysRevA.86.012125} {\bibfield  {journal} {\bibinfo  {journal} {Phys. Rev. A}\ }\textbf {\bibinfo {volume} {86}},\ \bibinfo {pages} {012125} (\bibinfo {year} {2012})}\BibitemShut {NoStop}%
\bibitem [{\citenamefont {Brunelli}\ \emph {et~al.}(2011)\citenamefont {Brunelli}, \citenamefont {Olivares},\ and\ \citenamefont {Paris}}]{PhysRevA.84.032105}%
  \BibitemOpen
  \bibfield  {author} {\bibinfo {author} {\bibfnamefont {Matteo}\ \bibnamefont {Brunelli}}, \bibinfo {author} {\bibfnamefont {Stefano}\ \bibnamefont {Olivares}}, \ and\ \bibinfo {author} {\bibfnamefont {Matteo G.~A.}\ \bibnamefont {Paris}},\ }\bibfield  {title} {\enquote {\bibinfo {title} {Qubit thermometry for micromechanical resonators},}\ }\href {\doibase 10.1103/PhysRevA.84.032105} {\bibfield  {journal} {\bibinfo  {journal} {Phys. Rev. A}\ }\textbf {\bibinfo {volume} {84}},\ \bibinfo {pages} {032105} (\bibinfo {year} {2011})}\BibitemShut {NoStop}%
\bibitem [{\citenamefont {Mancino}\ \emph {et~al.}(2017)\citenamefont {Mancino}, \citenamefont {Sbroscia}, \citenamefont {Gianani}, \citenamefont {Roccia},\ and\ \citenamefont {Barbieri}}]{PhysRevLett.118.130502}%
  \BibitemOpen
  \bibfield  {author} {\bibinfo {author} {\bibfnamefont {Luca}\ \bibnamefont {Mancino}}, \bibinfo {author} {\bibfnamefont {Marco}\ \bibnamefont {Sbroscia}}, \bibinfo {author} {\bibfnamefont {Ilaria}\ \bibnamefont {Gianani}}, \bibinfo {author} {\bibfnamefont {Emanuele}\ \bibnamefont {Roccia}}, \ and\ \bibinfo {author} {\bibfnamefont {Marco}\ \bibnamefont {Barbieri}},\ }\bibfield  {title} {\enquote {\bibinfo {title} {Quantum simulation of single-qubit thermometry using linear optics},}\ }\href {\doibase 10.1103/PhysRevLett.118.130502} {\bibfield  {journal} {\bibinfo  {journal} {Phys. Rev. Lett.}\ }\textbf {\bibinfo {volume} {118}},\ \bibinfo {pages} {130502} (\bibinfo {year} {2017})}\BibitemShut {NoStop}%
\bibitem [{\citenamefont {Feyles}\ \emph {et~al.}(2019)\citenamefont {Feyles}, \citenamefont {Mancino}, \citenamefont {Sbroscia}, \citenamefont {Gianani},\ and\ \citenamefont {Barbieri}}]{PhysRevA.99.062114}%
  \BibitemOpen
  \bibfield  {author} {\bibinfo {author} {\bibfnamefont {Michele~M.}\ \bibnamefont {Feyles}}, \bibinfo {author} {\bibfnamefont {Luca}\ \bibnamefont {Mancino}}, \bibinfo {author} {\bibfnamefont {Marco}\ \bibnamefont {Sbroscia}}, \bibinfo {author} {\bibfnamefont {Ilaria}\ \bibnamefont {Gianani}}, \ and\ \bibinfo {author} {\bibfnamefont {Marco}\ \bibnamefont {Barbieri}},\ }\bibfield  {title} {\enquote {\bibinfo {title} {Dynamical role of quantum signatures in quantum thermometry},}\ }\href {\doibase 10.1103/PhysRevA.99.062114} {\bibfield  {journal} {\bibinfo  {journal} {Phys. Rev. A}\ }\textbf {\bibinfo {volume} {99}},\ \bibinfo {pages} {062114} (\bibinfo {year} {2019})}\BibitemShut {NoStop}%
\bibitem [{\citenamefont {J\o{}rgensen}\ \emph {et~al.}(2020)\citenamefont {J\o{}rgensen}, \citenamefont {Potts}, \citenamefont {Paris},\ and\ \citenamefont {Brask}}]{PhysRevResearch.2.033394}%
  \BibitemOpen
  \bibfield  {author} {\bibinfo {author} {\bibfnamefont {Mathias~R.}\ \bibnamefont {J\o{}rgensen}}, \bibinfo {author} {\bibfnamefont {Patrick~P.}\ \bibnamefont {Potts}}, \bibinfo {author} {\bibfnamefont {Matteo G.~A.}\ \bibnamefont {Paris}}, \ and\ \bibinfo {author} {\bibfnamefont {Jonatan~B.}\ \bibnamefont {Brask}},\ }\bibfield  {title} {\enquote {\bibinfo {title} {Tight bound on finite-resolution quantum thermometry at low temperatures},}\ }\href {\doibase 10.1103/PhysRevResearch.2.033394} {\bibfield  {journal} {\bibinfo  {journal} {Phys. Rev. Res.}\ }\textbf {\bibinfo {volume} {2}},\ \bibinfo {pages} {033394} (\bibinfo {year} {2020})}\BibitemShut {NoStop}%
\bibitem [{\citenamefont {Bouton}\ \emph {et~al.}(2020)\citenamefont {Bouton}, \citenamefont {Nettersheim}, \citenamefont {Adam}, \citenamefont {Schmidt}, \citenamefont {Mayer}, \citenamefont {Lausch}, \citenamefont {Tiemann},\ and\ \citenamefont {Widera}}]{PhysRevX.10.011018}%
  \BibitemOpen
  \bibfield  {author} {\bibinfo {author} {\bibfnamefont {Quentin}\ \bibnamefont {Bouton}}, \bibinfo {author} {\bibfnamefont {Jens}\ \bibnamefont {Nettersheim}}, \bibinfo {author} {\bibfnamefont {Daniel}\ \bibnamefont {Adam}}, \bibinfo {author} {\bibfnamefont {Felix}\ \bibnamefont {Schmidt}}, \bibinfo {author} {\bibfnamefont {Daniel}\ \bibnamefont {Mayer}}, \bibinfo {author} {\bibfnamefont {Tobias}\ \bibnamefont {Lausch}}, \bibinfo {author} {\bibfnamefont {Eberhard}\ \bibnamefont {Tiemann}}, \ and\ \bibinfo {author} {\bibfnamefont {Artur}\ \bibnamefont {Widera}},\ }\bibfield  {title} {\enquote {\bibinfo {title} {Single-atom quantum probes for ultracold gases boosted by nonequilibrium spin dynamics},}\ }\href {\doibase 10.1103/PhysRevX.10.011018} {\bibfield  {journal} {\bibinfo  {journal} {Phys. Rev. X}\ }\textbf {\bibinfo {volume} {10}},\ \bibinfo {pages} {011018} (\bibinfo {year} {2020})}\BibitemShut {NoStop}%
\bibitem [{\citenamefont {Jevtic}\ \emph {et~al.}(2015)\citenamefont {Jevtic}, \citenamefont {Newman}, \citenamefont {Rudolph},\ and\ \citenamefont {Stace}}]{PhysRevA.91.012331}%
  \BibitemOpen
  \bibfield  {author} {\bibinfo {author} {\bibfnamefont {Sania}\ \bibnamefont {Jevtic}}, \bibinfo {author} {\bibfnamefont {David}\ \bibnamefont {Newman}}, \bibinfo {author} {\bibfnamefont {Terry}\ \bibnamefont {Rudolph}}, \ and\ \bibinfo {author} {\bibfnamefont {T.~M.}\ \bibnamefont {Stace}},\ }\bibfield  {title} {\enquote {\bibinfo {title} {Single-qubit thermometry},}\ }\href {\doibase 10.1103/PhysRevA.91.012331} {\bibfield  {journal} {\bibinfo  {journal} {Phys. Rev. A}\ }\textbf {\bibinfo {volume} {91}},\ \bibinfo {pages} {012331} (\bibinfo {year} {2015})}\BibitemShut {NoStop}%
\bibitem [{\citenamefont {Binder}\ \emph {et~al.}(2018)\citenamefont {Binder}, \citenamefont {Correa}, \citenamefont {Gogolin}, \citenamefont {Anders},\ and\ \citenamefont {Adesso}}]{binder2018thermodynamics}%
  \BibitemOpen
  \bibfield  {author} {\bibinfo {author} {\bibfnamefont {Felix}\ \bibnamefont {Binder}}, \bibinfo {author} {\bibfnamefont {Luis~A}\ \bibnamefont {Correa}}, \bibinfo {author} {\bibfnamefont {Christian}\ \bibnamefont {Gogolin}}, \bibinfo {author} {\bibfnamefont {Janet}\ \bibnamefont {Anders}}, \ and\ \bibinfo {author} {\bibfnamefont {Gerardo}\ \bibnamefont {Adesso}},\ }\bibfield  {title} {\enquote {\bibinfo {title} {Thermodynamics in the quantum regime},}\ }\href@noop {} {\bibfield  {journal} {\bibinfo  {journal} {Fundamental Theories of Physics}\ }\textbf {\bibinfo {volume} {195}} (\bibinfo {year} {2018})}\BibitemShut {NoStop}%
\bibitem [{\citenamefont {Campbell}\ \emph {et~al.}(2018)\citenamefont {Campbell}, \citenamefont {Genoni},\ and\ \citenamefont {Deffner}}]{Campbell_2018}%
  \BibitemOpen
  \bibfield  {author} {\bibinfo {author} {\bibfnamefont {Steve}\ \bibnamefont {Campbell}}, \bibinfo {author} {\bibfnamefont {Marco~G}\ \bibnamefont {Genoni}}, \ and\ \bibinfo {author} {\bibfnamefont {Sebastian}\ \bibnamefont {Deffner}},\ }\bibfield  {title} {\enquote {\bibinfo {title} {Precision thermometry and the quantum speed limit},}\ }\href {\doibase 10.1088/2058-9565/aaa641} {\bibfield  {journal} {\bibinfo  {journal} {Quantum Sci. Technol.}\ }\textbf {\bibinfo {volume} {3}},\ \bibinfo {pages} {025002} (\bibinfo {year} {2018})}\BibitemShut {NoStop}%
\bibitem [{\citenamefont {Campbell}\ \emph {et~al.}(2017)\citenamefont {Campbell}, \citenamefont {Mehboudi}, \citenamefont {Chiara},\ and\ \citenamefont {Paternostro}}]{Campbell_2017}%
  \BibitemOpen
  \bibfield  {author} {\bibinfo {author} {\bibfnamefont {Steve}\ \bibnamefont {Campbell}}, \bibinfo {author} {\bibfnamefont {Mohammad}\ \bibnamefont {Mehboudi}}, \bibinfo {author} {\bibfnamefont {Gabriele~De}\ \bibnamefont {Chiara}}, \ and\ \bibinfo {author} {\bibfnamefont {Mauro}\ \bibnamefont {Paternostro}},\ }\bibfield  {title} {\enquote {\bibinfo {title} {Global and local thermometry schemes in coupled quantum systems},}\ }\href {\doibase 10.1088/1367-2630/aa7fac} {\bibfield  {journal} {\bibinfo  {journal} {New J. Phys.}\ }\textbf {\bibinfo {volume} {19}},\ \bibinfo {pages} {103003} (\bibinfo {year} {2017})}\BibitemShut {NoStop}%
\bibitem [{\citenamefont {Sone}\ \emph {et~al.}(2019)\citenamefont {Sone}, \citenamefont {Zhuang}, \citenamefont {Li}, \citenamefont {Liu},\ and\ \citenamefont {Cappellaro}}]{PhysRevA.99.052318}%
  \BibitemOpen
  \bibfield  {author} {\bibinfo {author} {\bibfnamefont {Akira}\ \bibnamefont {Sone}}, \bibinfo {author} {\bibfnamefont {Quntao}\ \bibnamefont {Zhuang}}, \bibinfo {author} {\bibfnamefont {Changhao}\ \bibnamefont {Li}}, \bibinfo {author} {\bibfnamefont {Yi-Xiang}\ \bibnamefont {Liu}}, \ and\ \bibinfo {author} {\bibfnamefont {Paola}\ \bibnamefont {Cappellaro}},\ }\bibfield  {title} {\enquote {\bibinfo {title} {Nonclassical correlations for quantum metrology in thermal equilibrium},}\ }\href {\doibase 10.1103/PhysRevA.99.052318} {\bibfield  {journal} {\bibinfo  {journal} {Phys. Rev. A}\ }\textbf {\bibinfo {volume} {99}},\ \bibinfo {pages} {052318} (\bibinfo {year} {2019})}\BibitemShut {NoStop}%
\bibitem [{\citenamefont {Razavian}\ \emph {et~al.}(2019)\citenamefont {Razavian}, \citenamefont {Benedetti}, \citenamefont {Bina}, \citenamefont {Akbari-Kourbolagh},\ and\ \citenamefont {Paris}}]{Razavian2019}%
  \BibitemOpen
  \bibfield  {author} {\bibinfo {author} {\bibfnamefont {Sholeh}\ \bibnamefont {Razavian}}, \bibinfo {author} {\bibfnamefont {Claudia}\ \bibnamefont {Benedetti}}, \bibinfo {author} {\bibfnamefont {Matteo}\ \bibnamefont {Bina}}, \bibinfo {author} {\bibfnamefont {Yahya}\ \bibnamefont {Akbari-Kourbolagh}}, \ and\ \bibinfo {author} {\bibfnamefont {Matteo G.~A.}\ \bibnamefont {Paris}},\ }\bibfield  {title} {\enquote {\bibinfo {title} {Quantum thermometry by single-qubit dephasing},}\ }\href {\doibase 10.1140/epjp/i2019-12708-9} {\bibfield  {journal} {\bibinfo  {journal} {The European Physical Journal Plus}\ }\textbf {\bibinfo {volume} {134}},\ \bibinfo {pages} {284} (\bibinfo {year} {2019})}\BibitemShut {NoStop}%
\bibitem [{\citenamefont {Anto-Sztrikacs}\ \emph {et~al.}(2024)\citenamefont {Anto-Sztrikacs}, \citenamefont {Miller}, \citenamefont {Nazir},\ and\ \citenamefont {Segal}}]{PhysRevA.109.L060201}%
  \BibitemOpen
  \bibfield  {author} {\bibinfo {author} {\bibfnamefont {Nicholas}\ \bibnamefont {Anto-Sztrikacs}}, \bibinfo {author} {\bibfnamefont {Harry J.~D.}\ \bibnamefont {Miller}}, \bibinfo {author} {\bibfnamefont {Ahsan}\ \bibnamefont {Nazir}}, \ and\ \bibinfo {author} {\bibfnamefont {Dvira}\ \bibnamefont {Segal}},\ }\bibfield  {title} {\enquote {\bibinfo {title} {Bypassing thermalization timescales in temperature estimation using prethermal probes},}\ }\href {\doibase 10.1103/PhysRevA.109.L060201} {\bibfield  {journal} {\bibinfo  {journal} {Phys. Rev. A}\ }\textbf {\bibinfo {volume} {109}},\ \bibinfo {pages} {L060201} (\bibinfo {year} {2024})}\BibitemShut {NoStop}%
\bibitem [{\citenamefont {Stace}(2010)}]{PhysRevA.82.011611}%
  \BibitemOpen
  \bibfield  {author} {\bibinfo {author} {\bibfnamefont {Thomas~M.}\ \bibnamefont {Stace}},\ }\bibfield  {title} {\enquote {\bibinfo {title} {Quantum limits of thermometry},}\ }\href {\doibase 10.1103/PhysRevA.82.011611} {\bibfield  {journal} {\bibinfo  {journal} {Phys. Rev. A}\ }\textbf {\bibinfo {volume} {82}},\ \bibinfo {pages} {011611} (\bibinfo {year} {2010})}\BibitemShut {NoStop}%
\bibitem [{\citenamefont {Yuan}\ \emph {et~al.}(2023)\citenamefont {Yuan}, \citenamefont {Zhang}, \citenamefont {Song}, \citenamefont {Tang}, \citenamefont {Wang},\ and\ \citenamefont {Kuang}}]{PhysRevA.107.063317}%
  \BibitemOpen
  \bibfield  {author} {\bibinfo {author} {\bibfnamefont {Ji-Bing}\ \bibnamefont {Yuan}}, \bibinfo {author} {\bibfnamefont {Bo}~\bibnamefont {Zhang}}, \bibinfo {author} {\bibfnamefont {Ya-Ju}\ \bibnamefont {Song}}, \bibinfo {author} {\bibfnamefont {Shi-Qing}\ \bibnamefont {Tang}}, \bibinfo {author} {\bibfnamefont {Xin-Wen}\ \bibnamefont {Wang}}, \ and\ \bibinfo {author} {\bibfnamefont {Le-Man}\ \bibnamefont {Kuang}},\ }\bibfield  {title} {\enquote {\bibinfo {title} {Quantum sensing of temperature close to absolute zero in a bose-einstein condensate},}\ }\href {\doibase 10.1103/PhysRevA.107.063317} {\bibfield  {journal} {\bibinfo  {journal} {Phys. Rev. A}\ }\textbf {\bibinfo {volume} {107}},\ \bibinfo {pages} {063317} (\bibinfo {year} {2023})}\BibitemShut {NoStop}%
\bibitem [{\citenamefont {Adam}\ \emph {et~al.}(2022)\citenamefont {Adam}, \citenamefont {Bouton}, \citenamefont {Nettersheim}, \citenamefont {Burgardt},\ and\ \citenamefont {Widera}}]{PhysRevLett.129.120404}%
  \BibitemOpen
  \bibfield  {author} {\bibinfo {author} {\bibfnamefont {Daniel}\ \bibnamefont {Adam}}, \bibinfo {author} {\bibfnamefont {Quentin}\ \bibnamefont {Bouton}}, \bibinfo {author} {\bibfnamefont {Jens}\ \bibnamefont {Nettersheim}}, \bibinfo {author} {\bibfnamefont {Sabrina}\ \bibnamefont {Burgardt}}, \ and\ \bibinfo {author} {\bibfnamefont {Artur}\ \bibnamefont {Widera}},\ }\bibfield  {title} {\enquote {\bibinfo {title} {Coherent and dephasing spectroscopy for single-impurity probing of an ultracold bath},}\ }\href {\doibase 10.1103/PhysRevLett.129.120404} {\bibfield  {journal} {\bibinfo  {journal} {Phys. Rev. Lett.}\ }\textbf {\bibinfo {volume} {129}},\ \bibinfo {pages} {120404} (\bibinfo {year} {2022})}\BibitemShut {NoStop}%
\bibitem [{\citenamefont {Zhang}\ and\ \citenamefont {Tong}(2022)}]{Zhang2022}%
  \BibitemOpen
  \bibfield  {author} {\bibinfo {author} {\bibfnamefont {Da-Jian}\ \bibnamefont {Zhang}}\ and\ \bibinfo {author} {\bibfnamefont {D.~M.}\ \bibnamefont {Tong}},\ }\bibfield  {title} {\enquote {\bibinfo {title} {Approaching heisenberg-scalable thermometry with built-in robustness against noise},}\ }\href {\doibase 10.1038/s41534-022-00588-2} {\bibfield  {journal} {\bibinfo  {journal} {npj Quantum Information}\ }\textbf {\bibinfo {volume} {8}},\ \bibinfo {pages} {81} (\bibinfo {year} {2022})}\BibitemShut {NoStop}%
\bibitem [{\citenamefont {Ullah}\ \emph {et~al.}(2023)\citenamefont {Ullah}, \citenamefont {Naseem},\ and\ \citenamefont {M\"ustecapl\ifmmode \imath \else \i \fi{}o\ifmmode~\breve{g}\else \u{g}\fi{}lu}}]{PhysRevResearch.5.043184}%
  \BibitemOpen
  \bibfield  {author} {\bibinfo {author} {\bibfnamefont {A.}~\bibnamefont {Ullah}}, \bibinfo {author} {\bibfnamefont {M.~Tahir}\ \bibnamefont {Naseem}}, \ and\ \bibinfo {author} {\bibfnamefont {\"Ozg\"ur~E.}\ \bibnamefont {M\"ustecapl\ifmmode \imath \else \i \fi{}o\ifmmode~\breve{g}\else \u{g}\fi{}lu}},\ }\bibfield  {title} {\enquote {\bibinfo {title} {Low-temperature quantum thermometry boosted by coherence generation},}\ }\href {\doibase 10.1103/PhysRevResearch.5.043184} {\bibfield  {journal} {\bibinfo  {journal} {Phys. Rev. Res.}\ }\textbf {\bibinfo {volume} {5}},\ \bibinfo {pages} {043184} (\bibinfo {year} {2023})}\BibitemShut {NoStop}%
\bibitem [{\citenamefont {Aiache}\ \emph {et~al.}(2024{\natexlab{a}})\citenamefont {Aiache}, \citenamefont {El~Allati},\ and\ \citenamefont {El~Anouz}}]{PhysRevA.110.032605}%
  \BibitemOpen
  \bibfield  {author} {\bibinfo {author} {\bibfnamefont {Y.}~\bibnamefont {Aiache}}, \bibinfo {author} {\bibfnamefont {A.}~\bibnamefont {El~Allati}}, \ and\ \bibinfo {author} {\bibfnamefont {K.}~\bibnamefont {El~Anouz}},\ }\bibfield  {title} {\enquote {\bibinfo {title} {Harnessing coherence generation for precision single- and two-qubit quantum thermometry},}\ }\href {\doibase 10.1103/PhysRevA.110.032605} {\bibfield  {journal} {\bibinfo  {journal} {Phys. Rev. A}\ }\textbf {\bibinfo {volume} {110}},\ \bibinfo {pages} {032605} (\bibinfo {year} {2024}{\natexlab{a}})}\BibitemShut {NoStop}%
\bibitem [{\citenamefont {Seah}\ \emph {et~al.}(2019)\citenamefont {Seah}, \citenamefont {Nimmrichter}, \citenamefont {Grimmer}, \citenamefont {Santos}, \citenamefont {Scarani},\ and\ \citenamefont {Landi}}]{PhysRevLett.123.180602}%
  \BibitemOpen
  \bibfield  {author} {\bibinfo {author} {\bibfnamefont {Stella}\ \bibnamefont {Seah}}, \bibinfo {author} {\bibfnamefont {Stefan}\ \bibnamefont {Nimmrichter}}, \bibinfo {author} {\bibfnamefont {Daniel}\ \bibnamefont {Grimmer}}, \bibinfo {author} {\bibfnamefont {Jader~P.}\ \bibnamefont {Santos}}, \bibinfo {author} {\bibfnamefont {Valerio}\ \bibnamefont {Scarani}}, \ and\ \bibinfo {author} {\bibfnamefont {Gabriel~T.}\ \bibnamefont {Landi}},\ }\bibfield  {title} {\enquote {\bibinfo {title} {Collisional quantum thermometry},}\ }\href {\doibase 10.1103/PhysRevLett.123.180602} {\bibfield  {journal} {\bibinfo  {journal} {Phys. Rev. Lett.}\ }\textbf {\bibinfo {volume} {123}},\ \bibinfo {pages} {180602} (\bibinfo {year} {2019})}\BibitemShut {NoStop}%
\bibitem [{\citenamefont {Planella}\ \emph {et~al.}(2022)\citenamefont {Planella}, \citenamefont {Cenni}, \citenamefont {Ac\'{\i}n},\ and\ \citenamefont {Mehboudi}}]{PhysRevLett.128.040502}%
  \BibitemOpen
  \bibfield  {author} {\bibinfo {author} {\bibfnamefont {Guim}\ \bibnamefont {Planella}}, \bibinfo {author} {\bibfnamefont {Marina F.~B.}\ \bibnamefont {Cenni}}, \bibinfo {author} {\bibfnamefont {Antonio}\ \bibnamefont {Ac\'{\i}n}}, \ and\ \bibinfo {author} {\bibfnamefont {Mohammad}\ \bibnamefont {Mehboudi}},\ }\bibfield  {title} {\enquote {\bibinfo {title} {Bath-induced correlations enhance thermometry precision at low temperatures},}\ }\href {\doibase 10.1103/PhysRevLett.128.040502} {\bibfield  {journal} {\bibinfo  {journal} {Phys. Rev. Lett.}\ }\textbf {\bibinfo {volume} {128}},\ \bibinfo {pages} {040502} (\bibinfo {year} {2022})}\BibitemShut {NoStop}%
\bibitem [{\citenamefont {Ather}\ and\ \citenamefont {Chaudhry}(2021)}]{PhysRevA.104.012211}%
  \BibitemOpen
  \bibfield  {author} {\bibinfo {author} {\bibfnamefont {Hamza}\ \bibnamefont {Ather}}\ and\ \bibinfo {author} {\bibfnamefont {Adam~Zaman}\ \bibnamefont {Chaudhry}},\ }\bibfield  {title} {\enquote {\bibinfo {title} {Improving the estimation of environment parameters via initial probe-environment correlations},}\ }\href {\doibase 10.1103/PhysRevA.104.012211} {\bibfield  {journal} {\bibinfo  {journal} {Phys. Rev. A}\ }\textbf {\bibinfo {volume} {104}},\ \bibinfo {pages} {012211} (\bibinfo {year} {2021})}\BibitemShut {NoStop}%
\bibitem [{\citenamefont {Zhang}\ \emph {et~al.}(2022{\natexlab{a}})\citenamefont {Zhang}, \citenamefont {Chen}, \citenamefont {Bai}, \citenamefont {Wu},\ and\ \citenamefont {An}}]{PhysRevApplied.17.034073}%
  \BibitemOpen
  \bibfield  {author} {\bibinfo {author} {\bibfnamefont {Ning}\ \bibnamefont {Zhang}}, \bibinfo {author} {\bibfnamefont {Chong}\ \bibnamefont {Chen}}, \bibinfo {author} {\bibfnamefont {Si-Yuan}\ \bibnamefont {Bai}}, \bibinfo {author} {\bibfnamefont {Wei}\ \bibnamefont {Wu}}, \ and\ \bibinfo {author} {\bibfnamefont {Jun-Hong}\ \bibnamefont {An}},\ }\bibfield  {title} {\enquote {\bibinfo {title} {Non-markovian quantum thermometry},}\ }\href {\doibase 10.1103/PhysRevApplied.17.034073} {\bibfield  {journal} {\bibinfo  {journal} {Phys. Rev. Appl.}\ }\textbf {\bibinfo {volume} {17}},\ \bibinfo {pages} {034073} (\bibinfo {year} {2022}{\natexlab{a}})}\BibitemShut {NoStop}%
\bibitem [{\citenamefont {Wu}\ \emph {et~al.}(2021)\citenamefont {Wu}, \citenamefont {Bai},\ and\ \citenamefont {An}}]{PhysRevA.103.L010601}%
  \BibitemOpen
  \bibfield  {author} {\bibinfo {author} {\bibfnamefont {Wei}\ \bibnamefont {Wu}}, \bibinfo {author} {\bibfnamefont {Si-Yuan}\ \bibnamefont {Bai}}, \ and\ \bibinfo {author} {\bibfnamefont {Jun-Hong}\ \bibnamefont {An}},\ }\bibfield  {title} {\enquote {\bibinfo {title} {Non-markovian sensing of a quantum reservoir},}\ }\href {\doibase 10.1103/PhysRevA.103.L010601} {\bibfield  {journal} {\bibinfo  {journal} {Phys. Rev. A}\ }\textbf {\bibinfo {volume} {103}},\ \bibinfo {pages} {L010601} (\bibinfo {year} {2021})}\BibitemShut {NoStop}%
\bibitem [{\citenamefont {Xu}\ \emph {et~al.}(2023)\citenamefont {Xu}, \citenamefont {Yuan}, \citenamefont {Tang}, \citenamefont {Wu}, \citenamefont {Tan},\ and\ \citenamefont {Kuang}}]{PhysRevA.108.022608}%
  \BibitemOpen
  \bibfield  {author} {\bibinfo {author} {\bibfnamefont {Lan}\ \bibnamefont {Xu}}, \bibinfo {author} {\bibfnamefont {Ji-Bing}\ \bibnamefont {Yuan}}, \bibinfo {author} {\bibfnamefont {Shi-Qing}\ \bibnamefont {Tang}}, \bibinfo {author} {\bibfnamefont {Wei}\ \bibnamefont {Wu}}, \bibinfo {author} {\bibfnamefont {Qing-Shou}\ \bibnamefont {Tan}}, \ and\ \bibinfo {author} {\bibfnamefont {Le-Man}\ \bibnamefont {Kuang}},\ }\bibfield  {title} {\enquote {\bibinfo {title} {Non-markovian enhanced temperature sensing in a dipolar bose-einstein condensate},}\ }\href {\doibase 10.1103/PhysRevA.108.022608} {\bibfield  {journal} {\bibinfo  {journal} {Phys. Rev. A}\ }\textbf {\bibinfo {volume} {108}},\ \bibinfo {pages} {022608} (\bibinfo {year} {2023})}\BibitemShut {NoStop}%
\bibitem [{\citenamefont {Zhang}\ and\ \citenamefont {Wu}(2021)}]{PhysRevResearch.3.043039}%
  \BibitemOpen
  \bibfield  {author} {\bibinfo {author} {\bibfnamefont {Ze-Zhou}\ \bibnamefont {Zhang}}\ and\ \bibinfo {author} {\bibfnamefont {Wei}\ \bibnamefont {Wu}},\ }\bibfield  {title} {\enquote {\bibinfo {title} {Non-markovian temperature sensing},}\ }\href {\doibase 10.1103/PhysRevResearch.3.043039} {\bibfield  {journal} {\bibinfo  {journal} {Phys. Rev. Res.}\ }\textbf {\bibinfo {volume} {3}},\ \bibinfo {pages} {043039} (\bibinfo {year} {2021})}\BibitemShut {NoStop}%
\bibitem [{\citenamefont {Aiache}\ \emph {et~al.}(2024{\natexlab{b}})\citenamefont {Aiache}, \citenamefont {Seida}, \citenamefont {El~Anouz},\ and\ \citenamefont {El~Allati}}]{PhysRevE.110.024132}%
  \BibitemOpen
  \bibfield  {author} {\bibinfo {author} {\bibfnamefont {Y.}~\bibnamefont {Aiache}}, \bibinfo {author} {\bibfnamefont {C.}~\bibnamefont {Seida}}, \bibinfo {author} {\bibfnamefont {K.}~\bibnamefont {El~Anouz}}, \ and\ \bibinfo {author} {\bibfnamefont {A.}~\bibnamefont {El~Allati}},\ }\bibfield  {title} {\enquote {\bibinfo {title} {Non-markovian enhancement of nonequilibrium quantum thermometry},}\ }\href {\doibase 10.1103/PhysRevE.110.024132} {\bibfield  {journal} {\bibinfo  {journal} {Phys. Rev. E}\ }\textbf {\bibinfo {volume} {110}},\ \bibinfo {pages} {024132} (\bibinfo {year} {2024}{\natexlab{b}})}\BibitemShut {NoStop}%
\bibitem [{\citenamefont {Kiilerich}\ \emph {et~al.}(2018)\citenamefont {Kiilerich}, \citenamefont {De~Pasquale},\ and\ \citenamefont {Giovannetti}}]{PhysRevA.98.042124}%
  \BibitemOpen
  \bibfield  {author} {\bibinfo {author} {\bibfnamefont {Alexander~Holm}\ \bibnamefont {Kiilerich}}, \bibinfo {author} {\bibfnamefont {Antonella}\ \bibnamefont {De~Pasquale}}, \ and\ \bibinfo {author} {\bibfnamefont {Vittorio}\ \bibnamefont {Giovannetti}},\ }\bibfield  {title} {\enquote {\bibinfo {title} {Dynamical approach to ancilla-assisted quantum thermometry},}\ }\href {\doibase 10.1103/PhysRevA.98.042124} {\bibfield  {journal} {\bibinfo  {journal} {Phys. Rev. A}\ }\textbf {\bibinfo {volume} {98}},\ \bibinfo {pages} {042124} (\bibinfo {year} {2018})}\BibitemShut {NoStop}%
\bibitem [{\citenamefont {Ullah}\ \emph {et~al.}(2024)\citenamefont {Ullah}, \citenamefont {Tahir~Naseem},\ and\ \citenamefont {Müstecaplıoğlu}}]{Ullah_2025}%
  \BibitemOpen
  \bibfield  {author} {\bibinfo {author} {\bibfnamefont {A.}~\bibnamefont {Ullah}}, \bibinfo {author} {\bibfnamefont {M}~\bibnamefont {Tahir~Naseem}}, \ and\ \bibinfo {author} {\bibfnamefont {Özgür~E}\ \bibnamefont {Müstecaplıoğlu}},\ }\bibfield  {title} {\enquote {\bibinfo {title} {Mixing thermal coherent states for precision and range enhancement in quantum thermometry},}\ }\href {\doibase 10.1088/2058-9565/ad994a} {\bibfield  {journal} {\bibinfo  {journal} {Quantum Sci. Technol.}\ }\textbf {\bibinfo {volume} {10}},\ \bibinfo {pages} {015044} (\bibinfo {year} {2024})}\BibitemShut {NoStop}%
\bibitem [{\citenamefont {Tan}\ \emph {et~al.}(2024)\citenamefont {Tan}, \citenamefont {Liu}, \citenamefont {Xu}, \citenamefont {Wu},\ and\ \citenamefont {Kuang}}]{PhysRevA.109.042417}%
  \BibitemOpen
  \bibfield  {author} {\bibinfo {author} {\bibfnamefont {Qing-Shou}\ \bibnamefont {Tan}}, \bibinfo {author} {\bibfnamefont {Xulin}\ \bibnamefont {Liu}}, \bibinfo {author} {\bibfnamefont {Lan}\ \bibnamefont {Xu}}, \bibinfo {author} {\bibfnamefont {Wei}\ \bibnamefont {Wu}}, \ and\ \bibinfo {author} {\bibfnamefont {Le-Man}\ \bibnamefont {Kuang}},\ }\bibfield  {title} {\enquote {\bibinfo {title} {Enhancement of sensitivity in low-temperature quantum thermometry via reinforcement learning},}\ }\href {\doibase 10.1103/PhysRevA.109.042417} {\bibfield  {journal} {\bibinfo  {journal} {Phys. Rev. A}\ }\textbf {\bibinfo {volume} {109}},\ \bibinfo {pages} {042417} (\bibinfo {year} {2024})}\BibitemShut {NoStop}%
\bibitem [{\citenamefont {Mehboudi}\ \emph {et~al.}(2019{\natexlab{b}})\citenamefont {Mehboudi}, \citenamefont {Lampo}, \citenamefont {Charalambous}, \citenamefont {Correa}, \citenamefont {Garc\'{\i}a-March},\ and\ \citenamefont {Lewenstein}}]{PhysRevLett.122.030403}%
  \BibitemOpen
  \bibfield  {author} {\bibinfo {author} {\bibfnamefont {Mohammad}\ \bibnamefont {Mehboudi}}, \bibinfo {author} {\bibfnamefont {Aniello}\ \bibnamefont {Lampo}}, \bibinfo {author} {\bibfnamefont {Christos}\ \bibnamefont {Charalambous}}, \bibinfo {author} {\bibfnamefont {Luis~A.}\ \bibnamefont {Correa}}, \bibinfo {author} {\bibfnamefont {Miguel~\'Angel}\ \bibnamefont {Garc\'{\i}a-March}}, \ and\ \bibinfo {author} {\bibfnamefont {Maciej}\ \bibnamefont {Lewenstein}},\ }\bibfield  {title} {\enquote {\bibinfo {title} {Using polarons for sub-nk quantum nondemolition thermometry in a bose-einstein condensate},}\ }\href {\doibase 10.1103/PhysRevLett.122.030403} {\bibfield  {journal} {\bibinfo  {journal} {Phys. Rev. Lett.}\ }\textbf {\bibinfo {volume} {122}},\ \bibinfo {pages} {030403} (\bibinfo {year} {2019}{\natexlab{b}})}\BibitemShut {NoStop}%
\bibitem [{\citenamefont {Mitchison}\ \emph {et~al.}(2020)\citenamefont {Mitchison}, \citenamefont {Fogarty}, \citenamefont {Guarnieri}, \citenamefont {Campbell}, \citenamefont {Busch},\ and\ \citenamefont {Goold}}]{PhysRevLett.125.080402}%
  \BibitemOpen
  \bibfield  {author} {\bibinfo {author} {\bibfnamefont {Mark~T.}\ \bibnamefont {Mitchison}}, \bibinfo {author} {\bibfnamefont {Thom\'as}\ \bibnamefont {Fogarty}}, \bibinfo {author} {\bibfnamefont {Giacomo}\ \bibnamefont {Guarnieri}}, \bibinfo {author} {\bibfnamefont {Steve}\ \bibnamefont {Campbell}}, \bibinfo {author} {\bibfnamefont {Thomas}\ \bibnamefont {Busch}}, \ and\ \bibinfo {author} {\bibfnamefont {John}\ \bibnamefont {Goold}},\ }\bibfield  {title} {\enquote {\bibinfo {title} {In situ thermometry of a cold fermi gas via dephasing impurities},}\ }\href {\doibase 10.1103/PhysRevLett.125.080402} {\bibfield  {journal} {\bibinfo  {journal} {Phys. Rev. Lett.}\ }\textbf {\bibinfo {volume} {125}},\ \bibinfo {pages} {080402} (\bibinfo {year} {2020})}\BibitemShut {NoStop}%
\bibitem [{\citenamefont {Sab{\'i}n}\ \emph {et~al.}(2014)\citenamefont {Sab{\'i}n}, \citenamefont {White}, \citenamefont {Hackermuller},\ and\ \citenamefont {Fuentes}}]{Sabín2014}%
  \BibitemOpen
  \bibfield  {author} {\bibinfo {author} {\bibfnamefont {Carlos}\ \bibnamefont {Sab{\'i}n}}, \bibinfo {author} {\bibfnamefont {Angela}\ \bibnamefont {White}}, \bibinfo {author} {\bibfnamefont {Lucia}\ \bibnamefont {Hackermuller}}, \ and\ \bibinfo {author} {\bibfnamefont {Ivette}\ \bibnamefont {Fuentes}},\ }\bibfield  {title} {\enquote {\bibinfo {title} {Impurities as a quantum thermometer for a bose-einstein condensate},}\ }\href {\doibase 10.1038/srep06436} {\bibfield  {journal} {\bibinfo  {journal} {Sci. Rep.}\ }\textbf {\bibinfo {volume} {4}},\ \bibinfo {pages} {6436} (\bibinfo {year} {2014})}\BibitemShut {NoStop}%
\bibitem [{\citenamefont {Oghittu}\ and\ \citenamefont {Negretti}(2022)}]{PhysRevResearch.4.023069}%
  \BibitemOpen
  \bibfield  {author} {\bibinfo {author} {\bibfnamefont {Lorenzo}\ \bibnamefont {Oghittu}}\ and\ \bibinfo {author} {\bibfnamefont {Antonio}\ \bibnamefont {Negretti}},\ }\bibfield  {title} {\enquote {\bibinfo {title} {Quantum-limited thermometry of a fermi gas with a charged spin particle},}\ }\href {\doibase 10.1103/PhysRevResearch.4.023069} {\bibfield  {journal} {\bibinfo  {journal} {Phys. Rev. Res.}\ }\textbf {\bibinfo {volume} {4}},\ \bibinfo {pages} {023069} (\bibinfo {year} {2022})}\BibitemShut {NoStop}%
\bibitem [{\citenamefont {Khan}\ \emph {et~al.}(2022)\citenamefont {Khan}, \citenamefont {Mehboudi}, \citenamefont {Ter\ifmmode~\mbox{\c{c}}\else \c{c}\fi{}as}, \citenamefont {Lewenstein},\ and\ \citenamefont {Garcia-March}}]{PhysRevResearch.4.023191}%
  \BibitemOpen
  \bibfield  {author} {\bibinfo {author} {\bibfnamefont {Muhammad~Miskeen}\ \bibnamefont {Khan}}, \bibinfo {author} {\bibfnamefont {Mohammad}\ \bibnamefont {Mehboudi}}, \bibinfo {author} {\bibfnamefont {Hugo}\ \bibnamefont {Ter\ifmmode~\mbox{\c{c}}\else \c{c}\fi{}as}}, \bibinfo {author} {\bibfnamefont {Maciej}\ \bibnamefont {Lewenstein}}, \ and\ \bibinfo {author} {\bibfnamefont {Miguel~Angel}\ \bibnamefont {Garcia-March}},\ }\bibfield  {title} {\enquote {\bibinfo {title} {Subnanokelvin thermometry of an interacting $d$-dimensional homogeneous bose gas},}\ }\href {\doibase 10.1103/PhysRevResearch.4.023191} {\bibfield  {journal} {\bibinfo  {journal} {Phys. Rev. Res.}\ }\textbf {\bibinfo {volume} {4}},\ \bibinfo {pages} {023191} (\bibinfo {year} {2022})}\BibitemShut {NoStop}%
\bibitem [{\citenamefont {Brattegard}\ and\ \citenamefont {Mitchison}(2024)}]{PhysRevA.109.023309}%
  \BibitemOpen
  \bibfield  {author} {\bibinfo {author} {\bibfnamefont {Sindre}\ \bibnamefont {Brattegard}}\ and\ \bibinfo {author} {\bibfnamefont {Mark~T.}\ \bibnamefont {Mitchison}},\ }\bibfield  {title} {\enquote {\bibinfo {title} {Thermometry by correlated dephasing of impurities in a one-dimensional fermi gas},}\ }\href {\doibase 10.1103/PhysRevA.109.023309} {\bibfield  {journal} {\bibinfo  {journal} {Phys. Rev. A}\ }\textbf {\bibinfo {volume} {109}},\ \bibinfo {pages} {023309} (\bibinfo {year} {2024})}\BibitemShut {NoStop}%
\bibitem [{\citenamefont {Zhou}\ \emph {et~al.}(2023)\citenamefont {Zhou}, \citenamefont {Kong}, \citenamefont {Lan},\ and\ \citenamefont {Zhang}}]{PhysRevResearch.5.013087}%
  \BibitemOpen
  \bibfield  {author} {\bibinfo {author} {\bibfnamefont {Lu}~\bibnamefont {Zhou}}, \bibinfo {author} {\bibfnamefont {Jia}\ \bibnamefont {Kong}}, \bibinfo {author} {\bibfnamefont {Zhihao}\ \bibnamefont {Lan}}, \ and\ \bibinfo {author} {\bibfnamefont {Weiping}\ \bibnamefont {Zhang}},\ }\bibfield  {title} {\enquote {\bibinfo {title} {Dynamical quantum phase transitions in a spinor bose-einstein condensate and criticality enhanced quantum sensing},}\ }\href {\doibase 10.1103/PhysRevResearch.5.013087} {\bibfield  {journal} {\bibinfo  {journal} {Phys. Rev. Res.}\ }\textbf {\bibinfo {volume} {5}},\ \bibinfo {pages} {013087} (\bibinfo {year} {2023})}\BibitemShut {NoStop}%
\bibitem [{\citenamefont {Mukhopadhyay}\ and\ \citenamefont {Bayat}(2024)}]{PhysRevLett.133.120601}%
  \BibitemOpen
  \bibfield  {author} {\bibinfo {author} {\bibfnamefont {Chiranjib}\ \bibnamefont {Mukhopadhyay}}\ and\ \bibinfo {author} {\bibfnamefont {Abolfazl}\ \bibnamefont {Bayat}},\ }\bibfield  {title} {\enquote {\bibinfo {title} {Modular many-body quantum sensors},}\ }\href {\doibase 10.1103/PhysRevLett.133.120601} {\bibfield  {journal} {\bibinfo  {journal} {Phys. Rev. Lett.}\ }\textbf {\bibinfo {volume} {133}},\ \bibinfo {pages} {120601} (\bibinfo {year} {2024})}\BibitemShut {NoStop}%
\bibitem [{\citenamefont {Aybar}\ \emph {et~al.}(2022)\citenamefont {Aybar}, \citenamefont {Niezgoda}, \citenamefont {Mirkhalaf}, \citenamefont {Mitchell}, \citenamefont {Benedicto~Orenes},\ and\ \citenamefont {Witkowska}}]{Aybar2022criticalquantum}%
  \BibitemOpen
  \bibfield  {author} {\bibinfo {author} {\bibfnamefont {Enes}\ \bibnamefont {Aybar}}, \bibinfo {author} {\bibfnamefont {Artur}\ \bibnamefont {Niezgoda}}, \bibinfo {author} {\bibfnamefont {Safoura~S.}\ \bibnamefont {Mirkhalaf}}, \bibinfo {author} {\bibfnamefont {Morgan~W.}\ \bibnamefont {Mitchell}}, \bibinfo {author} {\bibfnamefont {Daniel}\ \bibnamefont {Benedicto~Orenes}}, \ and\ \bibinfo {author} {\bibfnamefont {Emilia}\ \bibnamefont {Witkowska}},\ }\bibfield  {title} {\enquote {\bibinfo {title} {Critical quantum thermometry and its feasibility in spin systems},}\ }\href {\doibase 10.22331/q-2022-09-19-808} {\bibfield  {journal} {\bibinfo  {journal} {{Quantum}}\ }\textbf {\bibinfo {volume} {6}},\ \bibinfo {pages} {808} (\bibinfo {year} {2022})}\BibitemShut {NoStop}%
\bibitem [{\citenamefont {Wu}\ and\ \citenamefont {Shi}(2021)}]{PhysRevA.104.022612}%
  \BibitemOpen
  \bibfield  {author} {\bibinfo {author} {\bibfnamefont {Wei}\ \bibnamefont {Wu}}\ and\ \bibinfo {author} {\bibfnamefont {Chuan}\ \bibnamefont {Shi}},\ }\bibfield  {title} {\enquote {\bibinfo {title} {Criticality-enhanced quantum sensor at finite temperature},}\ }\href {\doibase 10.1103/PhysRevA.104.022612} {\bibfield  {journal} {\bibinfo  {journal} {Phys. Rev. A}\ }\textbf {\bibinfo {volume} {104}},\ \bibinfo {pages} {022612} (\bibinfo {year} {2021})}\BibitemShut {NoStop}%
\bibitem [{\citenamefont {Mirkhalaf}\ \emph {et~al.}(2021)\citenamefont {Mirkhalaf}, \citenamefont {Benedicto~Orenes}, \citenamefont {Mitchell},\ and\ \citenamefont {Witkowska}}]{PhysRevA.103.023317}%
  \BibitemOpen
  \bibfield  {author} {\bibinfo {author} {\bibfnamefont {Safoura~S.}\ \bibnamefont {Mirkhalaf}}, \bibinfo {author} {\bibfnamefont {Daniel}\ \bibnamefont {Benedicto~Orenes}}, \bibinfo {author} {\bibfnamefont {Morgan~W.}\ \bibnamefont {Mitchell}}, \ and\ \bibinfo {author} {\bibfnamefont {Emilia}\ \bibnamefont {Witkowska}},\ }\bibfield  {title} {\enquote {\bibinfo {title} {Criticality-enhanced quantum sensing in ferromagnetic bose-einstein condensates: Role of readout measurement and detection noise},}\ }\href {\doibase 10.1103/PhysRevA.103.023317} {\bibfield  {journal} {\bibinfo  {journal} {Phys. Rev. A}\ }\textbf {\bibinfo {volume} {103}},\ \bibinfo {pages} {023317} (\bibinfo {year} {2021})}\BibitemShut {NoStop}%
\bibitem [{\citenamefont {Mihailescu}\ \emph {et~al.}(2024)\citenamefont {Mihailescu}, \citenamefont {Bayat}, \citenamefont {Campbell},\ and\ \citenamefont {Mitchell}}]{Mihailescu_2024}%
  \BibitemOpen
  \bibfield  {author} {\bibinfo {author} {\bibfnamefont {George}\ \bibnamefont {Mihailescu}}, \bibinfo {author} {\bibfnamefont {Abolfazl}\ \bibnamefont {Bayat}}, \bibinfo {author} {\bibfnamefont {Steve}\ \bibnamefont {Campbell}}, \ and\ \bibinfo {author} {\bibfnamefont {Andrew~K}\ \bibnamefont {Mitchell}},\ }\bibfield  {title} {\enquote {\bibinfo {title} {Multiparameter critical quantum metrology with impurity probes},}\ }\href {\doibase 10.1088/2058-9565/ad438d} {\bibfield  {journal} {\bibinfo  {journal} {Quantum Sci. Technol.}\ }\textbf {\bibinfo {volume} {9}},\ \bibinfo {pages} {035033} (\bibinfo {year} {2024})}\BibitemShut {NoStop}%
\bibitem [{\citenamefont {Mihailescu}\ \emph {et~al.}(2023)\citenamefont {Mihailescu}, \citenamefont {Campbell},\ and\ \citenamefont {Mitchell}}]{PhysRevA.107.042614}%
  \BibitemOpen
  \bibfield  {author} {\bibinfo {author} {\bibfnamefont {George}\ \bibnamefont {Mihailescu}}, \bibinfo {author} {\bibfnamefont {Steve}\ \bibnamefont {Campbell}}, \ and\ \bibinfo {author} {\bibfnamefont {Andrew~K.}\ \bibnamefont {Mitchell}},\ }\bibfield  {title} {\enquote {\bibinfo {title} {Thermometry of strongly correlated fermionic quantum systems using impurity probes},}\ }\href {\doibase 10.1103/PhysRevA.107.042614} {\bibfield  {journal} {\bibinfo  {journal} {Phys. Rev. A}\ }\textbf {\bibinfo {volume} {107}},\ \bibinfo {pages} {042614} (\bibinfo {year} {2023})}\BibitemShut {NoStop}%
\bibitem [{\citenamefont {Xie}\ and\ \citenamefont {Xu}(2024)}]{PhysRevResearch.6.033102}%
  \BibitemOpen
  \bibfield  {author} {\bibinfo {author} {\bibfnamefont {Dong}\ \bibnamefont {Xie}}\ and\ \bibinfo {author} {\bibfnamefont {Chunling}\ \bibnamefont {Xu}},\ }\bibfield  {title} {\enquote {\bibinfo {title} {Thermometry with a dissipative heavy impurity},}\ }\href {\doibase 10.1103/PhysRevResearch.6.033102} {\bibfield  {journal} {\bibinfo  {journal} {Phys. Rev. Res.}\ }\textbf {\bibinfo {volume} {6}},\ \bibinfo {pages} {033102} (\bibinfo {year} {2024})}\BibitemShut {NoStop}%
\bibitem [{\citenamefont {Tian}\ \emph {et~al.}(2017)\citenamefont {Tian}, \citenamefont {Wang}, \citenamefont {Jing},\ and\ \citenamefont {Dragan}}]{TIAN20171}%
  \BibitemOpen
  \bibfield  {author} {\bibinfo {author} {\bibfnamefont {Zehua}\ \bibnamefont {Tian}}, \bibinfo {author} {\bibfnamefont {Jieci}\ \bibnamefont {Wang}}, \bibinfo {author} {\bibfnamefont {Jiliang}\ \bibnamefont {Jing}}, \ and\ \bibinfo {author} {\bibfnamefont {Andrzej}\ \bibnamefont {Dragan}},\ }\bibfield  {title} {\enquote {\bibinfo {title} {Entanglement enhanced thermometry in the detection of the unruh effect},}\ }\href {\doibase https://doi.org/10.1016/j.aop.2017.01.011} {\bibfield  {journal} {\bibinfo  {journal} {Ann. Phys.}\ }\textbf {\bibinfo {volume} {377}},\ \bibinfo {pages} {1--9} (\bibinfo {year} {2017})}\BibitemShut {NoStop}%
\bibitem [{\citenamefont {Mancino}\ \emph {et~al.}(2020)\citenamefont {Mancino}, \citenamefont {Genoni}, \citenamefont {Barbieri},\ and\ \citenamefont {Paternostro}}]{PhysRevResearch.2.033498}%
  \BibitemOpen
  \bibfield  {author} {\bibinfo {author} {\bibfnamefont {Luca}\ \bibnamefont {Mancino}}, \bibinfo {author} {\bibfnamefont {Marco~G.}\ \bibnamefont {Genoni}}, \bibinfo {author} {\bibfnamefont {Marco}\ \bibnamefont {Barbieri}}, \ and\ \bibinfo {author} {\bibfnamefont {Mauro}\ \bibnamefont {Paternostro}},\ }\bibfield  {title} {\enquote {\bibinfo {title} {Nonequilibrium readiness and precision of gaussian quantum thermometers},}\ }\href {\doibase 10.1103/PhysRevResearch.2.033498} {\bibfield  {journal} {\bibinfo  {journal} {Phys. Rev. Res.}\ }\textbf {\bibinfo {volume} {2}},\ \bibinfo {pages} {033498} (\bibinfo {year} {2020})}\BibitemShut {NoStop}%
\bibitem [{\citenamefont {Glatthard}\ and\ \citenamefont {Correa}(2022)}]{Glatthard2022bendingrulesoflow}%
  \BibitemOpen
  \bibfield  {author} {\bibinfo {author} {\bibfnamefont {Jonas}\ \bibnamefont {Glatthard}}\ and\ \bibinfo {author} {\bibfnamefont {Luis~A.}\ \bibnamefont {Correa}},\ }\bibfield  {title} {\enquote {\bibinfo {title} {Bending the rules of low-temperature thermometry with periodic driving},}\ }\href {\doibase 10.22331/q-2022-05-03-705} {\bibfield  {journal} {\bibinfo  {journal} {{Quantum}}\ }\textbf {\bibinfo {volume} {6}},\ \bibinfo {pages} {705} (\bibinfo {year} {2022})}\BibitemShut {NoStop}%
\bibitem [{\citenamefont {Mukherjee}\ \emph {et~al.}(2019)\citenamefont {Mukherjee}, \citenamefont {Zwick}, \citenamefont {Ghosh}, \citenamefont {Chen},\ and\ \citenamefont {Kurizki}}]{Mukherjee2019}%
  \BibitemOpen
  \bibfield  {author} {\bibinfo {author} {\bibfnamefont {Victor}\ \bibnamefont {Mukherjee}}, \bibinfo {author} {\bibfnamefont {Analia}\ \bibnamefont {Zwick}}, \bibinfo {author} {\bibfnamefont {Arnab}\ \bibnamefont {Ghosh}}, \bibinfo {author} {\bibfnamefont {Xi}~\bibnamefont {Chen}}, \ and\ \bibinfo {author} {\bibfnamefont {Gershon}\ \bibnamefont {Kurizki}},\ }\bibfield  {title} {\enquote {\bibinfo {title} {Enhanced precision bound of low-temperature quantum thermometry via dynamical control},}\ }\href {\doibase 10.1038/s42005-019-0265-y} {\bibfield  {journal} {\bibinfo  {journal} {Commun. Phys.}\ }\textbf {\bibinfo {volume} {2}},\ \bibinfo {pages} {162} (\bibinfo {year} {2019})}\BibitemShut {NoStop}%
\bibitem [{\citenamefont {Brenes}\ and\ \citenamefont {Segal}(2023)}]{segalMultispin}%
  \BibitemOpen
  \bibfield  {author} {\bibinfo {author} {\bibfnamefont {Marlon}\ \bibnamefont {Brenes}}\ and\ \bibinfo {author} {\bibfnamefont {Dvira}\ \bibnamefont {Segal}},\ }\bibfield  {title} {\enquote {\bibinfo {title} {Multispin probes for thermometry in the strong-coupling regime},}\ }\href {\doibase 10.1103/PhysRevA.108.032220} {\bibfield  {journal} {\bibinfo  {journal} {Phys. Rev. A}\ }\textbf {\bibinfo {volume} {108}},\ \bibinfo {pages} {032220} (\bibinfo {year} {2023})}\BibitemShut {NoStop}%
\bibitem [{\citenamefont {Braun}(2002)}]{Braun2002}%
  \BibitemOpen
  \bibfield  {author} {\bibinfo {author} {\bibfnamefont {Daniel}\ \bibnamefont {Braun}},\ }\bibfield  {title} {\enquote {\bibinfo {title} {{Creation of Entanglement by Interaction with a Common Heat Bath}},}\ }\href {\doibase 10.1103/PhysRevLett.89.277901} {\bibfield  {journal} {\bibinfo  {journal} {Phys. Rev. Lett.}\ }\textbf {\bibinfo {volume} {89}},\ \bibinfo {pages} {277901} (\bibinfo {year} {2002})}\BibitemShut {NoStop}%
\bibitem [{\citenamefont {Benatti}\ \emph {et~al.}(2003)\citenamefont {Benatti}, \citenamefont {Floreanini},\ and\ \citenamefont {Piani}}]{Benatti2003a}%
  \BibitemOpen
  \bibfield  {author} {\bibinfo {author} {\bibfnamefont {Fabio}\ \bibnamefont {Benatti}}, \bibinfo {author} {\bibfnamefont {Roberto}\ \bibnamefont {Floreanini}}, \ and\ \bibinfo {author} {\bibfnamefont {Marco}\ \bibnamefont {Piani}},\ }\bibfield  {title} {\enquote {\bibinfo {title} {{Environment Induced Entanglement in Markovian Dissipative Dynamics}},}\ }\href {\doibase 10.1103/PhysRevLett.91.070402} {\bibfield  {journal} {\bibinfo  {journal} {Phys. Rev. Lett.}\ }\textbf {\bibinfo {volume} {91}},\ \bibinfo {pages} {070402} (\bibinfo {year} {2003})}\BibitemShut {NoStop}%
\bibitem [{\citenamefont {T\'oth}(2012)}]{PhysRevA.85.022322}%
  \BibitemOpen
  \bibfield  {author} {\bibinfo {author} {\bibfnamefont {G\'eza}\ \bibnamefont {T\'oth}},\ }\bibfield  {title} {\enquote {\bibinfo {title} {Multipartite entanglement and high-precision metrology},}\ }\href {\doibase 10.1103/PhysRevA.85.022322} {\bibfield  {journal} {\bibinfo  {journal} {Phys. Rev. A}\ }\textbf {\bibinfo {volume} {85}},\ \bibinfo {pages} {022322} (\bibinfo {year} {2012})}\BibitemShut {NoStop}%
\bibitem [{\citenamefont {Cattaneo}\ \emph {et~al.}(2019)\citenamefont {Cattaneo}, \citenamefont {Giorgi}, \citenamefont {Maniscalco},\ and\ \citenamefont {Zambrini}}]{Cattaneo_2019}%
  \BibitemOpen
  \bibfield  {author} {\bibinfo {author} {\bibfnamefont {Marco}\ \bibnamefont {Cattaneo}}, \bibinfo {author} {\bibfnamefont {Gian~Luca}\ \bibnamefont {Giorgi}}, \bibinfo {author} {\bibfnamefont {Sabrina}\ \bibnamefont {Maniscalco}}, \ and\ \bibinfo {author} {\bibfnamefont {Roberta}\ \bibnamefont {Zambrini}},\ }\bibfield  {title} {\enquote {\bibinfo {title} {Local versus global master equation with common and separate baths: superiority of the global approach in partial secular approximation},}\ }\href {\doibase 10.1088/1367-2630/ab54ac} {\bibfield  {journal} {\bibinfo  {journal} {New J. Phys.}\ }\textbf {\bibinfo {volume} {21}},\ \bibinfo {pages} {113045} (\bibinfo {year} {2019})}\BibitemShut {NoStop}%
\bibitem [{\citenamefont {Paris}(2009)}]{paris2009}%
  \BibitemOpen
  \bibfield  {author} {\bibinfo {author} {\bibfnamefont {Matteo G.~A.}\ \bibnamefont {Paris}},\ }\bibfield  {title} {\enquote {\bibinfo {title} {Quantum estimation for quantum technology},}\ }\href {\doibase 10.1142/S0219749909004839} {\bibfield  {journal} {\bibinfo  {journal} {Int. J. Quantum. Inf.}\ }\textbf {\bibinfo {volume} {07}},\ \bibinfo {pages} {125--137} (\bibinfo {year} {2009})}\BibitemShut {NoStop}%
\bibitem [{\citenamefont {Liu}\ \emph {et~al.}(2020)\citenamefont {Liu}, \citenamefont {Yuan}, \citenamefont {Lu},\ and\ \citenamefont {Wang}}]{Liu2020}%
  \BibitemOpen
  \bibfield  {author} {\bibinfo {author} {\bibfnamefont {Jing}\ \bibnamefont {Liu}}, \bibinfo {author} {\bibfnamefont {Haidong}\ \bibnamefont {Yuan}}, \bibinfo {author} {\bibfnamefont {Xiao~Ming}\ \bibnamefont {Lu}}, \ and\ \bibinfo {author} {\bibfnamefont {Xiaoguang}\ \bibnamefont {Wang}},\ }\bibfield  {title} {\enquote {\bibinfo {title} {{Quantum Fisher information matrix and multiparameter estimation}},}\ }\href {\doibase 10.1088/1751-8121/ab5d4d} {\bibfield  {journal} {\bibinfo  {journal} {J. Phys. A: Math. Theo.}\ }\textbf {\bibinfo {volume} {53}},\ \bibinfo {pages} {023001} (\bibinfo {year} {2020})}\BibitemShut {NoStop}%
\bibitem [{\citenamefont {Blais}\ \emph {et~al.}(2021)\citenamefont {Blais}, \citenamefont {Grimsmo}, \citenamefont {Girvin},\ and\ \citenamefont {Wallraff}}]{Blais2020}%
  \BibitemOpen
  \bibfield  {author} {\bibinfo {author} {\bibfnamefont {Alexandre}\ \bibnamefont {Blais}}, \bibinfo {author} {\bibfnamefont {Arne~L.}\ \bibnamefont {Grimsmo}}, \bibinfo {author} {\bibfnamefont {S.~M.}\ \bibnamefont {Girvin}}, \ and\ \bibinfo {author} {\bibfnamefont {Andreas}\ \bibnamefont {Wallraff}},\ }\bibfield  {title} {\enquote {\bibinfo {title} {{Circuit quantum electrodynamics}},}\ }\href {\doibase 10.1103/RevModPhys.93.025005} {\bibfield  {journal} {\bibinfo  {journal} {Rev. Mod. Phys.}\ }\textbf {\bibinfo {volume} {93}},\ \bibinfo {pages} {025005} (\bibinfo {year} {2021})}\BibitemShut {NoStop}%
\bibitem [{\citenamefont {Cattaneo}\ and\ \citenamefont {Paraoanu}(2021)}]{Cattaneo2021engineering}%
  \BibitemOpen
  \bibfield  {author} {\bibinfo {author} {\bibfnamefont {Marco}\ \bibnamefont {Cattaneo}}\ and\ \bibinfo {author} {\bibfnamefont {Gheorghe~Sorin}\ \bibnamefont {Paraoanu}},\ }\bibfield  {title} {\enquote {\bibinfo {title} {{Engineering Dissipation with Resistive Elements in Circuit Quantum Electrodynamics}},}\ }\href {\doibase 10.1002/qute.202100054} {\bibfield  {journal} {\bibinfo  {journal} {Adv. Quantum Technol.}\ }\textbf {\bibinfo {volume} {4}},\ \bibinfo {pages} {2100054} (\bibinfo {year} {2021})}\BibitemShut {NoStop}%
\bibitem [{\citenamefont {Lalumi{\`{e}}re}\ \emph {et~al.}(2013)\citenamefont {Lalumi{\`{e}}re}, \citenamefont {Sanders}, \citenamefont {van Loo}, \citenamefont {Fedorov}, \citenamefont {Wallraff},\ and\ \citenamefont {Blais}}]{Lalumiere2013}%
  \BibitemOpen
  \bibfield  {author} {\bibinfo {author} {\bibfnamefont {Kevin}\ \bibnamefont {Lalumi{\`{e}}re}}, \bibinfo {author} {\bibfnamefont {Barry~C.}\ \bibnamefont {Sanders}}, \bibinfo {author} {\bibfnamefont {A.~F.}\ \bibnamefont {van Loo}}, \bibinfo {author} {\bibfnamefont {A.}~\bibnamefont {Fedorov}}, \bibinfo {author} {\bibfnamefont {Andreas}\ \bibnamefont {Wallraff}}, \ and\ \bibinfo {author} {\bibfnamefont {Alexandre}\ \bibnamefont {Blais}},\ }\bibfield  {title} {\enquote {\bibinfo {title} {{Input-output theory for waveguide QED with an ensemble of inhomogeneous atoms}},}\ }\href {\doibase 10.1103/PhysRevA.88.043806} {\bibfield  {journal} {\bibinfo  {journal} {Phys. Rev. A}\ }\textbf {\bibinfo {volume} {88}},\ \bibinfo {pages} {043806} (\bibinfo {year} {2013})}\BibitemShut {NoStop}%
\bibitem [{\citenamefont {van Loo}\ \emph {et~al.}(2013)\citenamefont {van Loo}, \citenamefont {Fedorov}, \citenamefont {Lalumière}, \citenamefont {Sanders}, \citenamefont {Blais},\ and\ \citenamefont {Wallraff}}]{vanLoosuper}%
  \BibitemOpen
  \bibfield  {author} {\bibinfo {author} {\bibfnamefont {Arjan~F.}\ \bibnamefont {van Loo}}, \bibinfo {author} {\bibfnamefont {Arkady}\ \bibnamefont {Fedorov}}, \bibinfo {author} {\bibfnamefont {Kevin}\ \bibnamefont {Lalumière}}, \bibinfo {author} {\bibfnamefont {Barry~C.}\ \bibnamefont {Sanders}}, \bibinfo {author} {\bibfnamefont {Alexandre}\ \bibnamefont {Blais}}, \ and\ \bibinfo {author} {\bibfnamefont {Andreas}\ \bibnamefont {Wallraff}},\ }\bibfield  {title} {\enquote {\bibinfo {title} {Photon-mediated interactions between distant artificial atoms},}\ }\href {\doibase 10.1126/science.1244324} {\bibfield  {journal} {\bibinfo  {journal} {Science}\ }\textbf {\bibinfo {volume} {342}},\ \bibinfo {pages} {1494--1496} (\bibinfo {year} {2013})}\BibitemShut {NoStop}%
\bibitem [{\citenamefont {Mlynek}\ \emph {et~al.}(2014)\citenamefont {Mlynek}, \citenamefont {Abdumalikov}, \citenamefont {Eichler},\ and\ \citenamefont {Wallraff}}]{Mlynek2015}%
  \BibitemOpen
  \bibfield  {author} {\bibinfo {author} {\bibfnamefont {J.~A.}\ \bibnamefont {Mlynek}}, \bibinfo {author} {\bibfnamefont {A.~A.}\ \bibnamefont {Abdumalikov}}, \bibinfo {author} {\bibfnamefont {C.}~\bibnamefont {Eichler}}, \ and\ \bibinfo {author} {\bibfnamefont {Andreas}\ \bibnamefont {Wallraff}},\ }\bibfield  {title} {\enquote {\bibinfo {title} {{Observation of Dicke superradiance for two artificial atoms in a cavity with high decay rate}},}\ }\href {\doibase 10.1038/ncomms6186} {\bibfield  {journal} {\bibinfo  {journal} {Nat. Commun.}\ }\textbf {\bibinfo {volume} {5}},\ \bibinfo {pages} {5186} (\bibinfo {year} {2014})}\BibitemShut {NoStop}%
\bibitem [{\citenamefont {Sharafiev}\ \emph {et~al.}(2024)\citenamefont {Sharafiev}, \citenamefont {Juan}, \citenamefont {Cattaneo},\ and\ \citenamefont {Kirchmair}}]{Sharafiev2024}%
  \BibitemOpen
  \bibfield  {author} {\bibinfo {author} {\bibfnamefont {Aleksei}\ \bibnamefont {Sharafiev}}, \bibinfo {author} {\bibfnamefont {Mathieu}\ \bibnamefont {Juan}}, \bibinfo {author} {\bibfnamefont {Marco}\ \bibnamefont {Cattaneo}}, \ and\ \bibinfo {author} {\bibfnamefont {Gerhard}\ \bibnamefont {Kirchmair}},\ }\href {http://arxiv.org/abs/2407.05958} {\enquote {\bibinfo {title} {{Leveraging collective effects for thermometry in waveguide quantum electrodynamics}},}\ } (\bibinfo {year} {2024}),\ \bibinfo {note} {to appear in \textit{Phys. Rev. Lett.}},\ \Eprint {http://arxiv.org/abs/2407.0595} {arXiv:2407.0595 [quant-ph]} \BibitemShut {NoStop}%
\bibitem [{\citenamefont {Braunstein}\ and\ \citenamefont {Caves}(1994)}]{PhysRevLett.72.3439}%
  \BibitemOpen
  \bibfield  {author} {\bibinfo {author} {\bibfnamefont {Samuel~L.}\ \bibnamefont {Braunstein}}\ and\ \bibinfo {author} {\bibfnamefont {Carlton~M.}\ \bibnamefont {Caves}},\ }\bibfield  {title} {\enquote {\bibinfo {title} {Statistical distance and the geometry of quantum states},}\ }\href {\doibase 10.1103/PhysRevLett.72.3439} {\bibfield  {journal} {\bibinfo  {journal} {Phys. Rev. Lett.}\ }\textbf {\bibinfo {volume} {72}},\ \bibinfo {pages} {3439--3443} (\bibinfo {year} {1994})}\BibitemShut {NoStop}%
\bibitem [{\citenamefont {Dittmann}(1999)}]{Dittmann_1999}%
  \BibitemOpen
  \bibfield  {author} {\bibinfo {author} {\bibfnamefont {J}~\bibnamefont {Dittmann}},\ }\bibfield  {title} {\enquote {\bibinfo {title} {Explicit formulae for the bures metric},}\ }\href {\doibase 10.1088/0305-4470/32/14/007} {\bibfield  {journal} {\bibinfo  {journal} {J. Phys. A: Math. Gen.}\ }\textbf {\bibinfo {volume} {32}},\ \bibinfo {pages} {2663} (\bibinfo {year} {1999})}\BibitemShut {NoStop}%
\bibitem [{\citenamefont {Zhong}\ \emph {et~al.}(2013)\citenamefont {Zhong}, \citenamefont {Sun}, \citenamefont {Ma}, \citenamefont {Wang},\ and\ \citenamefont {Nori}}]{PhysRevA.87.022337}%
  \BibitemOpen
  \bibfield  {author} {\bibinfo {author} {\bibfnamefont {Wei}\ \bibnamefont {Zhong}}, \bibinfo {author} {\bibfnamefont {Zhe}\ \bibnamefont {Sun}}, \bibinfo {author} {\bibfnamefont {Jian}\ \bibnamefont {Ma}}, \bibinfo {author} {\bibfnamefont {Xiaoguang}\ \bibnamefont {Wang}}, \ and\ \bibinfo {author} {\bibfnamefont {Franco}\ \bibnamefont {Nori}},\ }\bibfield  {title} {\enquote {\bibinfo {title} {Fisher information under decoherence in bloch representation},}\ }\href {\doibase 10.1103/PhysRevA.87.022337} {\bibfield  {journal} {\bibinfo  {journal} {Phys. Rev. A}\ }\textbf {\bibinfo {volume} {87}},\ \bibinfo {pages} {022337} (\bibinfo {year} {2013})}\BibitemShut {NoStop}%
\bibitem [{\citenamefont {Breuer}\ and\ \citenamefont {Petruccione}(2007)}]{Breuer}%
  \BibitemOpen
  \bibfield  {author} {\bibinfo {author} {\bibfnamefont {Heinz-Peter}\ \bibnamefont {Breuer}}\ and\ \bibinfo {author} {\bibfnamefont {Francesco}\ \bibnamefont {Petruccione}},\ }\href {\doibase 10.1093/acprof:oso/9780199213900.001.0001} {\emph {\bibinfo {title} {{The Theory of Open Quantum Systems}}}}\ (\bibinfo  {publisher} {Oxford University Press},\ \bibinfo {year} {2007})\BibitemShut {NoStop}%
\bibitem [{\citenamefont {Jeske}\ \emph {et~al.}(2015)\citenamefont {Jeske}, \citenamefont {Ing}, \citenamefont {Plenio}, \citenamefont {Huelga},\ and\ \citenamefont {Cole}}]{Jeske2015}%
  \BibitemOpen
  \bibfield  {author} {\bibinfo {author} {\bibfnamefont {Jan}\ \bibnamefont {Jeske}}, \bibinfo {author} {\bibfnamefont {David~J.}\ \bibnamefont {Ing}}, \bibinfo {author} {\bibfnamefont {Martin~B.}\ \bibnamefont {Plenio}}, \bibinfo {author} {\bibfnamefont {Susana~F.}\ \bibnamefont {Huelga}}, \ and\ \bibinfo {author} {\bibfnamefont {Jared~H.}\ \bibnamefont {Cole}},\ }\bibfield  {title} {\enquote {\bibinfo {title} {{Bloch-Redfield equations for modeling light-harvesting complexes}},}\ }\href {\doibase 10.1063/1.4907370} {\bibfield  {journal} {\bibinfo  {journal} {J. Chem. Phys.}\ }\textbf {\bibinfo {volume} {142}},\ \bibinfo {pages} {064104} (\bibinfo {year} {2015})}\BibitemShut {NoStop}%
\bibitem [{\citenamefont {Farina}\ and\ \citenamefont {Giovannetti}(2019)}]{Farina2019}%
  \BibitemOpen
  \bibfield  {author} {\bibinfo {author} {\bibfnamefont {Donato}\ \bibnamefont {Farina}}\ and\ \bibinfo {author} {\bibfnamefont {Vittorio}\ \bibnamefont {Giovannetti}},\ }\bibfield  {title} {\enquote {\bibinfo {title} {{Open-quantum-system dynamics: Recovering positivity of the Redfield equation via the partial secular approximation}},}\ }\href {\doibase 10.1103/PhysRevA.100.012107} {\bibfield  {journal} {\bibinfo  {journal} {Phys. Rev. A}\ }\textbf {\bibinfo {volume} {100}},\ \bibinfo {pages} {012107} (\bibinfo {year} {2019})}\BibitemShut {NoStop}%
\bibitem [{\citenamefont {Ptaszy\ifmmode~\acute{n}\else \'{n}\fi{}ski}\ and\ \citenamefont {Esposito}(2019)}]{PhysRevLett.122.150603}%
  \BibitemOpen
  \bibfield  {author} {\bibinfo {author} {\bibfnamefont {Krzysztof}\ \bibnamefont {Ptaszy\ifmmode~\acute{n}\else \'{n}\fi{}ski}}\ and\ \bibinfo {author} {\bibfnamefont {Massimiliano}\ \bibnamefont {Esposito}},\ }\bibfield  {title} {\enquote {\bibinfo {title} {Thermodynamics of quantum information flows},}\ }\href {\doibase 10.1103/PhysRevLett.122.150603} {\bibfield  {journal} {\bibinfo  {journal} {Phys. Rev. Lett.}\ }\textbf {\bibinfo {volume} {122}},\ \bibinfo {pages} {150603} (\bibinfo {year} {2019})}\BibitemShut {NoStop}%
\bibitem [{\citenamefont {McCauley}\ \emph {et~al.}(2020)\citenamefont {McCauley}, \citenamefont {Cruikshank}, \citenamefont {Bondar},\ and\ \citenamefont {Jacobs}}]{McCauley2020a}%
  \BibitemOpen
  \bibfield  {author} {\bibinfo {author} {\bibfnamefont {Gavin}\ \bibnamefont {McCauley}}, \bibinfo {author} {\bibfnamefont {Benjamin}\ \bibnamefont {Cruikshank}}, \bibinfo {author} {\bibfnamefont {Denys~I.}\ \bibnamefont {Bondar}}, \ and\ \bibinfo {author} {\bibfnamefont {Kurt}\ \bibnamefont {Jacobs}},\ }\bibfield  {title} {\enquote {\bibinfo {title} {{Accurate Lindblad-form master equation for weakly damped quantum systems across all regimes}},}\ }\href {\doibase 10.1038/s41534-020-00299-6} {\bibfield  {journal} {\bibinfo  {journal} {npj Quantum Info.}\ }\textbf {\bibinfo {volume} {6}},\ \bibinfo {pages} {74} (\bibinfo {year} {2020})}\BibitemShut {NoStop}%
\bibitem [{\citenamefont {Nathan}\ and\ \citenamefont {Rudner}(2020)}]{Nathan2020a}%
  \BibitemOpen
  \bibfield  {author} {\bibinfo {author} {\bibfnamefont {Frederik}\ \bibnamefont {Nathan}}\ and\ \bibinfo {author} {\bibfnamefont {Mark~S.}\ \bibnamefont {Rudner}},\ }\bibfield  {title} {\enquote {\bibinfo {title} {{Universal Lindblad equation for open quantum systems}},}\ }\href {\doibase 10.1103/physrevb.102.115109} {\bibfield  {journal} {\bibinfo  {journal} {Phys. Rev. B}\ }\textbf {\bibinfo {volume} {102}},\ \bibinfo {pages} {115109} (\bibinfo {year} {2020})}\BibitemShut {NoStop}%
\bibitem [{\citenamefont {Trushechkin}(2021)}]{PhysRevA.103.062226}%
  \BibitemOpen
  \bibfield  {author} {\bibinfo {author} {\bibfnamefont {Anton}\ \bibnamefont {Trushechkin}},\ }\bibfield  {title} {\enquote {\bibinfo {title} {Unified gorini-kossakowski-lindblad-sudarshan quantum master equation beyond the secular approximation},}\ }\href {\doibase 10.1103/PhysRevA.103.062226} {\bibfield  {journal} {\bibinfo  {journal} {Phys. Rev. A}\ }\textbf {\bibinfo {volume} {103}},\ \bibinfo {pages} {062226} (\bibinfo {year} {2021})}\BibitemShut {NoStop}%
\bibitem [{\citenamefont {Zhang}\ \emph {et~al.}(2022{\natexlab{b}})\citenamefont {Zhang}, \citenamefont {Yu}, \citenamefont {Yuan}, \citenamefont {Wang}, \citenamefont {Demkowicz-Dobrza\ifmmode~\acute{n}\else \'{n}\fi{}ski},\ and\ \citenamefont {Liu}}]{PhysRevResearch.4.043057}%
  \BibitemOpen
  \bibfield  {author} {\bibinfo {author} {\bibfnamefont {Mao}\ \bibnamefont {Zhang}}, \bibinfo {author} {\bibfnamefont {Huai-Ming}\ \bibnamefont {Yu}}, \bibinfo {author} {\bibfnamefont {Haidong}\ \bibnamefont {Yuan}}, \bibinfo {author} {\bibfnamefont {Xiaoguang}\ \bibnamefont {Wang}}, \bibinfo {author} {\bibfnamefont {Rafa\l{}}\ \bibnamefont {Demkowicz-Dobrza\ifmmode~\acute{n}\else \'{n}\fi{}ski}}, \ and\ \bibinfo {author} {\bibfnamefont {Jing}\ \bibnamefont {Liu}},\ }\bibfield  {title} {\enquote {\bibinfo {title} {Quanestimation: An open-source toolkit for quantum parameter estimation},}\ }\href {\doibase 10.1103/PhysRevResearch.4.043057} {\bibfield  {journal} {\bibinfo  {journal} {Phys. Rev. Res.}\ }\textbf {\bibinfo {volume} {4}},\ \bibinfo {pages} {043057} (\bibinfo {year} {2022}{\natexlab{b}})}\BibitemShut {NoStop}%
\bibitem [{\citenamefont {Ficek}\ \emph {et~al.}(1987)\citenamefont {Ficek}, \citenamefont {Tana{\'{s}}},\ and\ \citenamefont {Kielich}}]{Ficek1987}%
  \BibitemOpen
  \bibfield  {author} {\bibinfo {author} {\bibfnamefont {Z.}~\bibnamefont {Ficek}}, \bibinfo {author} {\bibfnamefont {R.}~\bibnamefont {Tana{\'{s}}}}, \ and\ \bibinfo {author} {\bibfnamefont {S.}~\bibnamefont {Kielich}},\ }\bibfield  {title} {\enquote {\bibinfo {title} {{Quantum beats and superradiant effects in the spontaneous emission from two nonidentical atoms}},}\ }\href {\doibase 10.1016/0378-4371(87)90280-9} {\bibfield  {journal} {\bibinfo  {journal} {Physica A: Statistical Mechanics and its Applications}\ }\textbf {\bibinfo {volume} {146}},\ \bibinfo {pages} {452--482} (\bibinfo {year} {1987})}\BibitemShut {NoStop}%
\bibitem [{\citenamefont {Gebbia}\ \emph {et~al.}(2020)\citenamefont {Gebbia}, \citenamefont {Benedetti}, \citenamefont {Benatti}, \citenamefont {Floreanini}, \citenamefont {Bina},\ and\ \citenamefont {Paris}}]{PhysRevA.101.032112}%
  \BibitemOpen
  \bibfield  {author} {\bibinfo {author} {\bibfnamefont {Francesca}\ \bibnamefont {Gebbia}}, \bibinfo {author} {\bibfnamefont {Claudia}\ \bibnamefont {Benedetti}}, \bibinfo {author} {\bibfnamefont {Fabio}\ \bibnamefont {Benatti}}, \bibinfo {author} {\bibfnamefont {Roberto}\ \bibnamefont {Floreanini}}, \bibinfo {author} {\bibfnamefont {Matteo}\ \bibnamefont {Bina}}, \ and\ \bibinfo {author} {\bibfnamefont {Matteo G.~A.}\ \bibnamefont {Paris}},\ }\bibfield  {title} {\enquote {\bibinfo {title} {Two-qubit quantum probes for the temperature of an ohmic environment},}\ }\href {\doibase 10.1103/PhysRevA.101.032112} {\bibfield  {journal} {\bibinfo  {journal} {Phys. Rev. A}\ }\textbf {\bibinfo {volume} {101}},\ \bibinfo {pages} {032112} (\bibinfo {year} {2020})}\BibitemShut {NoStop}%
\bibitem [{\citenamefont {Solenov}\ \emph {et~al.}(2007)\citenamefont {Solenov}, \citenamefont {Tolkunov},\ and\ \citenamefont {Privman}}]{Solenov2007}%
  \BibitemOpen
  \bibfield  {author} {\bibinfo {author} {\bibfnamefont {Dmitry}\ \bibnamefont {Solenov}}, \bibinfo {author} {\bibfnamefont {Denis}\ \bibnamefont {Tolkunov}}, \ and\ \bibinfo {author} {\bibfnamefont {Vladimir}\ \bibnamefont {Privman}},\ }\bibfield  {title} {\enquote {\bibinfo {title} {{Exchange interaction, entanglement, and quantum noise due to a thermal bosonic field}},}\ }\href {\doibase 10.1103/PhysRevB.75.035134} {\bibfield  {journal} {\bibinfo  {journal} {Phys. Rev. B}\ }\textbf {\bibinfo {volume} {75}},\ \bibinfo {pages} {035134} (\bibinfo {year} {2007})}\BibitemShut {NoStop}%
\bibitem [{\citenamefont {Cattaneo}\ \emph {et~al.}(2021)\citenamefont {Cattaneo}, \citenamefont {Giorgi}, \citenamefont {Maniscalco}, \citenamefont {Paraoanu},\ and\ \citenamefont {Zambrini}}]{Cattaneo2021collective}%
  \BibitemOpen
  \bibfield  {author} {\bibinfo {author} {\bibfnamefont {Marco}\ \bibnamefont {Cattaneo}}, \bibinfo {author} {\bibfnamefont {Gian~Luca}\ \bibnamefont {Giorgi}}, \bibinfo {author} {\bibfnamefont {Sabrina}\ \bibnamefont {Maniscalco}}, \bibinfo {author} {\bibfnamefont {Gheorghe~Sorin}\ \bibnamefont {Paraoanu}}, \ and\ \bibinfo {author} {\bibfnamefont {Roberta}\ \bibnamefont {Zambrini}},\ }\bibfield  {title} {\enquote {\bibinfo {title} {{Bath‐Induced Collective Phenomena on Superconducting Qubits: Synchronization, Subradiance, and Entanglement Generation}},}\ }\href {\doibase 10.1002/andp.202100038} {\bibfield  {journal} {\bibinfo  {journal} {Ann. Phys. (Berlin)}\ }\textbf {\bibinfo {volume} {533}},\ \bibinfo {pages} {2100038} (\bibinfo {year} {2021})}\BibitemShut {NoStop}%
\end{thebibliography}%
\end{document}